\documentclass{article}
\usepackage[utf8]{inputenc}

\usepackage{newtxtext} 
\usepackage[scaled=.95]{cabin} 
\usepackage[varqu,varl]{inconsolata} 
\usepackage[title]{appendix}
\usepackage[T1]{fontenc}  

\usepackage{microtype}
\usepackage{geometry}
\usepackage{subfig}
\usepackage[
backend=biber,
style=apa,
citestyle=authoryear,
maxcitenames=2
]{biblatex}
\usepackage{graphicx}
\usepackage{csvsimple}
\usepackage{amsmath}
\usepackage{amssymb}
\usepackage{verbatim}
\usepackage{tabularx}
\usepackage{tikz}
\usetikzlibrary{positioning}
\usetikzlibrary{calc}
\usepackage[colorlinks,linkcolor=blue,citecolor=blue,urlcolor=blue,filecolor=blue,backref=false]{hyperref}

\usepackage{etoolbox}
\apptocmd{\appendices}{\apptocmd{\thesection}{: }{}{}}{}{}

\DeclareMathOperator{\Bin}{\mathrm{Bin}}
\DeclareMathOperator{\normal}{\mathrm{normal}}
\DeclareMathOperator{\MHyp}{\mathrm{MHyp}}
\DeclareMathOperator{\Hyp}{\mathrm{Hyp}}
\DeclareMathOperator{\given}{\;|\;}
\DeclareMathOperator*{\argmin}{arg\,min}

\title{Graphical Test for Discrete Uniformity and its Applications in Goodness of Fit Evaluation and Multiple Sample Comparison}
\author{Teemu Säilynoja$^{1 \star}$ , Paul-Christian Bürkner$^2$ , Aki Vehtari$^1$ }
\date{%
    $^1$ Department of Computer Science, Aalto University, Finland\\
    $^2$ Cluster of Excellence SimTech, University of Stuttgart, Germany \\
    $^\star$ Corresponding author, Email: teemu.sailynoja@aalto.fi
}

\begin{document}

\maketitle

\begin{abstract}
    Assessing goodness of fit to a given distribution plays an important role in computational statistics. The \emph{Probability integral transformation} (PIT) can be used to convert the question of whether a given sample originates from a reference distribution into a problem of testing for uniformity. We present new simulation and optimization based methods to obtain simultaneous confidence bands for the whole \emph{empirical cumulative distribution function} (ECDF) of the PIT values under the assumption of uniformity. Simultaneous confidence bands correspond to such confidence intervals at each point that jointly satisfy a desired coverage. These methods can also be applied in cases where the reference distribution is represented only by a finite sample, which is useful, for example, for simulation based calibration. The confidence bands provide an intuitive ECDF-based graphical test for uniformity, which also provides useful information on the quality of the discrepancy.
    We further extend the simulation and optimization methods to determine simultaneous confidence bands for testing whether multiple samples come from the same underlying distribution. This multiple sample comparison test is useful, for example, as a complementary diagnostic in multi-chain Markov chain Monte Carlo (MCMC) convergence diagnostics, where most currently used convergence diagnostics provide a single diagnostic value, but do not usually offer insight into the nature of the deviation.
    We provide numerical experiments to assess the properties of the tests using both simulated and real world data and give recommendations on their practical application in computational statistics workflows.
\end{abstract}

{\small \hspace{2.8mm} \textbf{Keywords:} PIT, ECDF, uniformity test, simulation based calibration, MCMC convergence diagnostic}

\section{Introduction}
    Tests for uniformity play an essential role in computational statistics when estimating goodness of fit to a given distribution \parencite{Marhuenda2005}. This is because, even when the distribution of interest is not uniform, there are methods to reduce the problem into testing for uniformity by transforming a sample from the given distribution to a (discrete or continuous) uniform distribution. Common use cases in Bayesian workflow \parencite{BayesianWorkflow2020} are simulation based calibration and Markov chain Monte Carlo convergence diagnostic, which we also use as examples in this paper.
    A graphical test can provide additional insight to the nature of discrepancy that goes beyond the dichotomy of the uniformity test.

\subsection{Probability integral transformation}
    
    \begin{figure}
        \centering
        \begin{tikzpicture}[
        node/.style={minimum size=7mm}
        ]
        \node[node] (y_hist){
        \includegraphics[width=0.18\textwidth]{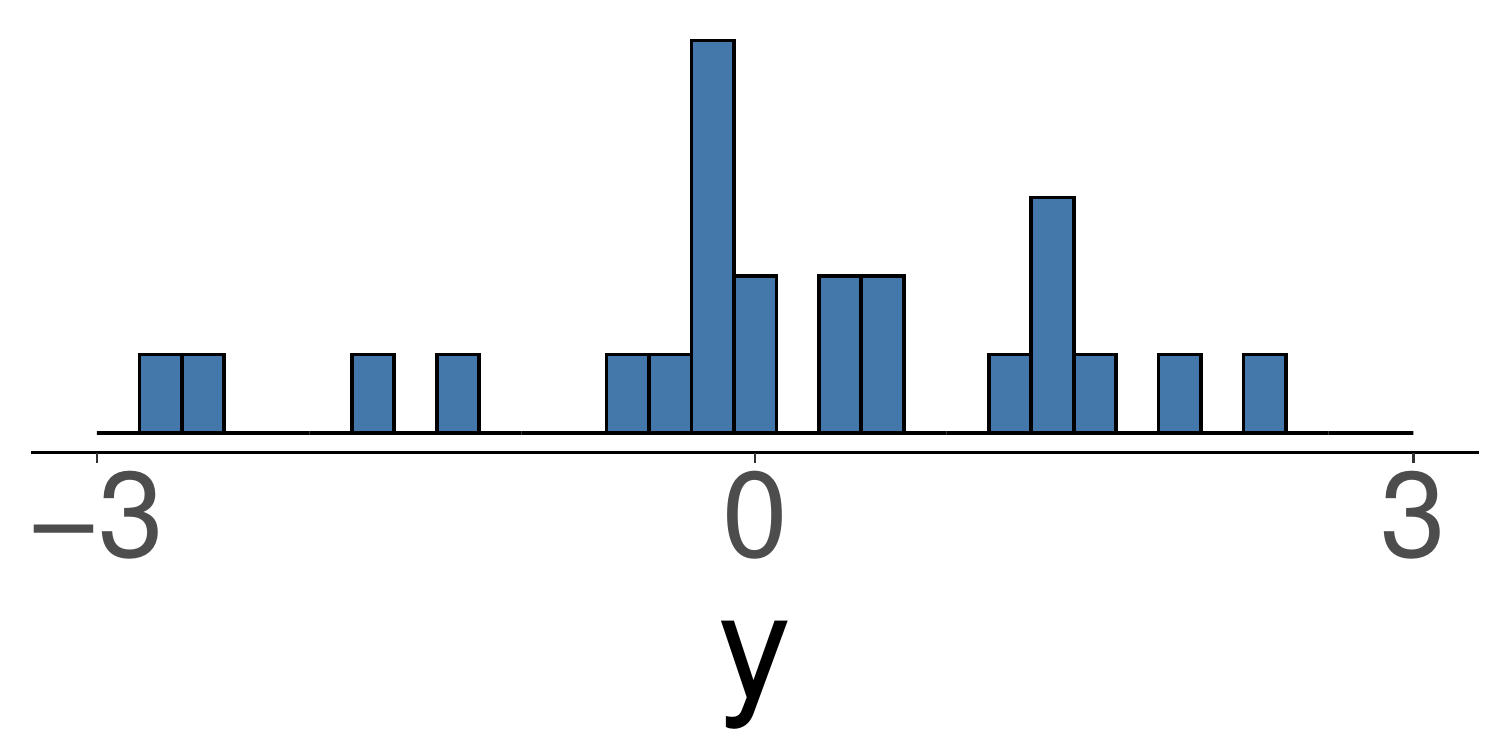}
        };
        \node[node] (density)   [above=0.0cm of y_hist] {
        \includegraphics[width=0.18\textwidth]{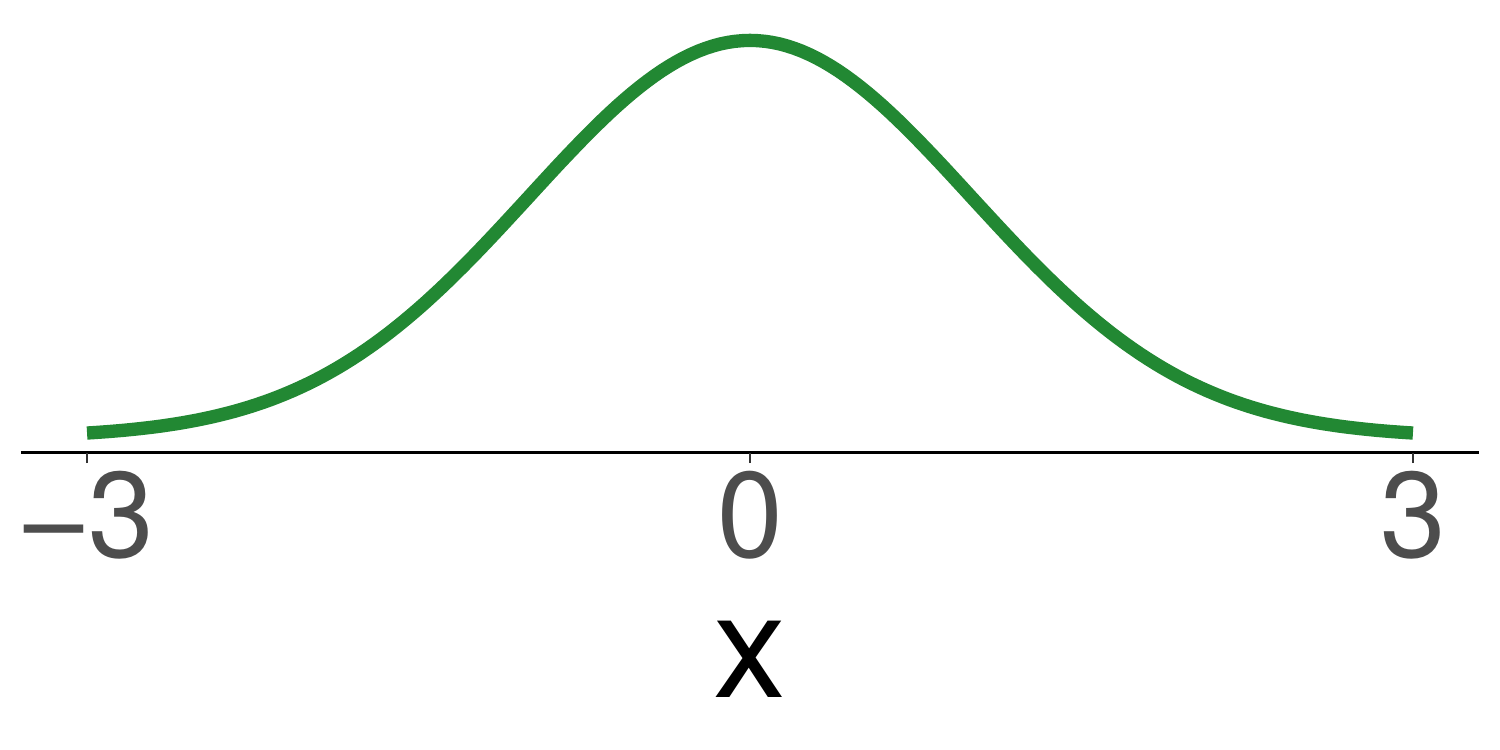}
        };
        \node[node] (x_hist1) [below=0.0cm of y_hist]{
        \includegraphics[width=0.18\textwidth]{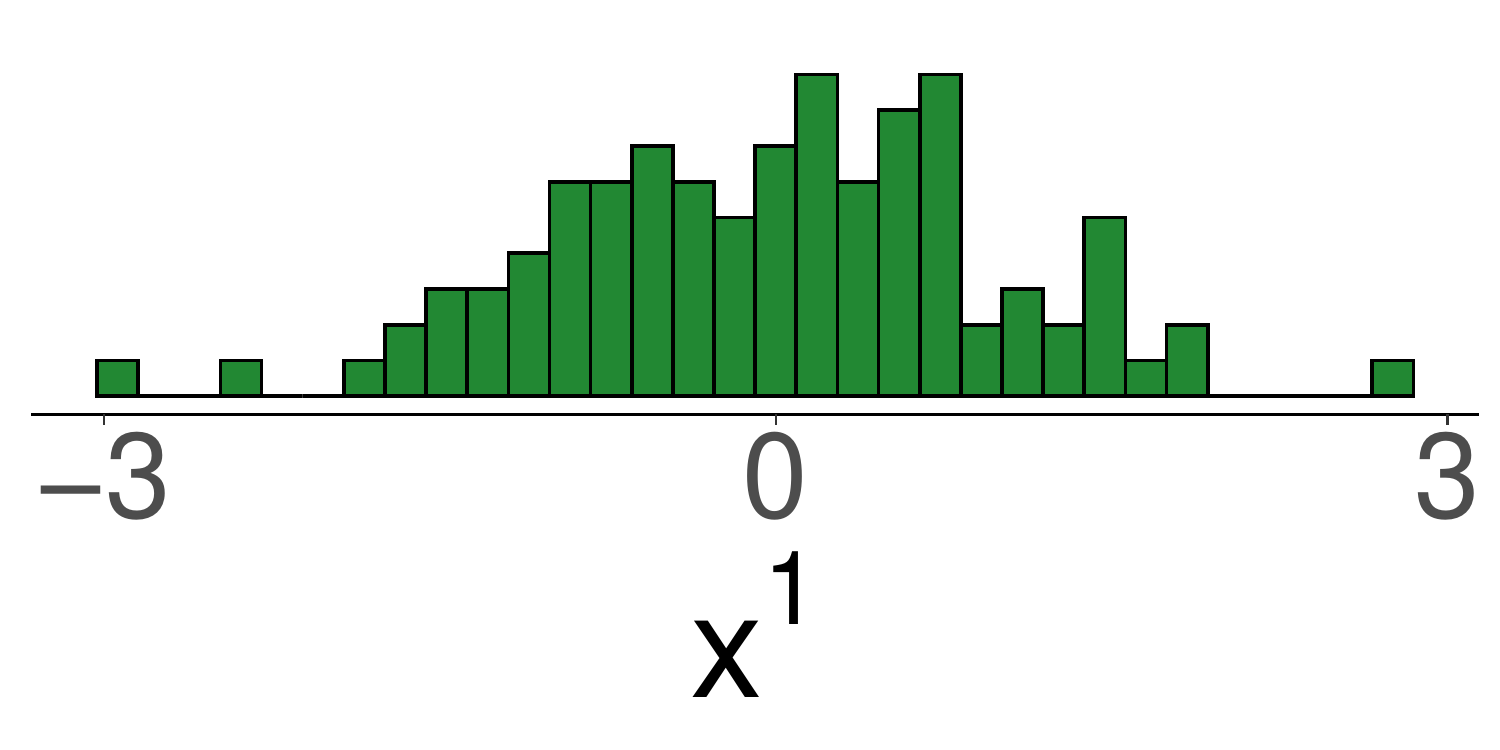}
        };
        \node[node] (x_hist2) [below=0.1cm of x_hist1]{
        \includegraphics[width=0.18\textwidth]{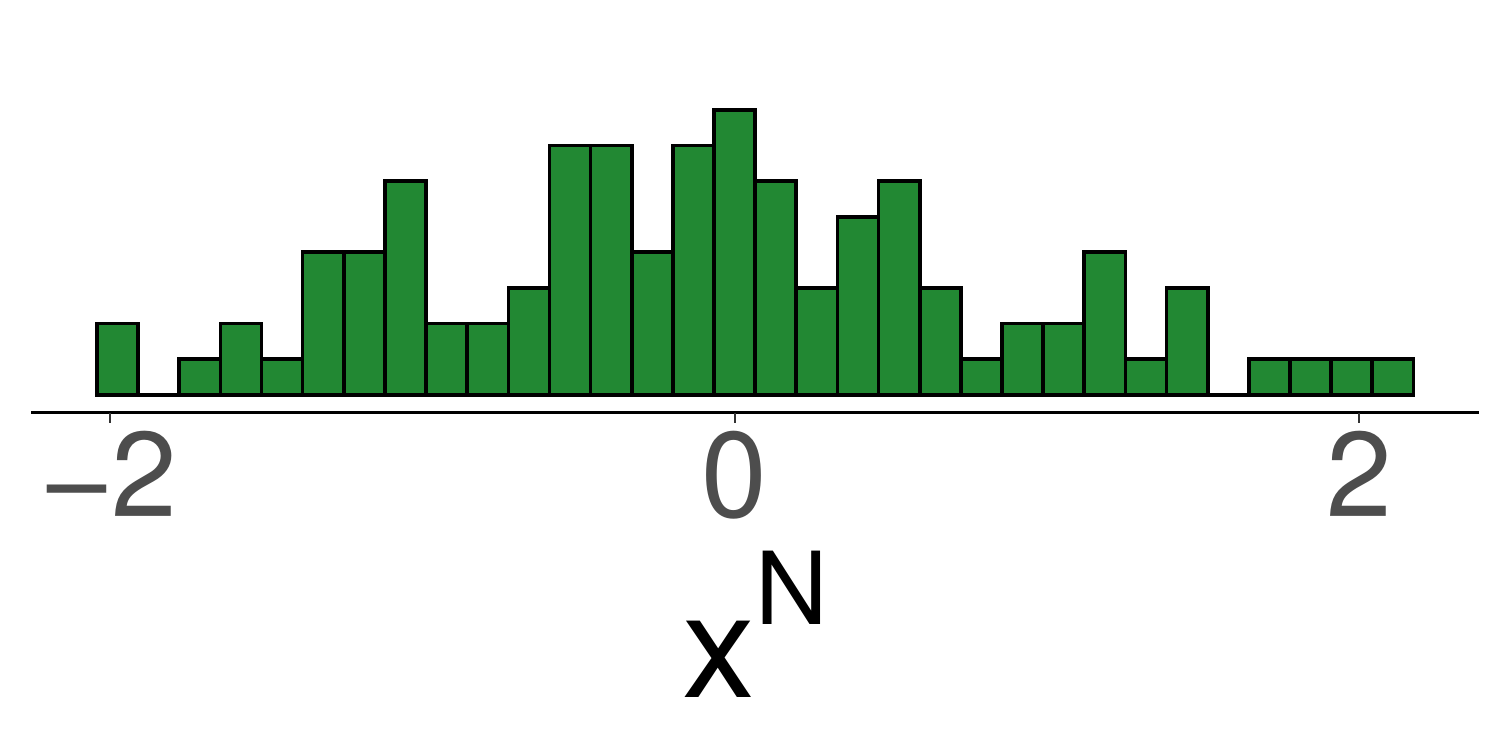}
        };
        \node at ($(x_hist1)!.5!(x_hist2)$) {\vdots};
        \node[node] (transformation) [above right=-0.4cm and 0.7cm of y_hist]{
        $\displaystyle \,u_i = \int_{-\infty}^{y_i}p(x)\,dx$
        };
        \node[node] (transformation_sum) [below right = 0.1cm and 0.7cm of  y_hist]{
        $\displaystyle \,u_i = \frac{1}{S}\sum_{j=1}^{S}\mathbb{I}(x^i_j \leq y_i)$
        };
        \node[node] (ecdf)  [right=13.5mm of transformation] {
        \includegraphics[width=0.2\textwidth]{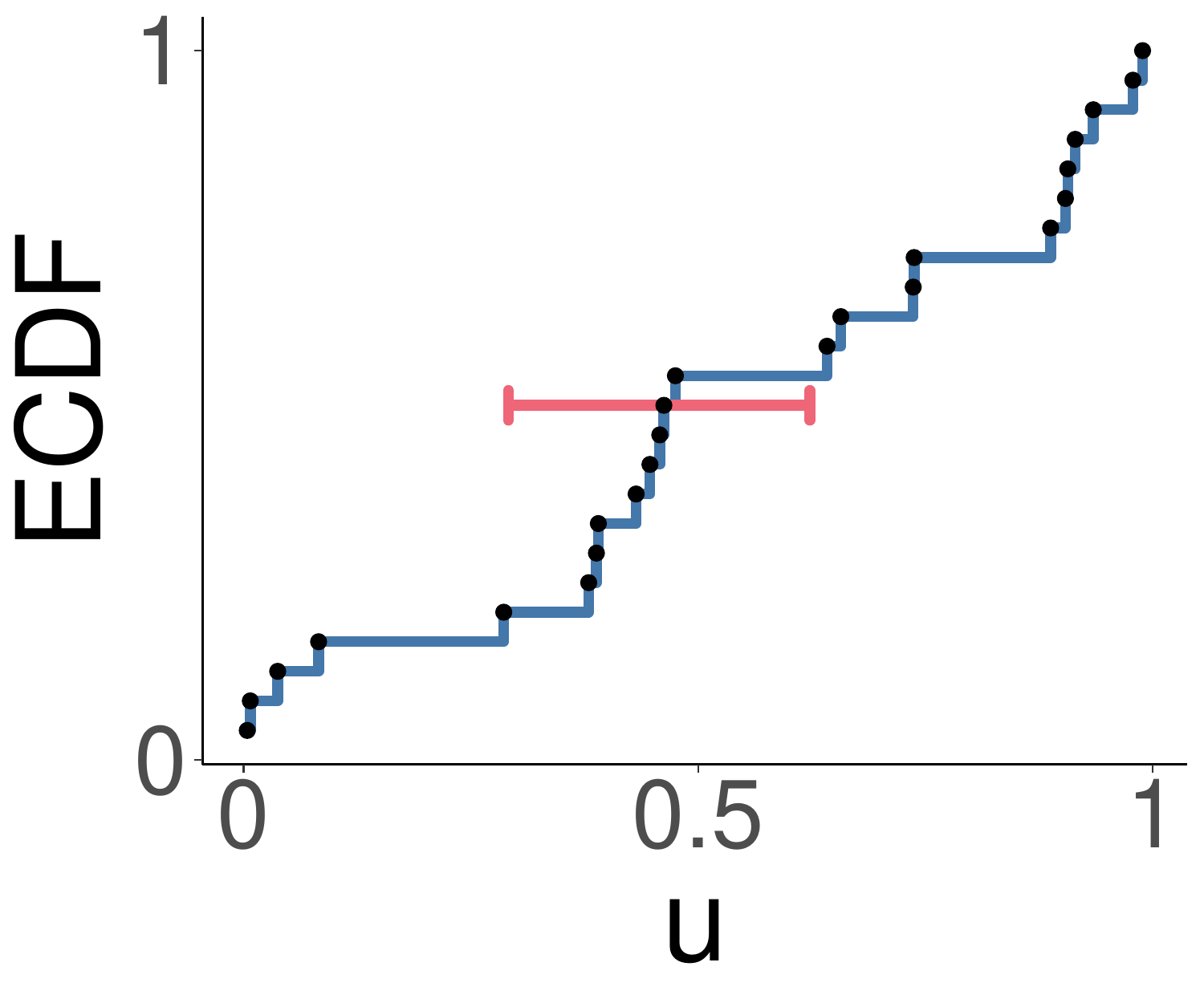}
        };
        \node[node] (ecdf_u) [below right=-1.8cm and 7mm of transformation_sum]{
        \includegraphics[width=0.2\textwidth]{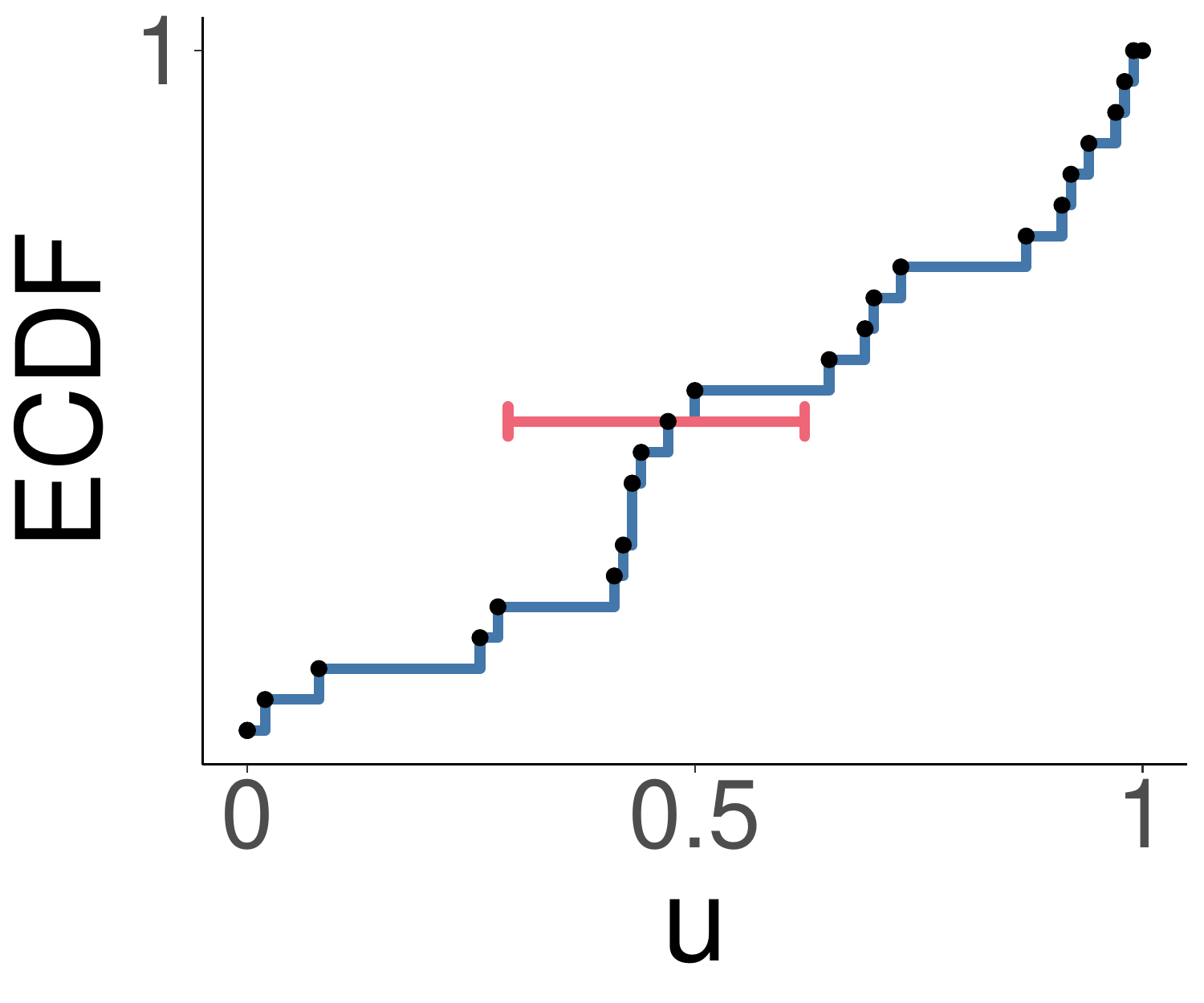}
        };
        \node[node] (ecdf_z) [right=.7cm of ecdf_u] {
        \includegraphics[width=0.2\textwidth]{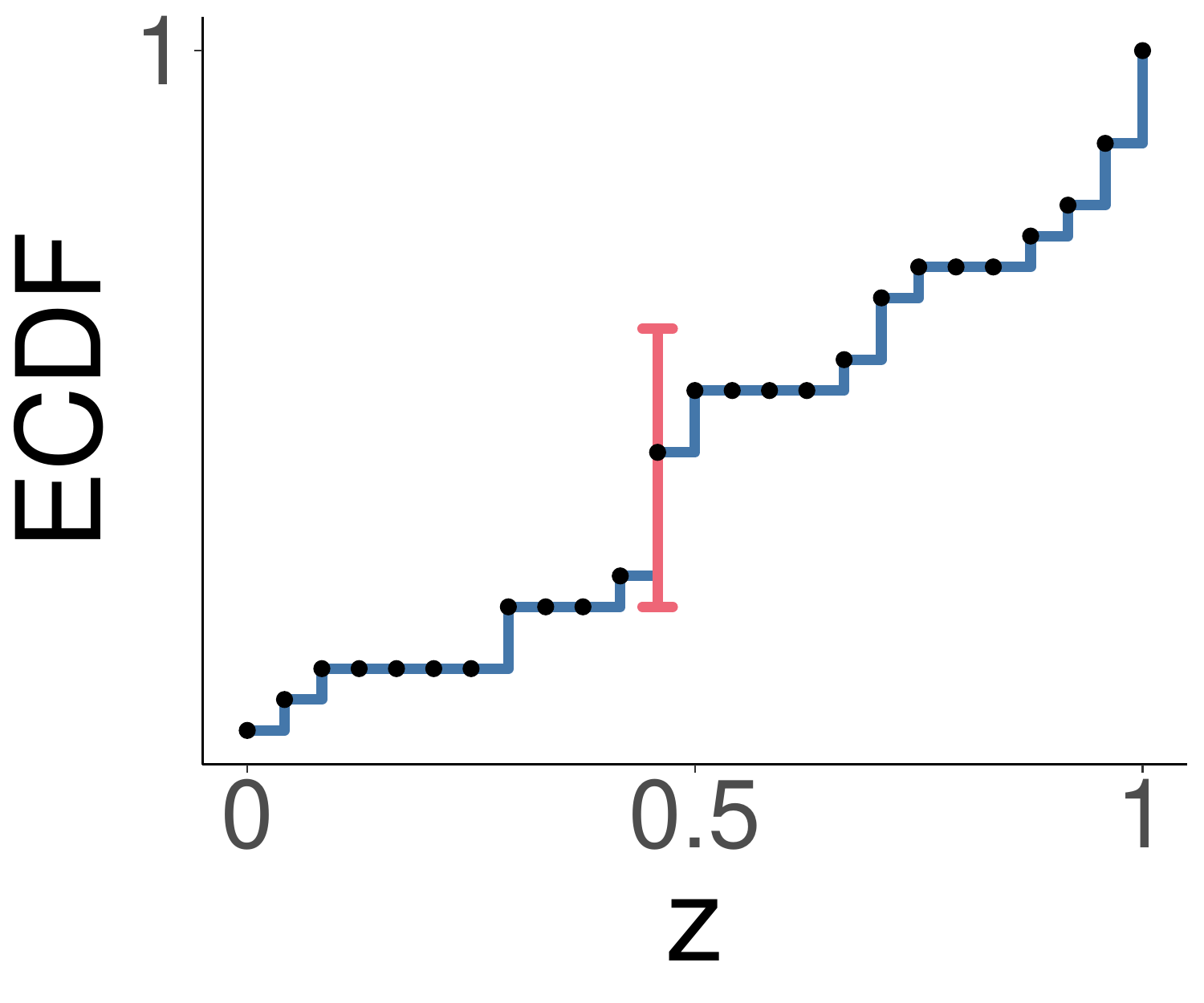}
        };
        \node[node] (1) [above left=-.5cm and -1.1cm of transformation]{(1)};
        \node[node] (2) [above left=-.6cm and -1.1cm of transformation_sum]{(2)};
        \node[node] (a) [above left=-.5cm and -1.3cm of ecdf]{(a)};
        \node[node] (b) [above left=-.5cm and -1.3cm of ecdf_u]{(b)};
        \node[node] (c) [above left=-.5cm and -1.3cm of ecdf_z]{(c)};
        \draw[->] (y_hist.10) .. controls +(right:7mm) and +(left:7mm) .. (transformation.185); 
        \draw[->] (density.east) .. controls +(right:7mm) and +(left:7mm) .. (transformation.175); 
        \draw[->] (transformation.2) -- (ecdf.178);
        \draw[->] (y_hist.350) .. controls +(right:7mm) and +(left:7mm) .. (transformation_sum.175);
        \draw[->] (x_hist2.25) .. controls +(right:7mm) and +(left:7mm) .. (transformation_sum.185);
        \draw[->] (transformation_sum.east) -- (ecdf_u.167);
        \draw[->] (ecdf_u.9) -- (ecdf_z.171);
        \end{tikzpicture}
        \caption{
        Given $y_1,\ldots, y_N \sim g(y)$, and a distribution $p(x)$, the hypothesis $g=p$ can be assessed in two ways. (1) If the CDF or the PDF of $p$ has a closed form, the PIT values $u_i$ are continuous and, if $g=p$, uniformly distributed. (a) The pointwise confidence intervals (red bars) for the ordered statistic $u_{(i)}$ are beta distributed and the simultaneous confidence intervals for the ECDF of $u_i$ are given by \textcite{Aldor-Noiman2013}. (2) If one can draw an independent sample, $x^i_1,\ldots, x^i_S \sim p(x)$, for each $y_i$, the empirical PIT values $u_i$ are discrete and, given $g=p$, uniformly distributed. (b) The pointwise confidence intervals of the discrete ordered statistic $u_{(i)}$ could be solved from Eq. \eqref{eq:discrete uniform ordered statistic CDF}.
        (c) The pointwise confidence intervals for the values of the ECDF of $u_i$ at evaluation points $z_i\in [0,1]$ are binomially distributed and the simultaneous confidence intervals are obtained by the method presented in this paper.
        }
        \label{fig:intro_pit_transforms}
     \end{figure}
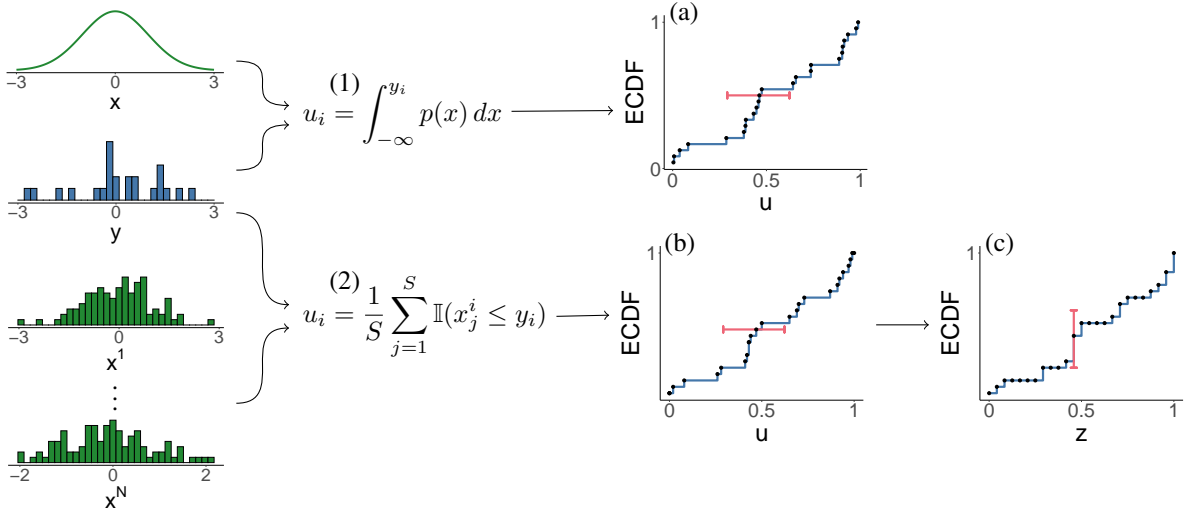
    
    Transforming sampled values to a uniform distribution is usually achieved via the \textit{probability integral transform} (PIT), provided the distribution of interest has a tractable \textit{cumulative distribution function} (CDF) \parencite{DAgostino1985}. Let $y_1,\ldots, y_N \sim g(y)$ be an independent sample from an unknown \textit{continuous} distribution with \textit{probability density function} (PDF) $g$. We want to know whether $g = p$, where $p$ is the PDF of a known distribution with a tractable CDF. The PIT of the sampled value $y_i$ with respect to $p$ is
    \begin{align}\label{eq: probability integral transformation}
        u_i = \int_{-\infty}^{y_i}p(x)\,dx.
    \end{align}
    If $g = p$, the transformed values $u_i$ are continuously, independently, and uniformly distributed on the unit interval $[0,1]$, reducing the evaluation of the hypothesis into testing for uniformity of the transformed sample $u_1, \ldots, u_N$. If the integral \eqref{eq: probability integral transformation} does not have closed from, the CDF (and hence the PIT values) can still be computed with sufficient accuracy through numerical integration (e.g., quadrature), if at least the corresponding PDF is tractable. 
    
    If neither the CDF nor the PDF have closed form, but a comparison sample of independent values $x^i_1, \ldots, x^i_S \sim p(x)$ can be drawn separately for each $y_i$, the hypothesis $g = p$ can be evaluated through the empirical PIT values
    \begin{align}\label{eq: rank statistic}
        u_i = \frac{1}{S}\sum_{j=1}^S\mathbb{I}(x^i_j\leq y_i),
    \end{align}
    where $\mathbb{I}$ is the indicator function. 
    Now, given $g = p$, the transformed values $u = u_1, \ldots, u_N$ are independently distributed according to a discrete uniform distribution with $S+1$ values
    $(0,1/S,\ldots,(S-1)/S,1)$. Accordingly, we can still apply uniformity tests to assess $g = p$, just that this time, we need to test for discrete uniformity.
    
    If either sample has dependencies, like an autocorrelated sample from a Markov chain, the ordered statistics is affected by the dependencies and the empirical PIT values \eqref{eq: rank statistic} are not distributed uniformly even if $g=p$ (unless the sample size goes to infinity). For Markov chains, the usual remedy is to thin the chain to obtain an approximately independent sample. This issue is illustrated, and thinning recommendations are provided in Appendix \hyperref[appendix: autocorrelated samples]{A}.
    
    Figure~\ref{fig:intro_pit_transforms} shows an example of the
    \emph{empirical cumulative distribution function} (ECDF) for $u$ obtained through both Eq.~\eqref{eq: probability integral transformation} and Eq.~\eqref{eq: rank statistic}. The figure also shows an example of a pointwise confidence interval for each ECDF. For the continuous integral of Eq.~\eqref{eq: probability integral transformation}, the pointwise confidence interval can be computed from the continuous uniform ordered statistics distribution which is a common beta distribution. For the discrete sum of Eq.~\eqref{eq: rank statistic}, the pointwise confidence interval can be computed from the discrete uniform ordered statistics distribution, with the cumulative distribution function of the $i$th ordered statistic $u_{(i)}$ given as
    \begin{equation}\label{eq:discrete uniform ordered statistic CDF}
    F_{i}(z) = \sum_{k=i}^N {N \choose k} z^k\left(1 - z\right)^{N-k},
    \end{equation}
    for $z \in (0,1/S,\ldots,(S-1)/S,1)$ \parencite[][Example 3.1]{arnold2008}. The corresponding pointwise intervals do not have a nice form in general and, more importantly, the discrete ordered statistics do not exhibit Markovian structure (exploited by our new optimization based approach) if there are possible ties in $u$ \parencite[][Theorem 3.4.1]{arnold2008}.
 
    To make the computation of the simultaneous confidence bands more straightforward and efficient, we propose making an additional transformation by computing the ECDF of $u$ at chosen evaluation points $z_i$:
    \begin{equation}\label{eq:empirical_cumulative_distribution}
    F(z_i) = \frac{1}{N}\sum_{j=1}^N\mathbb{I}(u_j \leq z_i).
    \end{equation}
    We recommend choosing $z_i$ as the ordered fractional ranks $\tilde{r}_i$ of $y_i$, defined as
    \begin{equation}\label{eq:fractional_ranks_def}
    \tilde{r}_i = \frac{1}{N}\sum_{j=1}^N\mathbb{I}(y_j\leq y_i).
    \end{equation}
    The ordered fractional ranks form a uniform partition of the unit interval independent of the distribution of $y_i$. Thus, they provide an ECDF that is easier to interpret than the corresponding ECDF based directly on the original sample $y_i$. The resulting ECDF is illustrated in Figure~\ref{fig:intro_pit_transforms}(c). As we will show, useful properties of this ECDF are that 1) its pointwise confidence intervals can be computed easily from the binomial distribution, with a quantile function already implemented in most widely used environments for statistical computing, and 2) the distribution of the ECDF trajectories is Markovian, which is exploited in Section~\ref{subsec:single_sample_optimization_method}.

\subsection{Simultaneous confidence bands}

    The major challenge that arises when developing a uniformity test based on the ECDF is to obtain simultaneous confidence bands with the desired overall coverage. For this purpose, one needs to take into account the inter-dependency in the ECDF values and adjust the coverage parameter accordingly (we will discuss this in more detail in Section \ref{sec:confidence_bands}).

When considering whether a given ECDF could present a sample from a uniform distribution, we need to jointly consider all pointwise uncertainties. 
For a set of evaluation points $(z_i)_{i=1}^K$, we provide lower and upper confidence bands $L_i$ and $U_i$ respectively, that jointly satisfy
    \begin{equation}
        \Pr \left(L_i \leq F(z_i) \leq U_i \text{ for all } i \in \{1,\dots,K\}\right) = 1 - \alpha,
    \end{equation}
    where $F(z_i)$ is the ECDF of a sample from either the standard uniform distribution or discrete uniform distribution on the unit interval evaluated at $z_i \in (0, 1)$ and $1 - \alpha$ is the desired simultaneous confidence level.
In addition to offering a numerical test for uniformity, the simultaneous confidence bands provide an intuitive graphical representation of possible discrepancies from uniformity.

\begin{figure}[t]
    \centering
    \subfloat[Method by Aldor-Noiman et al.]{\includegraphics[width=0.28\textwidth]{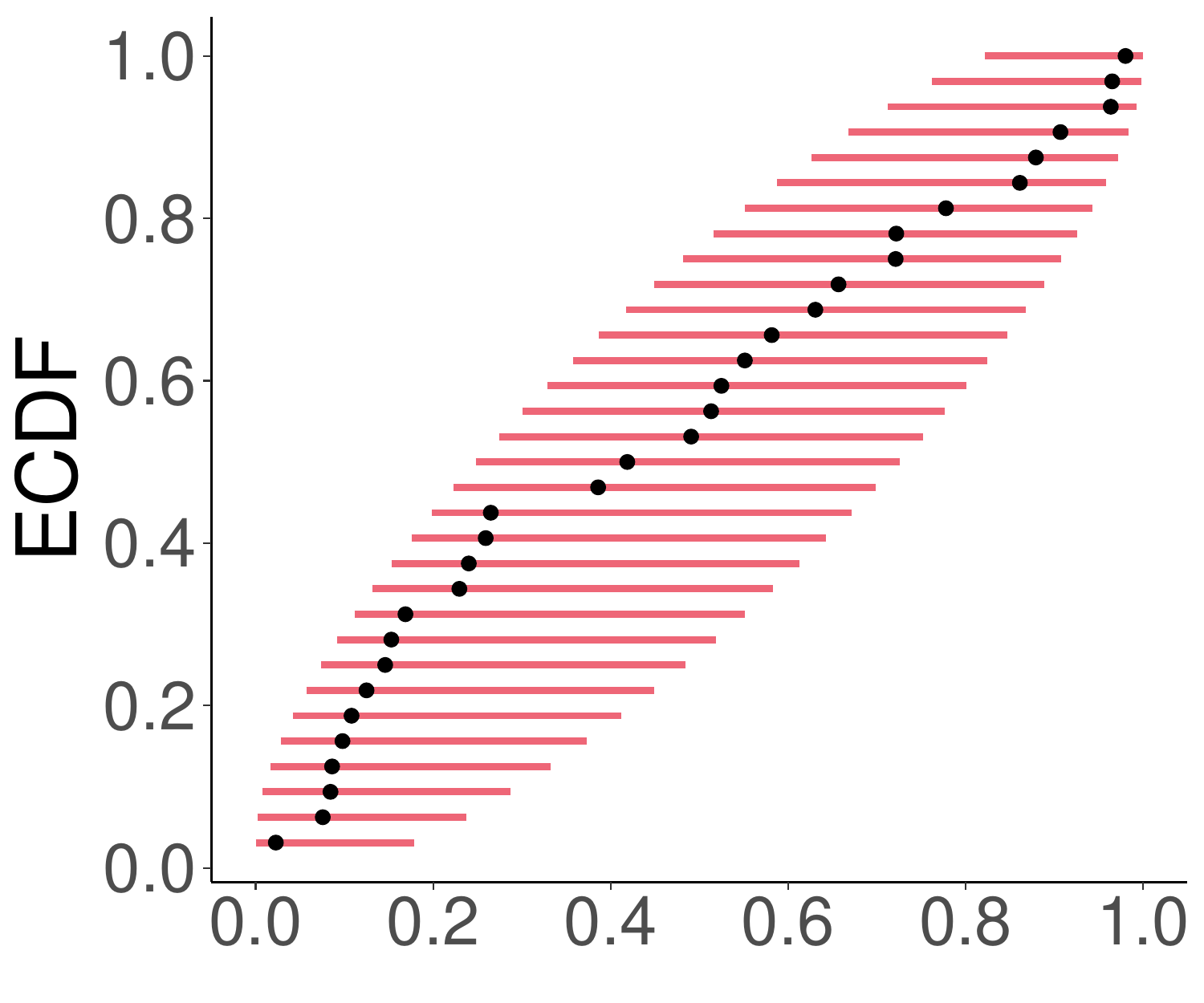}}
          \hspace{20mm}
         \subfloat[Our method]{\includegraphics[width=0.28\textwidth]{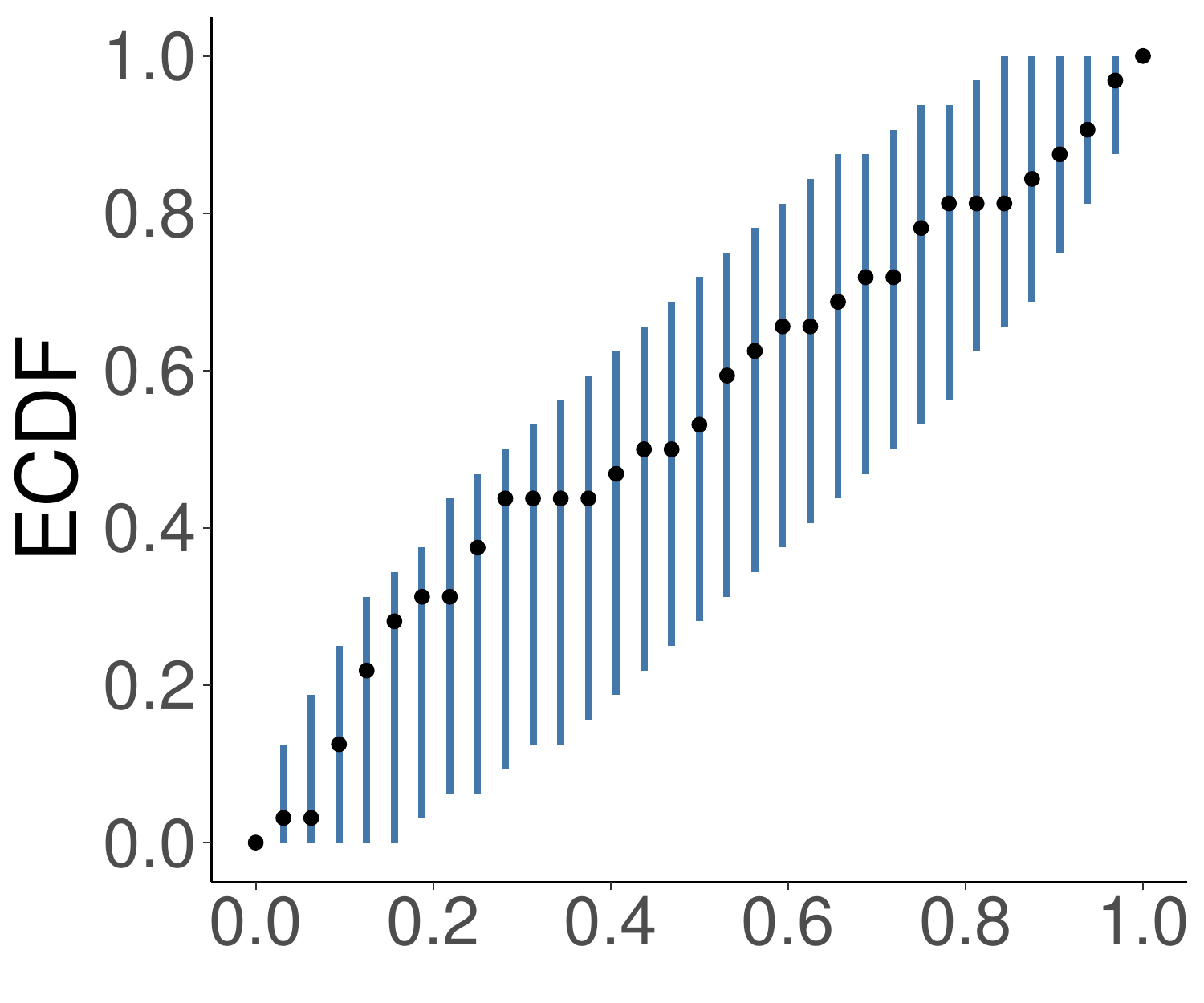}}
            \caption{Simultaneous confidence bands: (a) Method by Aldor-Noiman et al.: Beta distribution-based 95\% simultaneous confidence bands for quantiles are provided for reaching a set of ECDF values (along the x-axis).
            (b) Our method: For a set of evaluation quantiles, we provide binomial distribution-based 95\% simultaneous confidence intervals for the ECDF value (along the y-axis).}
            \label{fig:test_approach_differences}
        \end{figure}

    \textcite{Aldor-Noiman2013} presented a simulation-based approach for computing simultaneous confidence band for the ECDF of the transformed sample acquired from Eq.~\eqref{eq: probability integral transformation} under the assumption of uniformity. In this paper, we present a simulation method inspired by \textcite{Aldor-Noiman2013} as well as a new, faster optimization method for computing simultaneous confidence bands under uniformity, when the ECDF is computed from the empirical PIT values using Eqs.~\eqref{eq: rank statistic} and \eqref{eq:empirical_cumulative_distribution}. Figure \ref{fig:test_approach_differences} contrasts the simultaneous confidence bands by \textcite{Aldor-Noiman2013} against those obtained from our proposed method.
    Furthermore, we generalize our method and simultaneous confidence bands to test whether multiple samples originate from the same underlying distribution.
    
    \subsection{Related work}
    
    The idea of utilizing the ECDF to test uniformity is not new, but its potential has not yet been realized in full. For example,
    the well known Kolmogorov-Smirnov (KS) test, first introduced by Kolmogorov (see e.g. \textcite{massey1951kolmogorov}, original article in Italian is \textcite{kolmogorov1933sulla}), is based on evaluating the maximum deviation of the sample ECDF from the theoretical CDF of the distribution to be tested against.
    Unfortunately, the KS test is relatively insensitive to deviations in the tails of the distribution \parencite{Aldor-Noiman2013}, and numerous test have been proposed to replace the KS test. An extensive comparison of more than thirty tests of uniformity of a single sample is provided by \textcite{Marhuenda2005}.
    
\begin{figure}[t]
        \centering
        \subfloat[Histogram with 20 bins]{\label{fig:intro a}\includegraphics[width=0.245\textwidth]{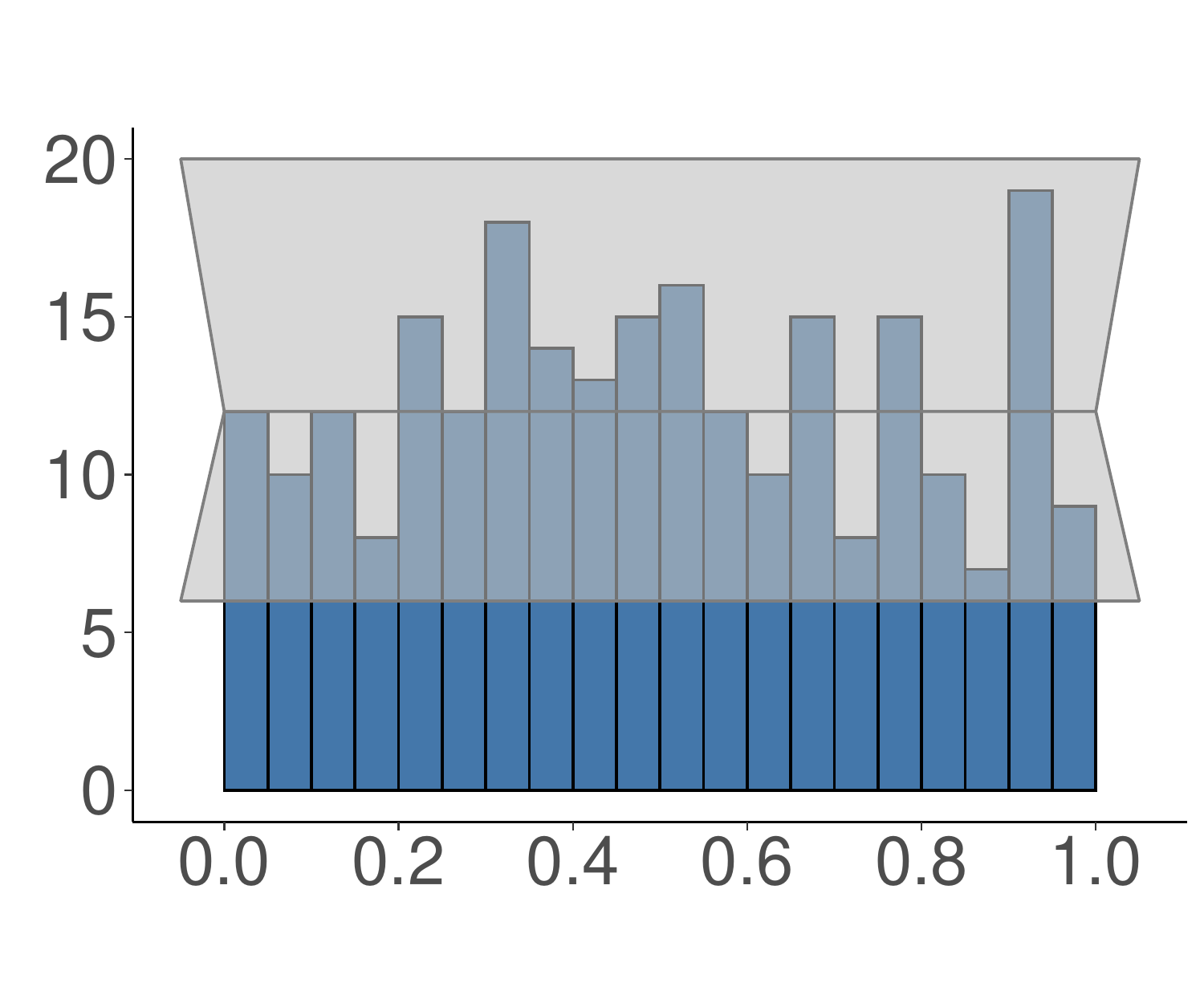}}
        \subfloat[Histogram with 50 bins]{\label{fig:intro b}%
        \includegraphics[width=0.245\textwidth]{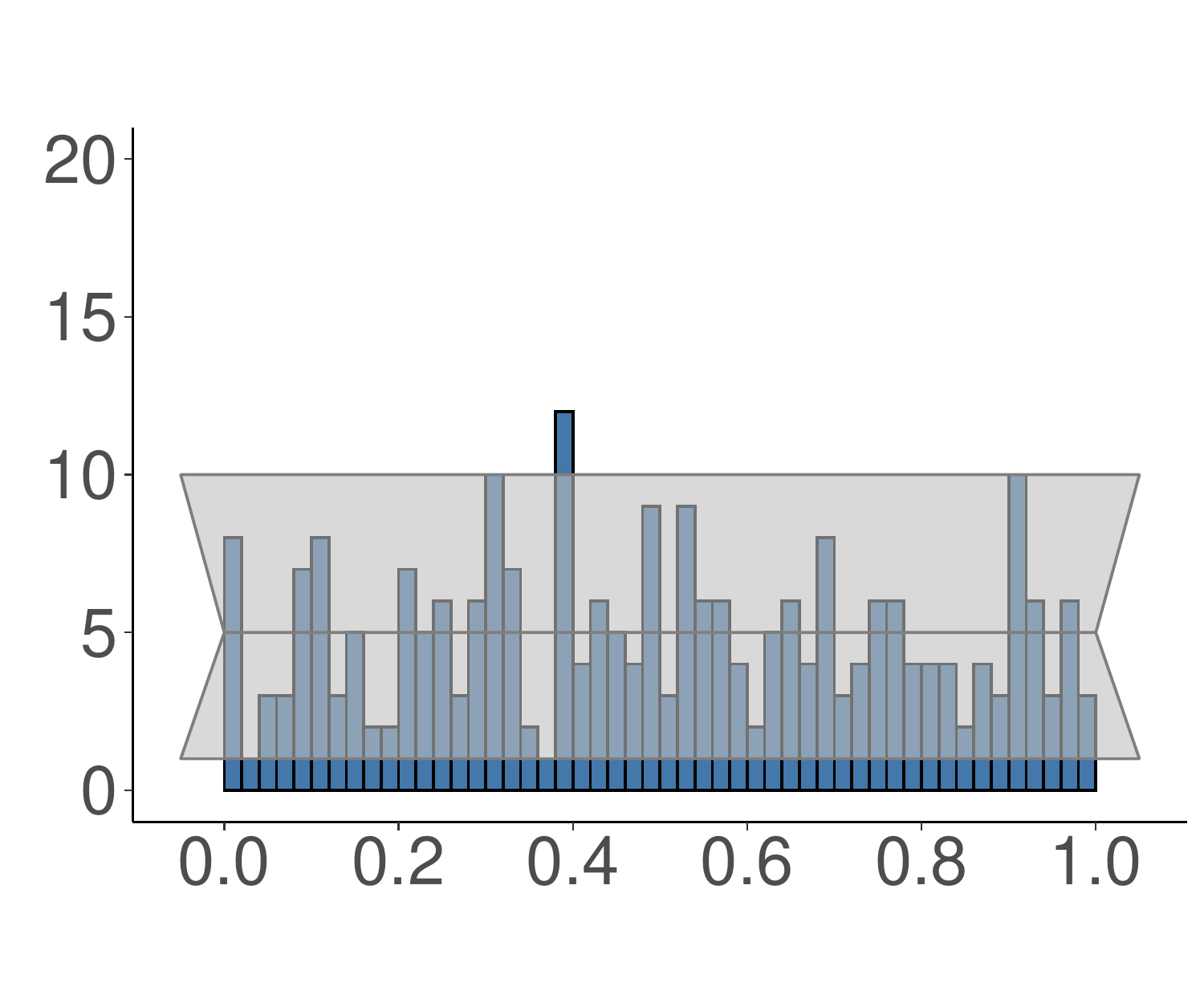}}%
        \subfloat[ECDF plot]{\label{fig:intro c}%
        \includegraphics[width=0.245\textwidth]{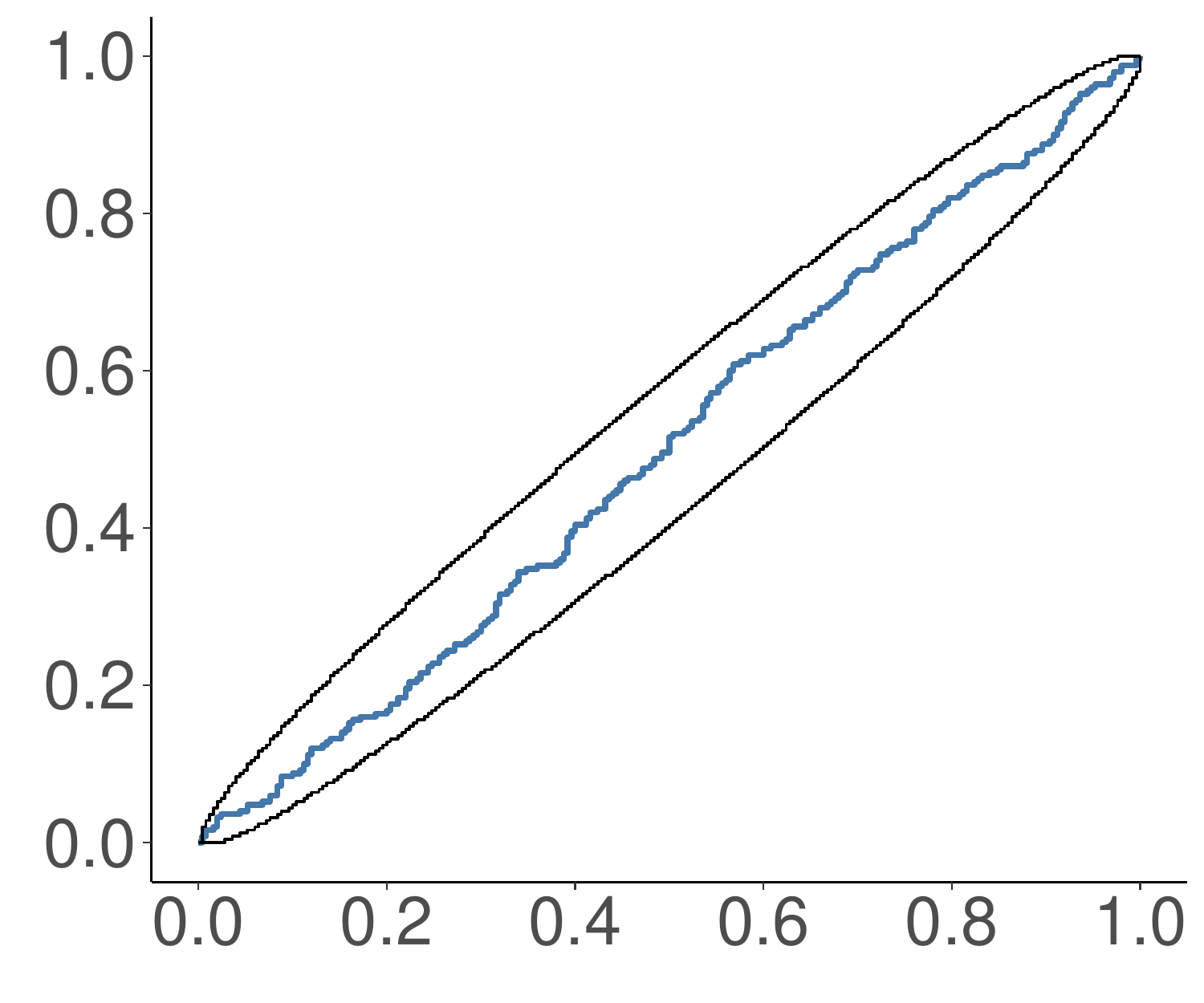}}%
        \subfloat[ECDF difference plot]{\label{fig:intro d}%
        \includegraphics[width=0.245\textwidth]{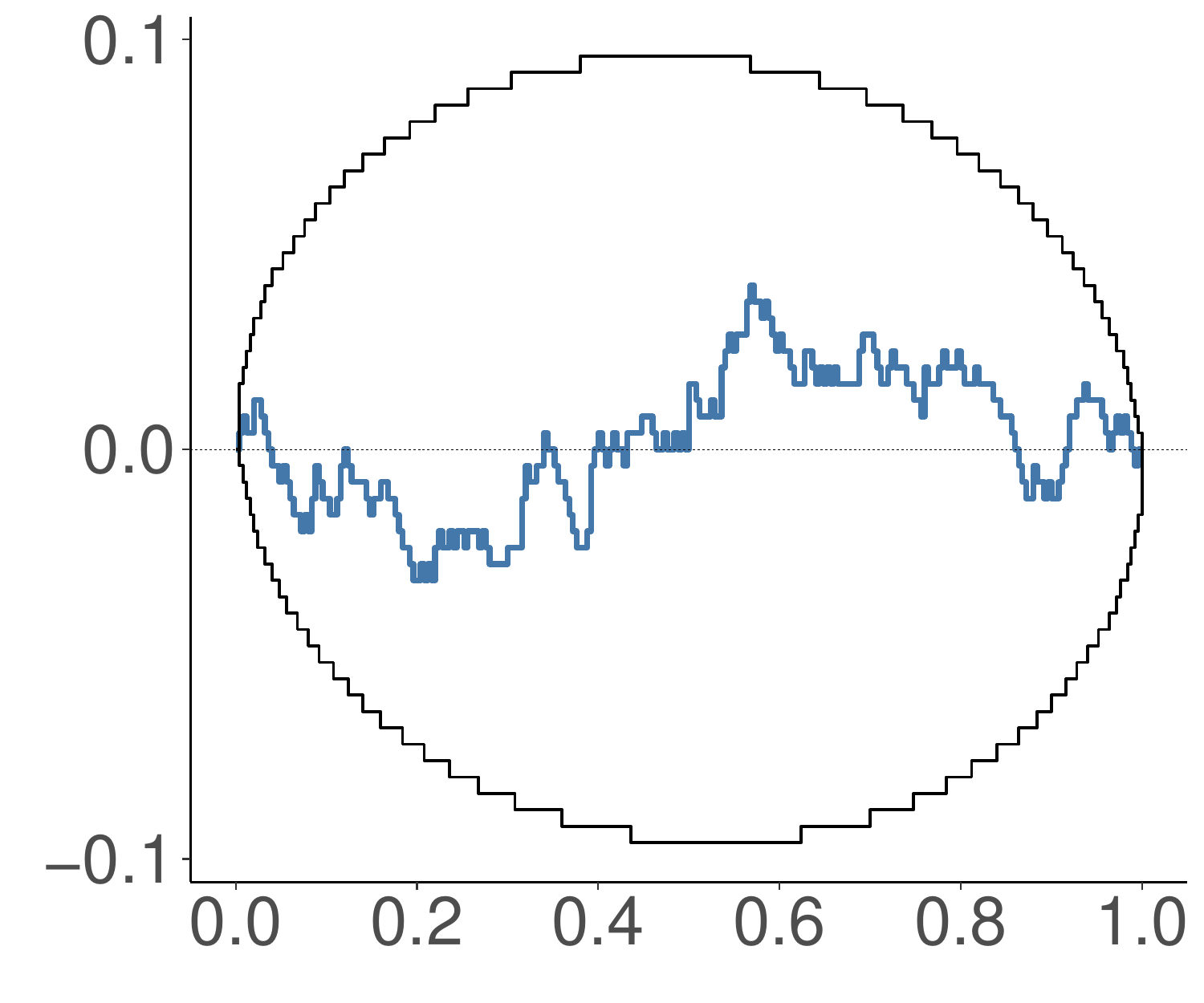}}%
        \caption{Four visualizations depicting the same random sample of 250 values. To asses uniformity of the sample, histograms (a) and (b) show a 95\% confidence interval for each bin. Histograms can be sensitive to the number and placement of the bins selected, and the confidence intervals do not take into account possible inter-dependencies between the bin heights. For example, given the same sample, a 20 bin histogram stays within the confidence interval (a), but a 50 bin histogram exceeds the confidence interval (b). The ECDF plot (c) and ECDF difference plot (d) with 95\% simultaneous confidence bands for the ECDF both show the sample staying within the given limits with the ECDF difference plot providing a more dynamic range for the visualization.}
        \label{fig:intro_hist_and_ecdfs}
    \end{figure}

    Due to its ease of interpretation and familiarity to people even with basic statistical knowledge, a graphical method for assessing uniformity commonly used as part of many statistical workflows is plotting histograms. This can even be turned into a formal test of uniformity with confidence intervals for the individual bins \parencite[e.g.,][]{Talts2018}. Drawbacks of histograms are that binning discards information, there can be binning artifacts depending on the choice of bin width and placement, and they ignore the dependency between bins. The proposed ECDF-based method doesn't require binning or smoothing, provides intuitive visual interpretation, and works for continuous Eq.~\eqref{eq: probability integral transformation} and discrete Eq.~\eqref{eq: rank statistic} values. An illustration and comparison of histograms with two binning choices and our new method is given in Figure \ref{fig:intro_hist_and_ecdfs}. The visual range between the simultaneous confidence bands for the ECDF is often narrow when visualizing a sample with a large number of observations. Thus, to achieve a more dynamic range for the visualization, we recommend to show ECDF difference plots, instead, as illustrated in Figure \ref{fig:intro_hist_and_ecdfs}(d). The ECDF difference plot is obtained by subtracting the values of the expected theoretical CDF (i.e., the identity function in [0,1] in case of standard uniformity) from the observed ECDF values.

    \subsection{Summary of contributions}
    
    In this article, we focus on use case examples arising from inference validation and {Markov chain Monte Carlo} (MCMC) convergence diagnostics as part of a Bayesian workflow \parencite{BayesianWorkflow2020}, but our developed methods are applicable more generally. Our use cases can be divided into two main categories: a single sample test for uniformity, and a multiple sample comparison where the hypothesis is tested that the samples are drawn from the same underlying (potentially non-uniform) distribution. We discuss both cases in more detail below.

    We offer a graphical test for uniformity by providing simultaneous confidence bands for one or more ECDF trajectories obtained through the empirical probability integral transformation.
    As our first contribution, we modify an existing ECDF-based approach proposed by \textcite{Aldor-Noiman2013} to take into account the discreteness of the fractional rank-based PIT values. This forms the basis for our proposed single and multi-sample tests.
    
    As our second contribution, we 
    provide both a simulation and optimization method to determine the adjustment needed to achieve a desired simultaneous confidence level for the ECDF trajectory given the fractional rank-based PIT values. In addition to presenting a simulation-based adjustment following the method of \textcite{Aldor-Noiman2013},
    we introduce a new optimization method that is computationally considerably more efficient in determining the needed adjustment, especially when bands with high resolution are desired for a large sample size. 
    Although our focus is on providing a test with an intuitive graphical representation, we show that our method performs competitively when compared to existing uniformity tests with state-of-the-art performance. We demonstrate the usefulness of this graphical test in context of simulation based calibration approach for assessing inference methods \parencite{Talts2018}.
    
    Finally, as our third contribution, we generalize the graphical test as well as both the simulation and the optimization method  to evaluate the hypothesis that two or more samples are drawn from the same underlying distribution. We demonstrate the usefulness of this graphical test in MCMC convergence diagnostics, where the currently most common graphical tools for assessing convergence are trace plots of the individual sampled chains.
        
    \subsection{Outline of the paper}

    In Section \ref{sec:confidence_bands}, we first provide a simulation-based method to determine simultaneous confidence bands for the ECDF of a single uniform sample and then present new more efficient optimization-based method.
    In Section \ref{sec:multi_sample_test}, we extend the test to multiple sample comparison, and follow a similar structure by offering both a simulation and an optimization-based methods. 
    We continue in Section \ref{sec:experiments} with simulated and real-world examples illustrating the application of our proposed method, and end with a discussion in Section \ref{sec:discussion}.
    
\section{Simultaneous Confidence Bands for the Empirical Cumulative Distribution}\label{sec:confidence_bands}

We propose simulation and optimization based approaches to providing the ECDF of a uniform sample with $1-\alpha$ level simultaneous confidence bands that are compatible with empirical PIT values, that is, confidence bands with a type-1-error rate of $\alpha$.  Our approach is similar to that presented by \textcite{Aldor-Noiman2013} with one central distinction illustrated in Figure \ref{fig:test_approach_differences}. The method by \textcite{Aldor-Noiman2013} obtains simultaneous confidence bands for the evaluation quantiles with fixed ECDF values based on beta distributions, that is, it obtains confidence bands along the horizontal axis (Figure \ref{fig:test_approach_differences}(a)). In contrast, our new method provides simultaneous confidence bands for the ECDF values at fixed evaluation quantiles based on binomial distributions, that is, it obtains confidence bands along the vertical axis (Figure \ref{fig:test_approach_differences}(b)). In the limit, as the sample size approaches infinity, there is no practical difference between the methods. However, when the number of possible unique ranks is small, our proposed method behaves better for smallest and largest ranks, and consistently if the ranks are further binned.

\subsection{Pointwise confidence bands}\label{subsec:single_sample_test}

Determining the pointwise confidence interval for the ECDF value of a sample from the continuous uniform distribution at a given evaluation point $z_i\in (0,1)$ is rather straightforward. By definition, given a sample $u = u_1, \ldots, u_N$, the ECDF value is
\begin{equation}\label{eq:ecdf_at_zi}
    F(z_i) = \frac{1}{N}\sum_{j=1}^N \mathbb{I}(u_j\leq z_i).
\end{equation}
As the sampled values, $u_j\in (0,1)$, are expected to be continuously uniformly distributed, $\Pr(u_j \leq z_i) = z_i$ for each $j=1,\ldots,N$. Thus, the values resulting from scaling the ECDF with the sample size $N$ are binomially distributed as
\begin{equation}\label{eq:marginal_distribution_of_ecdf}
    NF(z_i) \sim \Bin\left(N, z_i\right).
\end{equation}

If we instead expect $u$ to be sampled from a discrete uniform distribution with $S$ distinct equally spaced values, $s_j = j/S$, by choosing the partition points to form a subset of these category values, we again have $\Pr(u_j \leq z_i) = z_i$ for $j=1,\ldots,N$, and the marginal distribution of the scaled ECDF follows Eq.~\eqref{eq:marginal_distribution_of_ecdf}. Therefore, the methods introduced in sections \ref{subsubsec:single_sample_simulated_method} and \ref{subsec:single_sample_optimization_method} can be used to determine simultaneous confidence bands for both continuous and discrete uniform samples, allowing for testing uniformity of both the continuous PIT values of Eq.~\eqref{eq: probability integral transformation} and the discrete empirical PIT values in Eq.~\eqref{eq: rank statistic}.

\begin{figure}
\centering
    \subfloat{\includegraphics[width=0.3\textwidth]{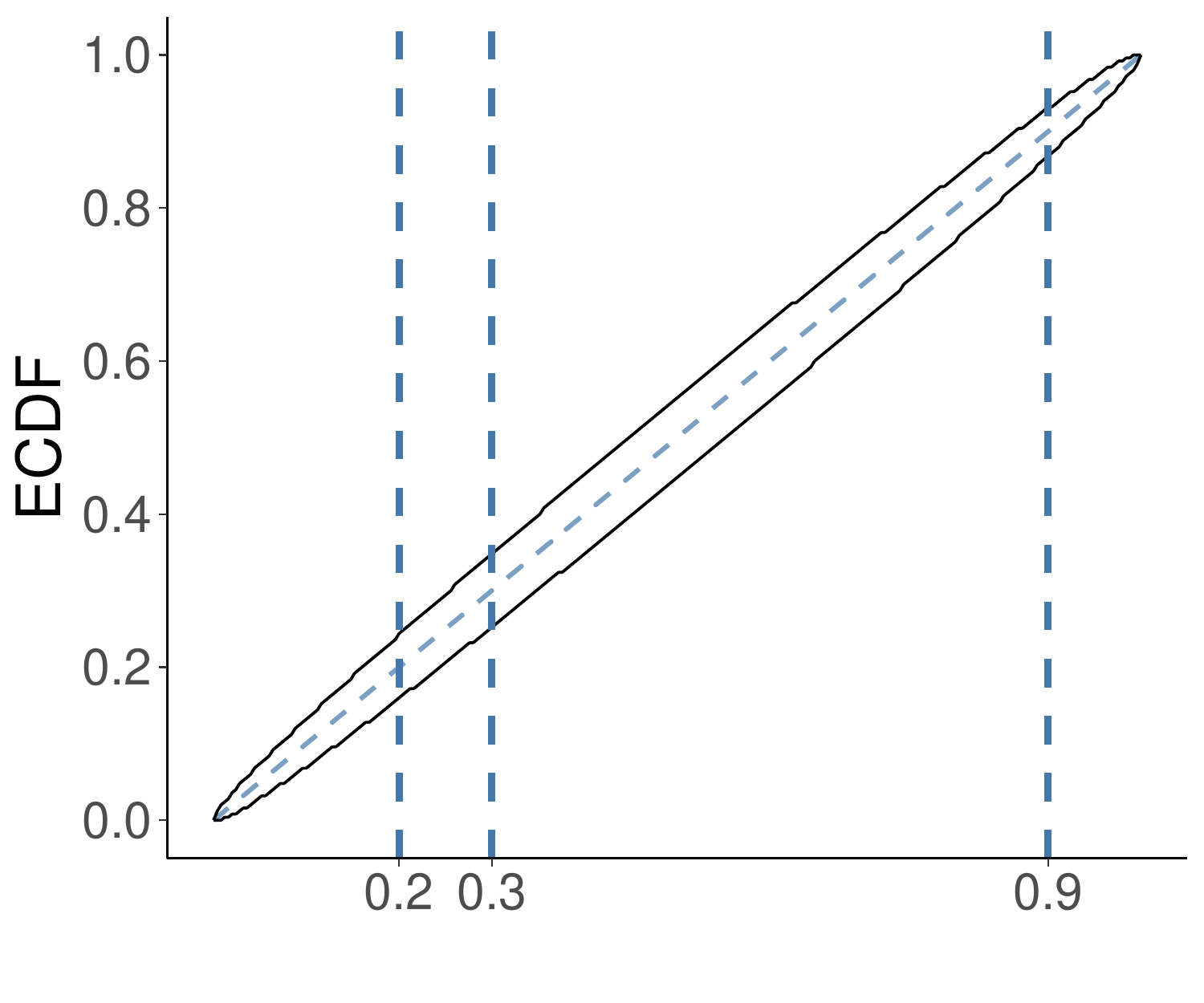}} 
    \subfloat{\includegraphics[width=0.3\textwidth]{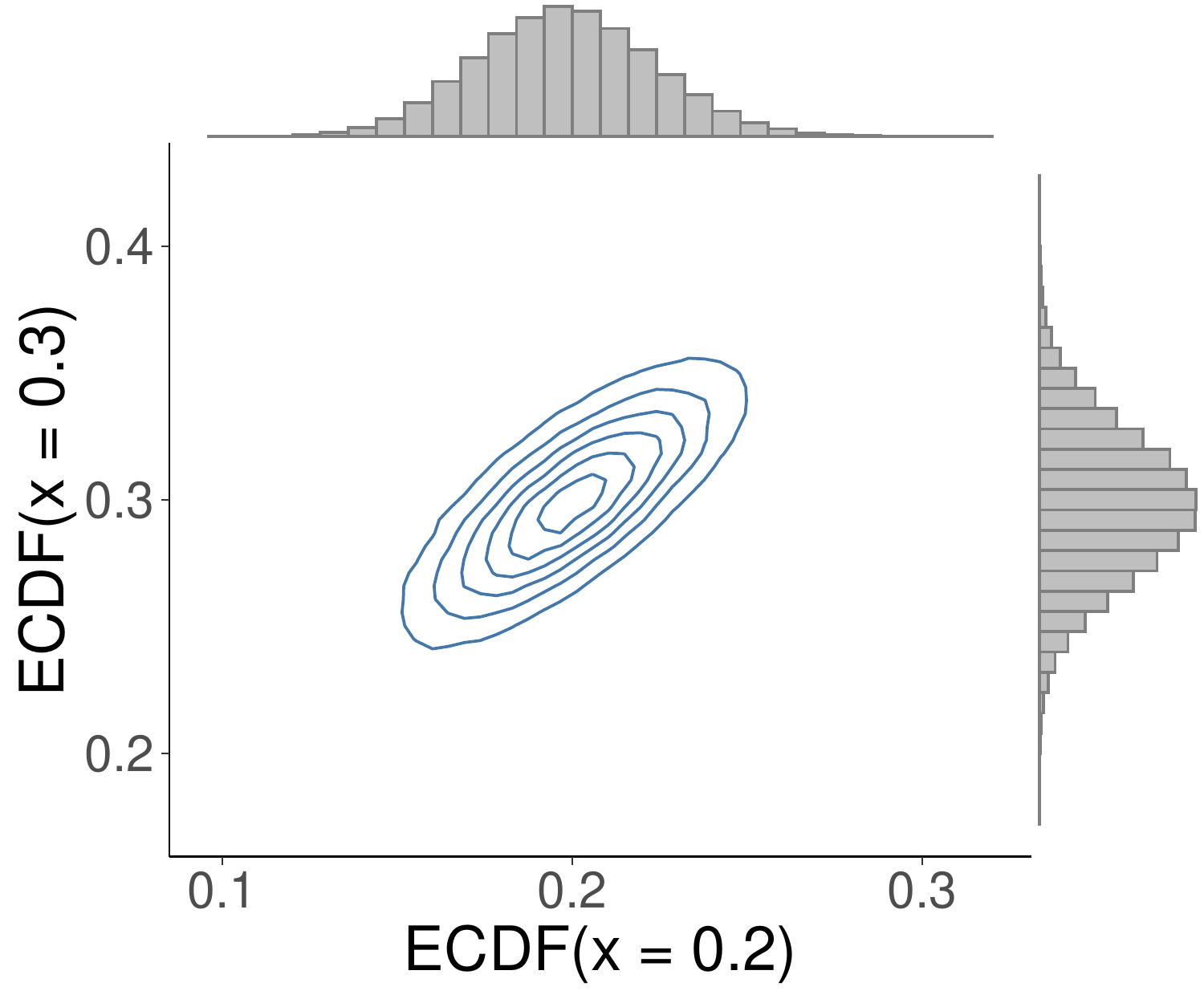}} 
    \subfloat{\includegraphics[width=0.3\textwidth]{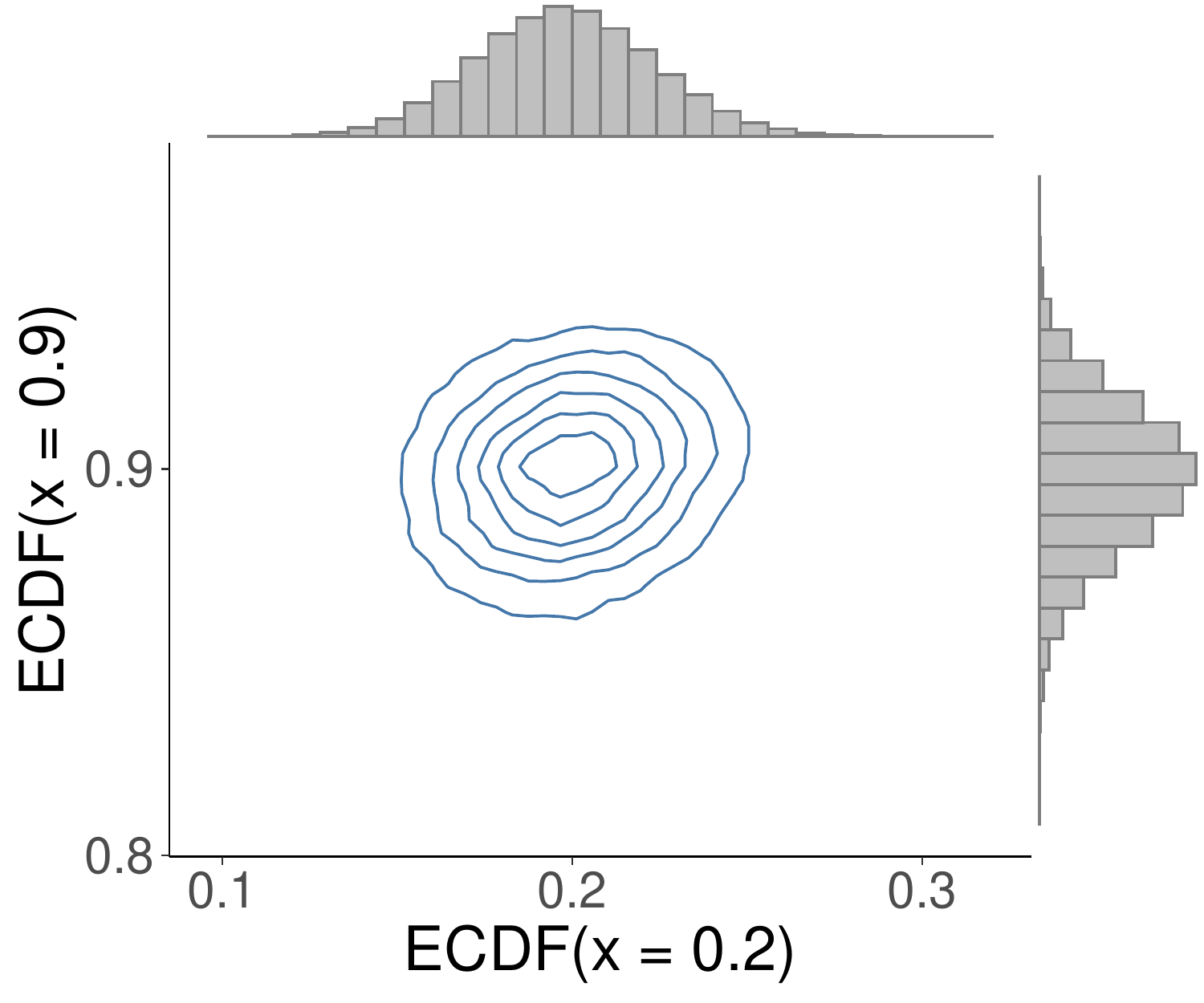}}
    \caption{Dependence between the ECDF values of standard uniform samples evaluated at three distinct points. Simultaneous 95\% confidence bands for the ECDF computed via either of the new methods introduced in this paper are shown on the left. In the middle, one can see a stronger dependency between the ECDF values obtained at evaluation points close to each other whereas on the right the ECDF values are only weakly dependent as the evaluation points far away from each other.}
    \label{fig:ecdf_height_dependency_with_marginals}
\end{figure}

From Equation \eqref{eq:marginal_distribution_of_ecdf}, it is straightforward to determine the $1 - \alpha$ level pointwise lower and upper confidence bands, $L_i$ and $U_i$ respectively, satisfying for all $i = 1,\ldots,N$ individually
\begin{equation}\label{eq:pointwise_confidence_bands_single_sample}
    \Pr\left( L_i \leq F(z_i) \leq U_i \right) = 1- \alpha.
\end{equation}
In contrast, determining the simultaneous confidence bands for ECDF trajectories (i.e., sets of ECDF values) is more complicated. In Figure \ref{fig:ecdf_height_dependency_with_marginals}, we illustrate the dependency between ECDF values at distinct evaluation quantiles, together with simultaneous confidence bands computed via either of the new methods described in the following sections. As is illustrated in the figure, ECDF values evaluated at two quantiles close to each other are strongly dependent while ECDF values evaluated at two quantiles far away from each other are only weakly dependent. In any case, these dependencies need to be taken into account when constructing simultaneous confidence bands.

Another important remark is that, as the marginal distribution of the scaled ECDF is discrete, the simultaneous confidence intervals do not in all cases meet the desired coverage level exactly. \textcite{brown2001} provide a thorough exploration of the effect discreteness plays in the coverage level of various interval estimations for binomial proportion, with listings of what the authors call lucky and unlucky sample lengths. In our experience, even though discreteness plays a role in the coverage level of the pointwise confidence intervals, this effect is reduced to deviation of under $\pm 1\%$ for $N\in [50,2000]$ in the coverage level of the resulting simultaneous confidence bands we introduce next.

\subsection{Simultaneous confidence bands through simulation}\label{subsubsec:single_sample_simulated_method}
 
 Our goal is to define simultaneous confidence bands for the ECDF of a sample of $N$ values drawn from the standard uniform distribution so that the interior of the confidence bands contains trajectories induced by that distribution with rate $1 - \alpha$, where $\alpha \in (0,1)$.

    In this section we describe a simulation based method for determining the simultaneous confidence bands for the ECDF trajectory.
     We follow steps similar to those introduced by \textcite{Aldor-Noiman2013}; with the exception that instead of determining limits for the Q-Q plot, we now determine the upper and lower limits of the ECDF values at the evaluation points $z_i$:
    \begin{enumerate}
        \item Choose a partition $\left(z_i\right)_{i=1}^K$ of the unit interval.
        \item Determine coverage parameter $\gamma$ to account for multiplicity in order to obtain the $1-\alpha$ level simultaneous confidence bands:
        \begin{equation}
            \Pr\left( L_i(\gamma)\leq F(z_i)\leq U_i(\gamma) \text{ for all } i\in \{1,\ldots, K\}\right) = 1-\alpha.
        \end{equation}
    \end{enumerate}
    
    In determining these confidence bands, we use the knowledge from that the values of the scaled ECDF at each point $z_i$ follow a binomial distribution 
    and  denote the value of the cumulative binomial distribution function with parameters $N$ and $z_i$ at $k\in\mathbb{N}$ by $\Bin(k \given N,z_i)$ and its inverse by $\Bin^{-1}(q \given N,z_i)$ for quantile $q \in [0, 1]$.
    
    To find the desired coverage value $\gamma$, we simulate $M$ draws of size $N$ from the standard uniform distribution. Let $F^m$ denote the ECDF of the $m$th sample, $u_1^m,\ldots u_N^m \sim {\rm uniform}(0,1)$. For each sample, we find the value of $\gamma$ such that the equal tail quantiles
    \begin{equation}
        L_i(\gamma) = \frac{1}{N}\Bin^{-1}\left(\frac{\gamma}{2}\given N,z_i\right),
    \end{equation}
    and
    \begin{equation}
        U_i(\gamma) = \frac{1}{N}\Bin^{-1}\left(1-\frac{\gamma}{2} \given N,z_i\right)
    \end{equation}
    provide the tightest possible lower and upper limits respectively to the sample ECDF, $F^m$, at each $z_i$. This value of $\gamma$ for the $m$th sample is
    \begin{equation}
        \gamma^m = 2\min_i\left\lbrace\min\left(\Bin(NF^m(z_i) \given N,z_i), 1-\Bin(NF^m(z_i)-1 \given N,z_i)\right)\right\rbrace
    \end{equation}
    As now we have for $\gamma^m$ equally that
    \begin{equation}\label{eq:gamma_m}
        \gamma^m = \arg\max_\gamma \left\lbrace\frac{\gamma}{2} \leq \Bin(NF^m(z_i)\given N,z_i) \leq 1-\frac{\gamma}{2}\mid \forall i\right\rbrace
    \end{equation}
    and as it holds that
    \begin{align}
        &\frac{\gamma}{2} \leq \Bin(NF^m(z_i)\given N,z_i) \leq 1-\frac{\gamma}{2}\nonumber\\
        \Rightarrow & \Bin^{-1}\left(\frac{\gamma}{2} \given N,z_i\right) \leq NF^m(z_i)\leq \Bin^{-1}\left(1-\frac{\gamma}{2} \given N,z_i\right)\nonumber\\
        \Rightarrow & L_i(\gamma) \leq F^m(z_i)\leq U_i(\gamma)  \nonumber,
    \end{align}
    $\gamma^m$ defines a set of upper and lower limits to the ECDF which is by Eq. \eqref{eq:gamma_m} the tightest possible pair of limits defining equal tail quantiles for the ECDF at each $z_i$.
    To obtain bands covering a $1-\alpha$ fraction of the ECDFs of the simulated samples, we set $\gamma$ to the $\alpha$ quantile of the values $\lbrace \gamma^1,\ldots, \gamma^M\rbrace$.
    Since $\gamma^m > 0$ by construction, we also have $\gamma > 0$.

    The following steps summarize the algorithm for simulating the adjusted coverage parameter $\gamma$ and determining the $1-\gamma$ level simultaneous confidence bands:
    \begin{enumerate}
        \item For $m=1,\ldots,M$:
            \begin{enumerate}
                \item Simulate $u_1^m,\ldots,u_N^m \sim {\rm uniform}(0,1)$.
                \item For $i=1,\ldots,K$, compute $F^m(z_i)$.
                \item For $i=1,\ldots,K$, compute 
                $$\Bin(NF^m(z_i) \given N,z_i) \text{ and } \Bin(NF^m(z_i)-1 \given N,z_i).$$
                \item Find the minimum probability
                $$\gamma^m = 2\min_i\left\lbrace\min\left(\Bin(NF^m(z_i) \given N,z_i), 1-\Bin(NF^m(z_i)-1 \given N,z_i)\right)\right\rbrace.$$
            \end{enumerate}
        \item Set $\gamma$ to be the $100\alpha$ percentile of $\lbrace\gamma^1,\ldots,\gamma^M\rbrace$.
        \item Form the confidence bands
        $$\left[L_i\left(\gamma\right),U_i\left(\gamma\right)\right] = \left[\frac{1}{N}\Bin^{-1}\left(\frac{\gamma}{2} \given N,z_i\right),\frac{1}{N}\Bin^{-1}\left(1-\frac{\gamma}{2} \given N,z_i\right)\right]$$
        for $i=1,\ldots,K$.
    \end{enumerate}

\subsection{Simultaneous confidence bands through optimization}\label{subsec:single_sample_optimization_method}

We also propose a computationally more efficient optimization based method for determining the simultaneous confidence bands.

In the following derivation of the optimization method, we denote the interior of the confidence bands for the ECDF at quantile $z_i$ as $\tilde{I}_i(\gamma)$.
By denoting $r_i = N F(z_i)$, the scaled interior $I_i(\gamma)$ for $r_i$ is given by
\begin{equation}
I_i(\gamma) = \left\{ r \in \{0, \ldots, N\} \given \Bin^{-1} \left(\frac{\gamma}{2} \given N,z_i \right) \leq r \leq \Bin^{-1} \left(1 - \frac{\gamma}{2} \given N,z_i \right)  \right\}.
\end{equation}
As is common for discrete statistical tests, we treat the borders between the interior and exterior as being within the confidence bands. Based on $I_i(\gamma)$, we can easily obtain $\tilde{I}_i(\gamma)$ as $r \in I_i(\gamma)$ is equivalent to $r / N \in \tilde{I}_i(\gamma)$. 

A scaled ECDF trajectory defined as
\begin{equation}
t_0^K = \left((z_i)_{i=0}^K, (r_i)_{i=0}^K\right)
\end{equation}
with $z_0 = 0$ and $z_K = 1$ stays within the simultaneous confidence bands completely if and only if $r_i \in I_i(\gamma)$ for all $i \in \{0, \ldots, K\}$. If we denote the set of trajectories fulfilling $r_i \in I_i$ as $T_i$, we can write the set of trajectories which are completely within the simultaneous confidence bands as 
\begin{equation}
T(\gamma) = \bigcap_{i = 0}^K T_i(\gamma).
\end{equation}
In order for the simultaneous confidence bands to have confidence level $1-\alpha$, we must have
\begin{equation}
\label{find-gamma}
\Pr\left(T(\gamma) \right) = 1 - \alpha.
\end{equation}

Due to the pairwise independence of the original draws $u_{i}$ (by assumption), the distribution of the ECDF values within a single trajectory is Markovian in the sense that the ECDF value $F(z_{i+1})$ only depends on the observed value at the previous evaluation point, $F(z_i)$ and not on the earlier behaviour of the ECDF trajectory.

This implies that, under uniformity of the original distribution, the remaining $N - NF(z_i) = N - r_i$ samples are uniformly distributed on the interval $[z_i, 1]$, and thus the growth of the scaled ECDF from $r_{i}$ to $r_{i+1}$, between $z_{i}$ and $z_{i+1}$ is binomially distributed with $N - r_{i}$ trials and the success probability
\begin{equation}
\tilde{z}_{i+1} = \frac{z_{i+1} - z_{i}}{1 - z_{i}}.
\end{equation}
And so we have
\begin{equation} 
\Pr(r_{i+1} \given r_{i}) =  \Bin\left(r_{i+1} - r_{i} \given N - r_{i}, \tilde{z}_{i+1}  \right).
\end{equation}
The probability for $r_{i+1} = k \in I_{i+1}$ to occur in a scaled ECDF trajectory $t_0^K$ which stayed within the simultaneous confidence bands until point $i$, that is, for which we have
\begin{equation}
t_0^i \in \bigcap_{j = 0}^{i} T_j(\gamma) ,
\end{equation}
can thus be written recursively as
\begin{equation}
\Pr \left(r_{i+1} = k \cap \bigcap_{j = 0}^{i} T_j(\gamma) \right) = \sum_{m \in I_{i}} \; \Pr \left(r_{i} = m \cap \bigcap_{n = 0}^{i-1} T_n(\gamma) \right) \; \Pr(r_{i+1} = k \given r_{i} = m).
\end{equation}
The recursion is initialized at $z_0 = 0$ with $\Pr(r_{0} = 0) = 1$ so that $\Pr(T_0(\gamma)) = 1$ for all $\gamma \in [0, 1]$. At any point $i \in \{0, \ldots, K\}$, we can obtain
\begin{equation}
\label{pcapTi}
\Pr \left(\bigcap_{j = 0}^{i} T_j(\gamma) \right) = \sum_{m \in I_{i}} \; \Pr \left(r_{i} = m \cap \bigcap_{n = 0}^{i-1} T_n(\gamma) \right),
\end{equation}
which is equal to $\Pr(T(\gamma))$ when arriving at $i = K$. Clearly, $\Pr(T(\gamma))$ is monotonically decreasing but not continuous in $\gamma$ due to the discrete nature of the binomial distribution. Thus, Equation \eqref{find-gamma} will not have an exact solution in general and so we will not be able to meet the simultaneous confidence level $1-\alpha$ exactly. We can, however, try to get as close as possible by computing
\begin{equation} 
\hat{\gamma} = \argmin_{\gamma \in [0, \alpha]} \; |1 - \alpha - \Pr \left(T(\gamma) \right)| 
\end{equation}
with a unidimensional derivative-free optimizer. In our experiments, the optimizer proposed by \textcite{brent1973} (which is implemented, e.g., in the R function \texttt{optimize}) converged quickly in all cases to $\hat{\gamma}$ values implying a simultaneous confidence level very close to the nominal $1-\alpha$.

With a 2015 laptop equipped with a 2.90GHz Intel® Core™ i5-5287U processor, the optimization method reduces the time required to compute the adjustment parameter $\gamma$ from 10s to 600ms for a sample of length 250 when compared against the time required for 10,000 steps of the simulation method. With $N = 1000$ this reduction is from 75s to 10s.

Both of the implementations used for this article only use a single computation thread, but would benefit from parallelization,  as both methods include independent iterations.

The computation time required can be further reduced by using a grid of pre-computed values as the adjustment parameters, and interpolate for different values of $N$ in log-log scale.

\section{Comparison of multiple samples}\label{sec:multi_sample_test}

\begin{figure}
        \centering
        \begin{tikzpicture}[
        node/.style={minimum size=1cm}
        ]
        \node[node] (chain_1){
        \includegraphics[width=0.18\textwidth]{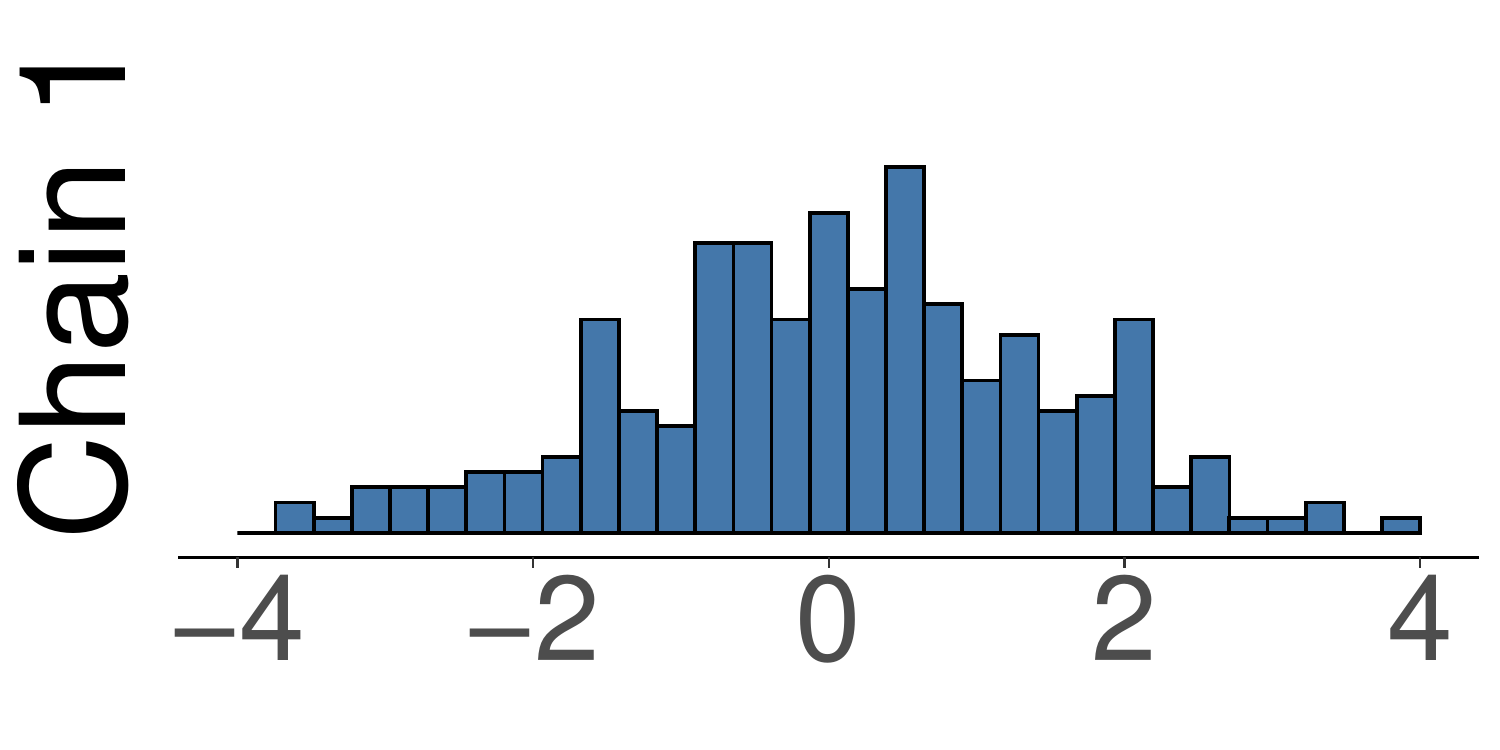}
        };
        \node[node] (chain_2) [below=0cm of chain_1] {
        \includegraphics[width=0.18\textwidth]{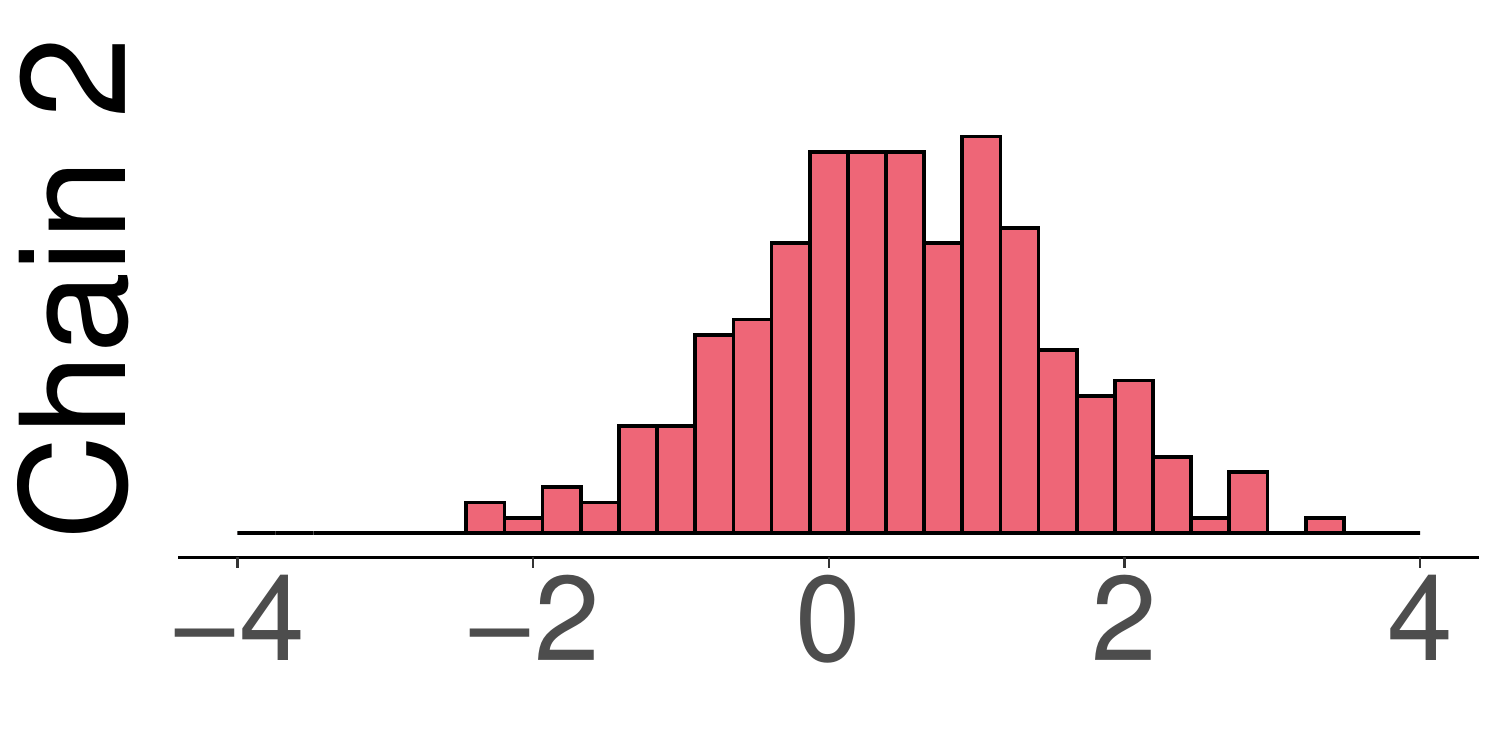}
        };
        \node[node] (chain_3) [below=0cm of chain_2] {
        \includegraphics[width=0.18\textwidth]{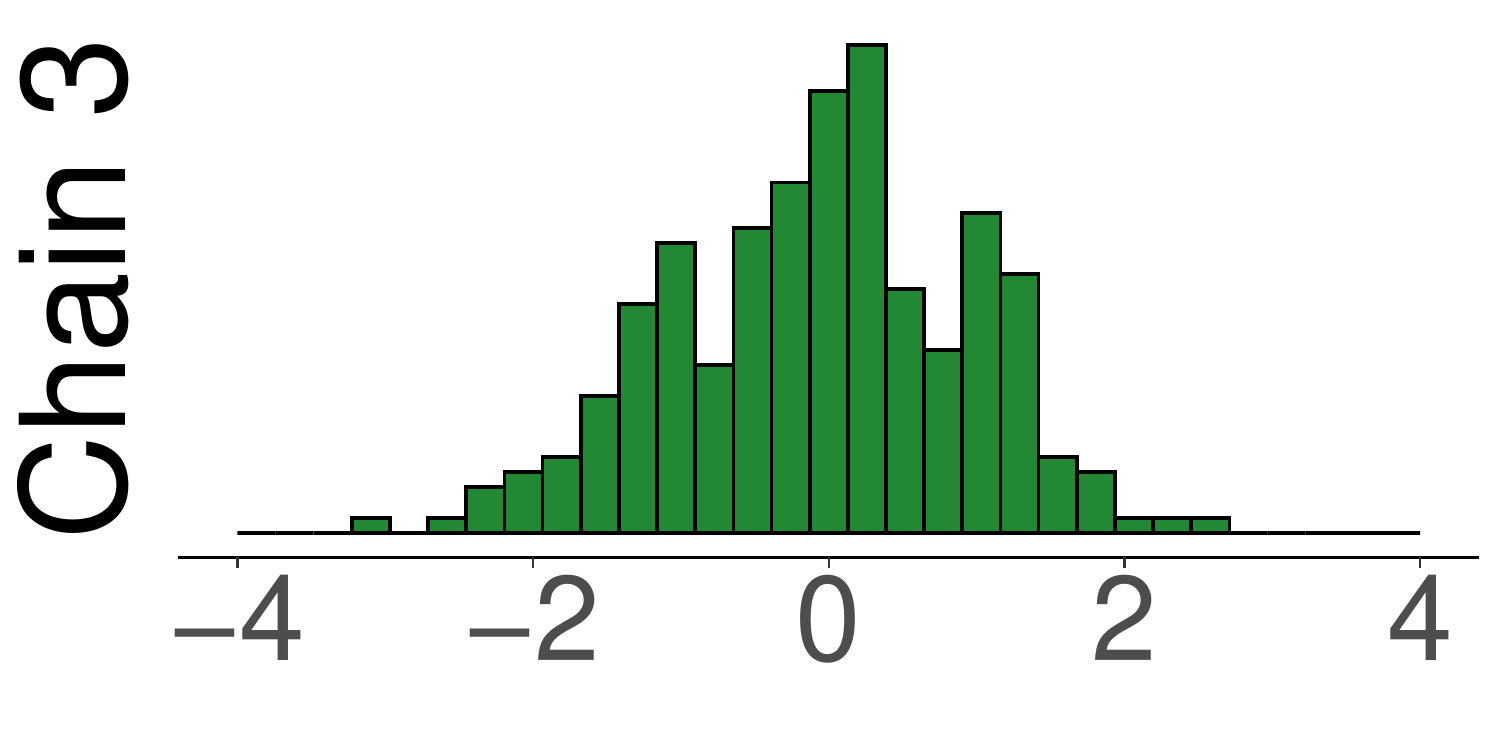}
        };
        \node[node] (chain_4) [below=0cm of chain_3] {
        \includegraphics[width=0.18\textwidth]{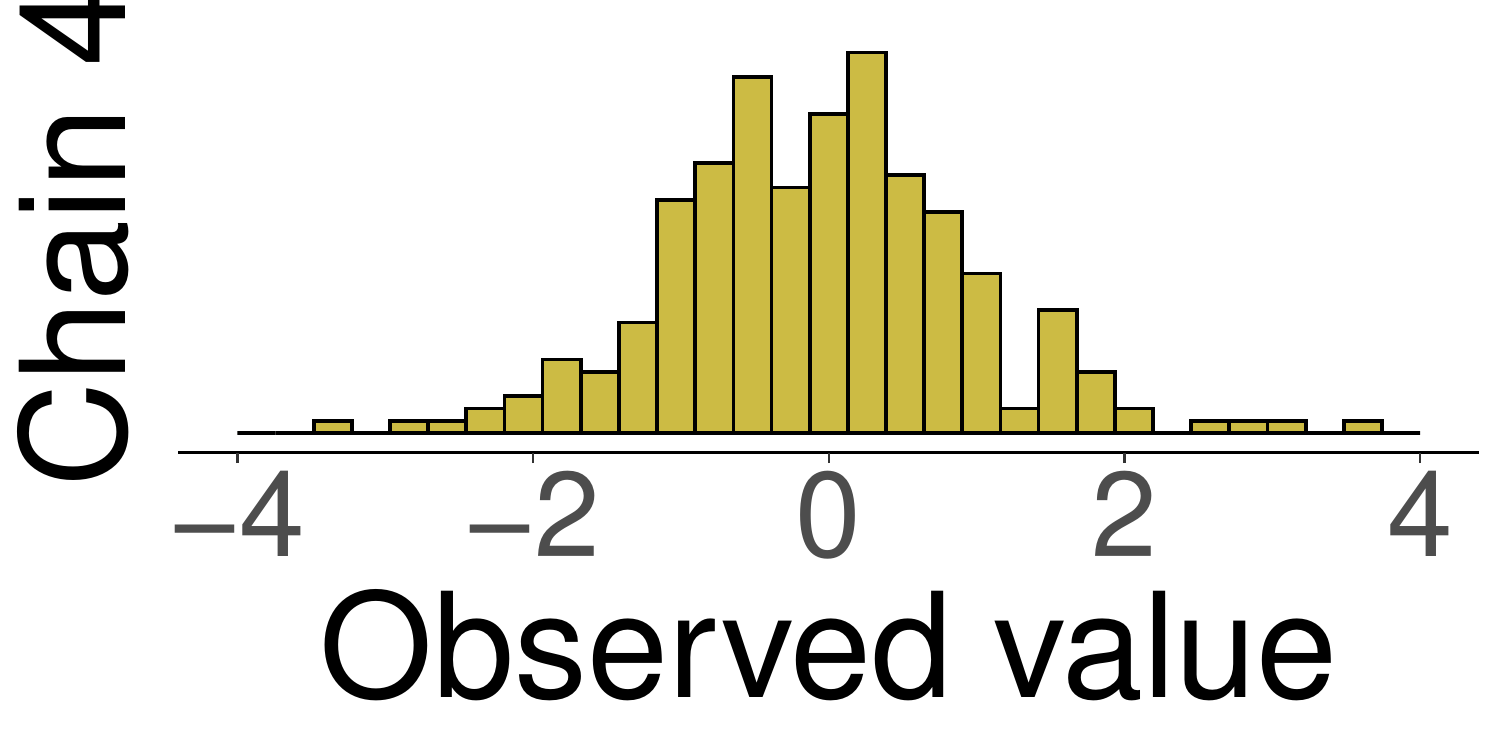}
        };
        \node[node] (tchain_1) [right=1cm of chain_1]{
        \includegraphics[width=0.18\textwidth]{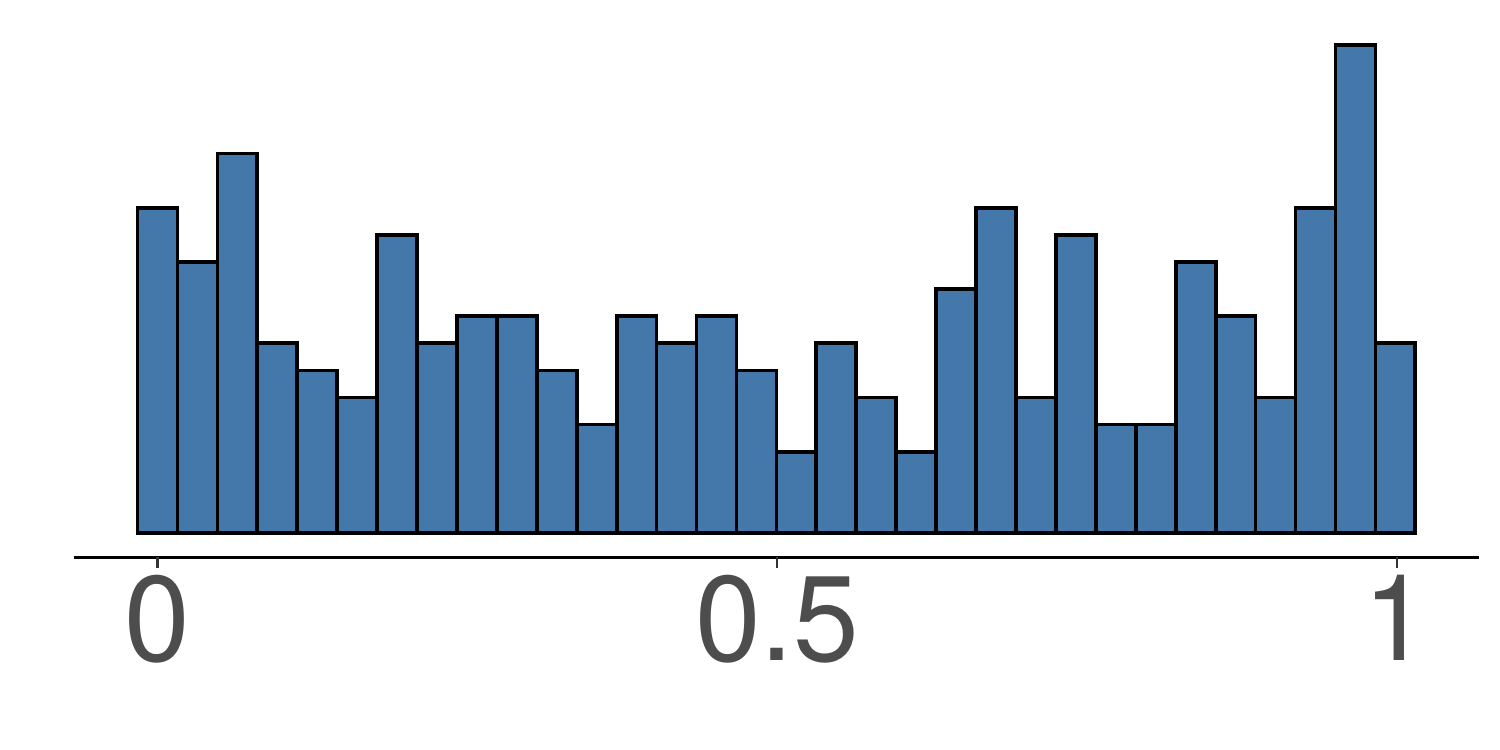}
        };
        \node[node] (tchain_2) [below=0cm of tchain_1] {
        \includegraphics[width=0.18\textwidth]{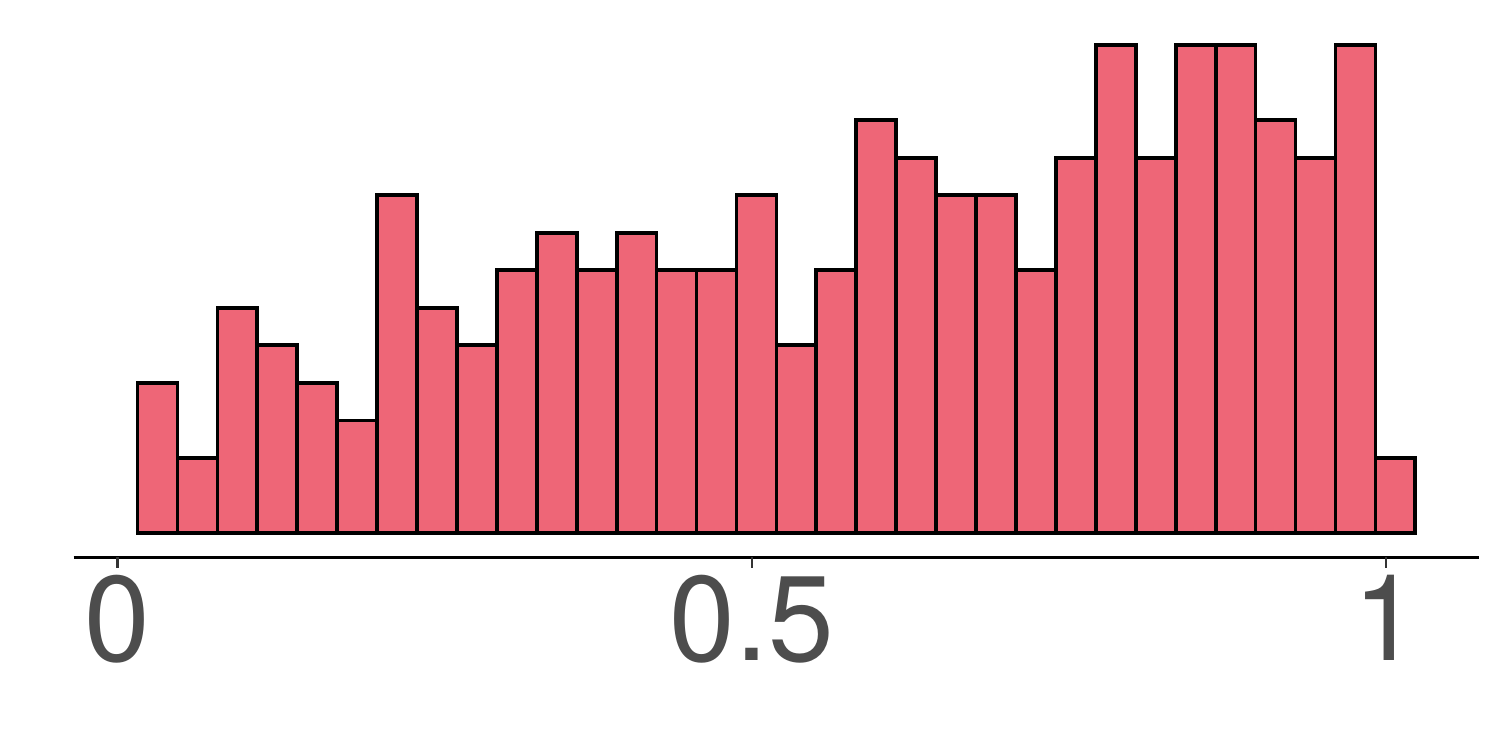}
        };
        \node[node] (tchain_3) [below=0cm of tchain_2] {
        \includegraphics[width=0.18\textwidth]{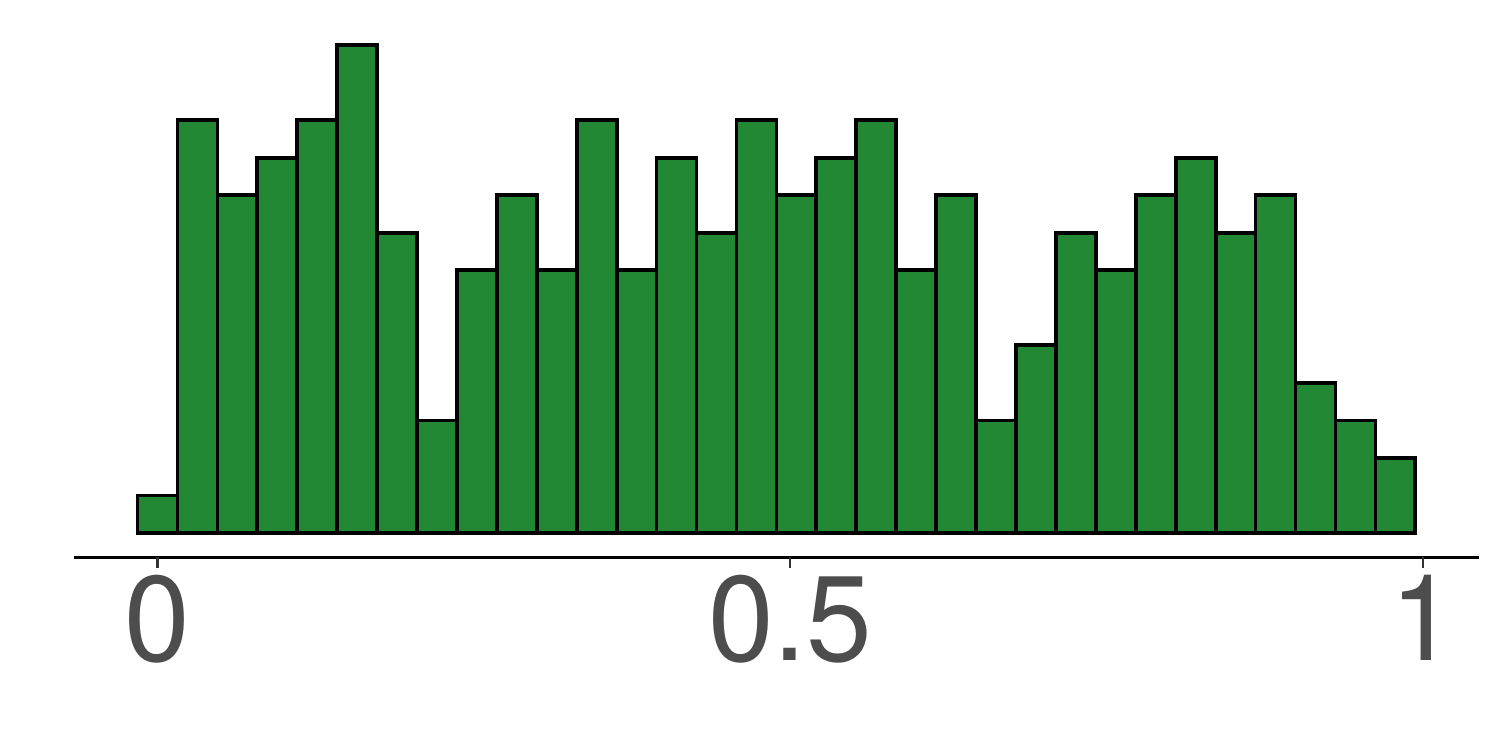}
        };
        \node[node] (tchain_4) [below=0cm of tchain_3] {
        \includegraphics[width=0.18\textwidth]{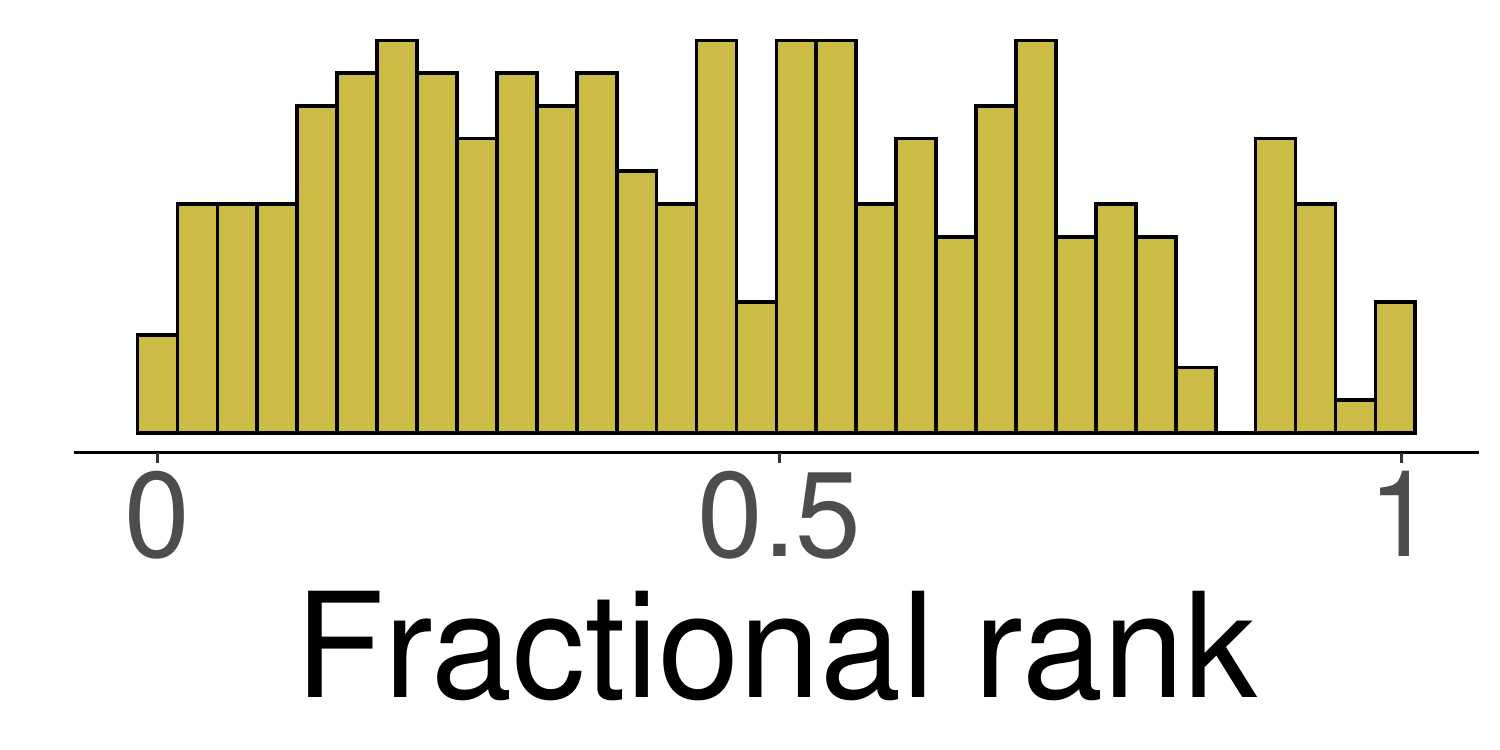}
        };
        \node[node] (ecdf)   [below right=-1.6cm and 1cm of tchain_1] {
        \includegraphics[width=0.24\textwidth]{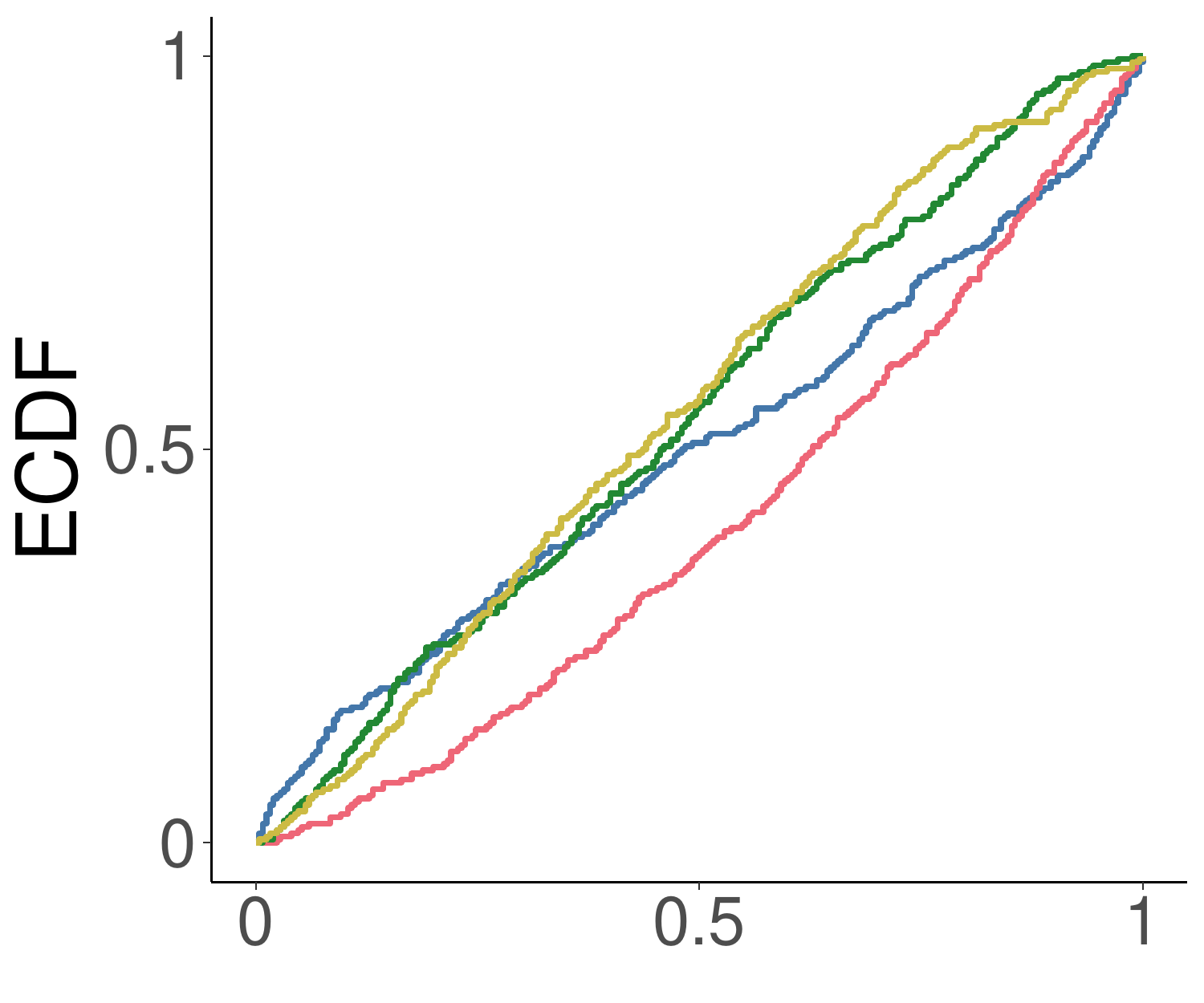}
        };
        \node[node] (diff)   [below right=-1.8cm and .9cm of tchain_3] {
        \includegraphics[width=0.24\textwidth]{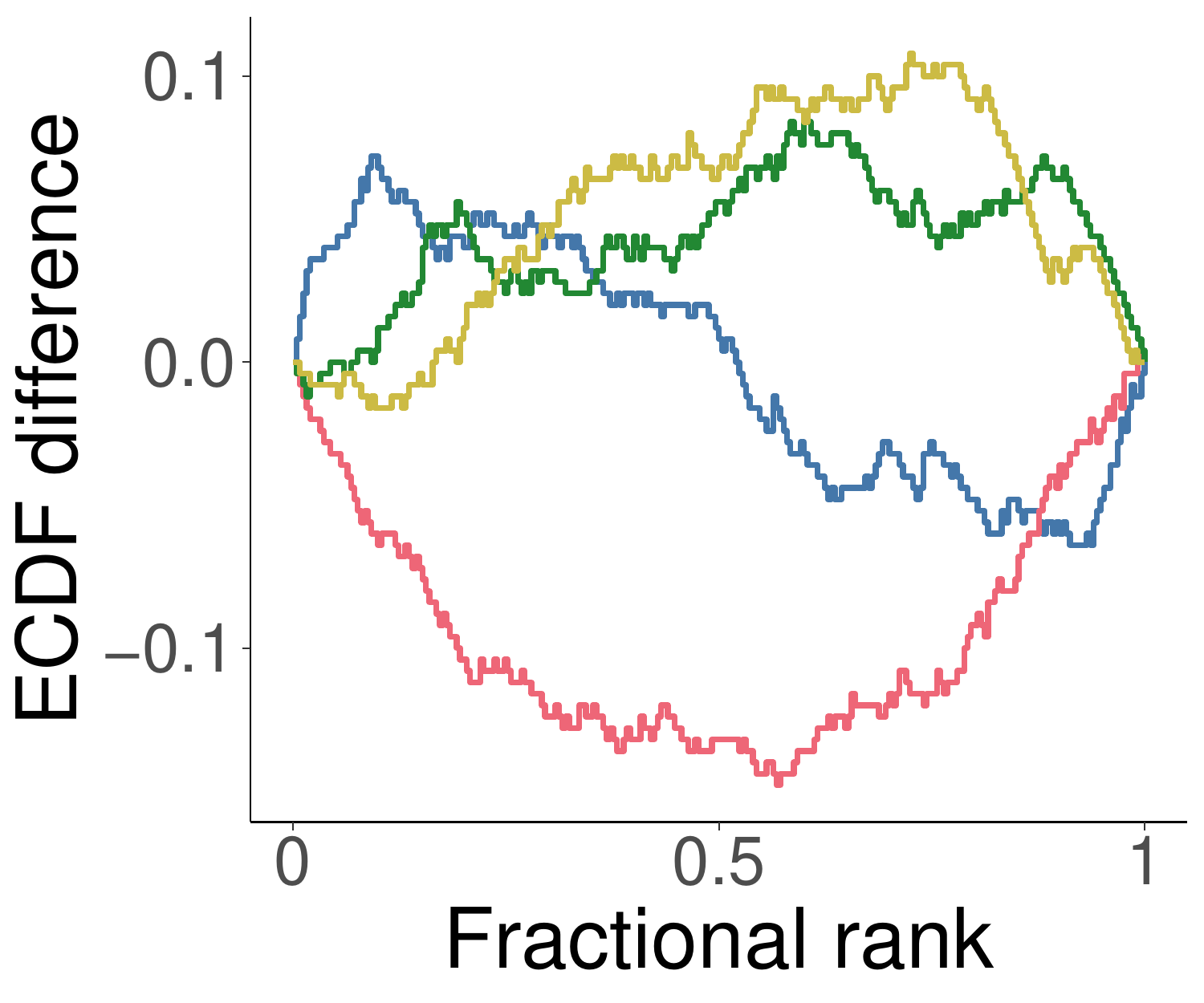}
        };
        \end{tikzpicture}
        \caption{
        We evaluate the hypothesis of the four samples on the left originating from the same underlying distribution by inspecting the distribution of fractional ranks among the joint rank-transformed, Eq.~\eqref{eq:fractional_ranks}, samples presented in the middle. When comparing multiple samples, we would like to take into account the within sample dependency, but also the between sample dependency introduced by the joint transformation. In Sections \ref{subsec: multisample simulated confidence bands} and \ref{subseq:simultaneous_confidence_bands_through_optimization} we extend our methods in order to provide simultaneous confidence bands for the ECDF and the ECDF difference plots shown on the right.}
        \label{fig:multi_chain_demo}
     \end{figure}

In this section, we extend the uniformity test of section \ref{sec:confidence_bands} to test whether multiple samples originate from the same underlying distribution.
In the case of multiple samples sharing the same distribution, the rank statistics of the values within each sample, when ranked jointly across all samples, are uniformly distributed on the interval $(1,\tilde{N})$, where $\tilde{N}$ is the total length of the combined sample \parencite{Vehtari2021}. Thus, instead of considering the sampled values directly, we consider the implied jointly rank-transformed values.

Due to this joint rank-transformation, the resulting chains are dependent on each other and the confidence intervals we construct in the following two sections are used to answer whether all the two or more samples originate from the same underlying distribution. In other words, in the case one or more of the ECDF trajectories leaves the confidence bands, we conclude that at least one of the samples exhibits larger than expected deviance from the other samples at hand.

An illustration of the connection between the sampled values, the corresponding fractional rank statistics and the two ECDF plots of these rank statistics are displayed in Figure \ref{fig:multi_chain_demo}. 

\subsection{Pointwise confidence bands}\label{sec:multi_sample_test_pointwise_bands}

An important distinction to the ECDF case considered in section \ref{sec:confidence_bands}, is the form of the marginal distribution at quantile $z_i$ when determining the adjusted coverage parameter $\gamma$. As our main application is the comparison of distributions induced by MCMC chains, we speak of the $L$ different samples as chains and assume all chains to have the same length $N$. We define $r_i$ as the vector (of length $L$) of joint ranks across chains smaller than or equal to the sample size $s_i = \lfloor z_i N L \rfloor$. That is, for each of the $L$ elements $r_{il}$ of $r_i$, we have
    \begin{equation}
        r_{il} = \left\{ \sum_{j=1}^N \mathbb{I}_{\{1, \ldots, s_i\}} \left(R(u_{lj} \given u) \right) \right\},
    \end{equation}
    where $u_{lj}$ is the $j$th draw of the $l$th chain before transformation, $R(u_{lj} \given u)$ is the rank of $u_{lj}$ within the vector $u$ of all draws across all chains, and $\mathbb{I}$ is the indicator function. Clearly, because of the definition of ranks, we know for all $i$ that
    \begin{equation}
        \label{hyp-constraint}
        \sum_{l=1}^L r_{il} = s_i,
    \end{equation}
    and we define the set of all $r_i$ satisfying \eqref{hyp-constraint} as $R_i$. Due to the pairwise independence of the original draws $u_{lj}$ (by assumption), the marginal distribution of $r_i$ at quantile $z_i$ is multivariate hypergeometric
    \begin{equation}
        r_{i} \sim \MHyp(\tilde{N}, s_i),
    \end{equation}
    where $\tilde{N} = (N_1, \ldots N_L)$ is the vector chain lengths (i.e., population sizes) and $N_1 = \ldots = N_L = N$ as we assume chains to have equal length. It is well known that, in this case, the marginal distribution of $r_{il}$, and thus the distribution defining the pointwise confidence bands, is hypergeometric
    \begin{equation}
        r_{il} \sim \Hyp(N, N (L - 1), s_i).
    \end{equation}

    \subsection{Simultaneous confidence bands through simulation}\label{subsec: multisample simulated confidence bands}
    In this section, we extend the simulation method presented in Section~\ref{subsubsec:single_sample_simulated_method} to comparison of multiple samples.
    Our aim is to define simultaneous confidence bands for the ECDFs of multiple, jointly rank-transformed distributions so that the interior of the simultaneous confidence bands jointly contains all trajectories induced by the rank-transformed distributions with rate $1 - \alpha$. To this end, we define $r_i$ and $s_i$ as in Section~\ref{sec:multi_sample_test_pointwise_bands} and denote the interior of the simultaneous confidence bands at quantile $z_i$ as $\tilde{I}_i(\gamma)$, with $\gamma$ being the adjusted coverage parameter to be determined.

    We continue the use of fractional ranks in the ECDF plots to provide illustrations independent of the length of the sampled chains. Suppose we have $L$ chains of length $N$. The fractional rank score $\tilde{r}_{il}$  corresponding to the $i$th value of the $l$th chain, $u_{li}$, is 
    \begin{equation}\label{eq:fractional_ranks}
        \tilde{r}_{il} = \frac{R(u_{lj} \given u)}{LN}.
    \end{equation}
    Instead of using the adjusted value of $\gamma$ to obtain the $1-\alpha$ level simultaneous confidence bands for a single ECDF trajectory, we adjust $\gamma$
    to account for the dependence between the samples introduced in the transformation into fractional ranks. That is, after choosing the evaluation quantiles $z_i$, we adjust $\gamma$ to find upper and lower simultaneous confidence bands satisfying
    \begin{equation}
        \Pr(L_i(\gamma)\leq F_l(z_i) \leq U_i(\gamma) \text{ for all } i\in \{1,\dots, K\},l \in \{1,\dots, L\}) = 1-\alpha,
    \end{equation}
    where $F_l$ is the ECDF of the fractional rank scores of the $l$th chain.
    
    We denote the CDF of the hypergeometric distribution as $\Hyp$ and its inverse as $\Hyp^{-1}$.
    The algorithm to approximate the adjusted coverage parameter $\gamma$ when comparing $L$ samples is as follows:
    
    \begin{enumerate}
        \item For $m=1,\ldots,M$:
            \begin{enumerate}
                \item For $l=1,\dots,L$, simulate $u_{l1}^m,\ldots,u_{lN}^m \sim {\rm uniform}(0,1)$.
                \item For $j=1,\ldots,N$ and $l=1,\ldots,L$, compute  $\tilde{r}_{jl}^m$.
                \item For $i=1,\ldots,K$ and $l=1,\ldots,L$, compute $F_l^m(z_i)$.
                \item For $i=1,\ldots,K$ and $l=1,\ldots,L$, compute $$\Hyp\left(NF_l^m(z_i) \given N, (L-1)N, s_i\right) \text{ and } \Hyp\left(NF_l^m(z_i)-1 \given N, (L-1)N, s_i\right),$$ where $s_i = \lfloor z_i N L \rfloor$.
                \item Find the minimum probability
                $$\gamma^m = 2\min_{i,l}\left\lbrace\min\left(\Hyp\left(NF_l^m(z_i) \given N, (L-1)N, s_i\right), 1-\Hyp\left(NF_l^m(z_i)-1 \given N, (L-1)N, s_i\right)\right)\right\rbrace.$$
            \end{enumerate}
        \item Set $\gamma$ to be the $100\alpha$ percentile of $\lbrace\gamma^1,\ldots,\gamma^M\rbrace$.
        \item Form the confidence bands 
        $$\left[L_i(\gamma),U_i(\gamma)\right] = \left[\Hyp^{-1} \left(\frac{\gamma}{2} \given N, N (L - 1), s_i \right),\Hyp^{-1} \left(1 - \frac{\gamma}{2} \given N, N (L - 1), s_i \right)\right],$$
        for $i=1,\dots,K$.
    \end{enumerate}
    
\subsection{Simultaneous confidence bands through optimization}\label{subseq:simultaneous_confidence_bands_through_optimization}

    In this section, we extend the optimization method presented in Section~\ref{subsec:single_sample_optimization_method} to comparison of multiple samples.
    With the marginal distribution of $r_{il}$ being hypergeometric, the rank interior $I_i(\gamma)$ for $z_i$ is given by
    \begin{equation}
        I_i(\gamma) = \left\{ r \in R_i \given \forall r_l \in r : \; \Hyp^{-1} \left(\frac{\gamma}{2} \given N, N (L - 1), s_i \right) \leq r_l \leq \Hyp^{-1} \left(1 - \frac{\gamma}{2} \given N, N (L - 1), s_i \right) \right\}.
    \end{equation}
    We treat the borders between interior and exterior as belonging to the interior. Based on $I_i(\gamma)$, we can again easily obtain $\tilde{I}_i(\gamma)$, as $r \in I_i(\gamma)$ is equivalent to $r/N \in \tilde{I}_i(\gamma)$. 
    
    The remainder of the proof proceeds similar to the one-sample case, except that we replace the binomial distribution with the (multivariate) hypergeometric distribution. A (multivariate) rank ECDF trajectory, defined as
    \begin{equation}
        t_0^K = \left((z_i)_{i=0}^K, (r_i)_{i=0}^K\right),
    \end{equation}
    where $z_0 = 0$ and $z_K = 1$, stays within the simultaneous confidence bands completely if and only if $r_i \in I_i(\gamma)$ for all $i \in \{0, \ldots, K\}$. If we denote the set of trajectories fulfilling $r_i \in I_i$ as $T_i$, we can write the set of trajectories which are completely in the interior of the simultaneous confidence bands as 
    \begin{equation}
        T(\gamma) = \bigcap_{i = 0}^K T_i(\gamma).
    \end{equation}
    In order for the simultaneous confidence bands to have a confidence level $1-\alpha$, we must satisfy
    \begin{equation}
        \Pr\left(T(\gamma) \right) = 1 - \alpha.
    \end{equation}
    
    Due to the pairwise independence of the original draws $u_{lj}$ (by assumption), the distribution of the rank ECDF trajectories again exhibits a similar Markovian property as in the single sample case. That is, any ECDF value $F(z_{i+1})$ beyond a given point $z_i$ only depends on $F(z_i)$ but not on the earlier history of the ECDF trajectory. This implies that, under the assumption of all chains coming from the same underlying distribution, the growth $r_{i+1} - r_{i}$ of the ECDF from $z_{i}$ to $z_{i+1}$ is multivariate hypergeometric with $\tilde{N}_i = \tilde{N} - r_{i}$ and sample size $\tilde{s}_{i+1} = s_{i+1} - s_i$.
    Accordingly, we have
    \begin{equation} 
        \Pr(r_{i+1} \given r_{i}) = p_{\MHyp}\left(r_{i+1} - r_{i} \given \tilde{N}_i, \tilde{s}_{i+1}  \right),
    \end{equation}
    where $p_{\MHyp}$ denotes the discrete PDF of the multivariate hypergeometric distribution. The probability for $r_{i+1} = k \in I_{i+1}$ to occur in a rank ECDF trajectory $t_0^K$ which stayed in the simultaneous confidence bands until point $i$, that is, for which we have
    \begin{equation}
        t_0^i \in \bigcap_{j = 0}^{i} T_j(\gamma) 
    \end{equation}
    can thus be written recursively as
    \begin{equation}
    \label{rec-hyp}
        \Pr \left(r_{i+1} = k \cap \bigcap_{j = 0}^{i} T_j(\gamma) \right) = \sum_{m \in I_{i}} \; \Pr \left(r_{i} = m \cap \bigcap_{n = 0}^{i-1} T_n(\gamma) \right) \; \Pr(r_{i+1} = k \given r_{i} = m).
    \end{equation}
    The recursion is initialized at $z_0 = 0$ with $\Pr(x_{0} = (0, \ldots, 0)) = 1$ so that $\Pr(T_0(\gamma)) = 1$ for all $\gamma \in [0, 1]$. At any point $i \in \{0, \ldots, K\}$, we can obtain
    \begin{equation}
        \Pr \left(\bigcap_{j = 0}^{i} T_j(\gamma) \right) = \sum_{m \in I_{i}} \; \Pr \left(r_{i} = m \cap \bigcap_{n = 0}^{i-1} T_n(\gamma) \right),
    \end{equation}
    which is equal to $\Pr(T(\gamma))$ when arriving at $i = K$. Clearly, $\Pr(T(\gamma))$ is monotonically decreasing but not continuous in $\gamma$ due to the discrete nature of the (multivariate) hypergeometric distribution. We can compute
    \begin{equation} 
        \hat{\gamma} = \argmin_{\gamma \in [0, \alpha]} \; |1 - \alpha - \Pr \left(T(\gamma) \right)| 
    \end{equation}
    using a unidimensional derivative-free optimizer. In our experiments, the optimizer proposed by \textcite{brent1973} converged in all cases to $\hat{\gamma}$ values implying a simultaneous confidence level very close to the nominal $1-\alpha$.
    
    Unfortunately, evaluating Eq. \eqref{rec-hyp} suffers from combinatorial explosion as the $R_i$ are $L$-dimensional sets constraint only by Equation \eqref{hyp-constraint} and as $\Pr(r_{i+1} = k \given r_{i} = m)$ has to be computed for all combinations of elements $k \in I_{i+1}$ and $m \in I_{i+1}$ at each point $i$. Several measures can be taken to reduce the complexity of the computation. First, the ranks of one of the $L$ chains are redundant as they follow deterministically from Equation \eqref{hyp-constraint} based on the ranks of the other $L-1$ chains. This implies in particular that the $2$-chain case has the same computational complexity as the one-sample case as only one of the two chains needs to be evaluated. Second, due to a-priori symmetry of the chains, we can, without loss of generality, assume at the first non-zero quantile $z_1$ that the elements $r_{1l}$ of $r_1$ are ordered such that $r_{11} \leq r_{12} \leq \ldots \leq r_{1L}$. This reduces the number of trajectories to be evaluated by a factor of $L (L + 1) / 2$. Still even with these measures in place, computation will scale badly with $L$, and the simulation based method, which scales almost linearly, or grid-based interpolation from pre-computed values is faster for larger number of chains.
  
\section{Numerical Experiments and Power Analysis}\label{sec:experiments}
In this section, we provide insights into how the plots produced by our proposed methods should be interpreted. In each of the following cases, we link together the histogram, ECDF plot, and the ECDF difference plot. The code for the experiments and plots is available at \url{https://github.com/TeemuSailynoja/simultaneous-confidence-bands}.

    \subsection{Uniformity of a Single Sample}\label{subsec:single_sample_examples}
        We begin by providing two examples connecting the shape of the histogram of the transformed sample to the characteristics of the corresponding ECDF and ECDF difference plots with basic discrepancies between the sample and the comparison distribution. After this we illustrate an application of our method as part of a workflow to detect issues in model implementation or the computation of the posterior distribution. Lastly we provide power analysis comparing the performance of our proposed method to existing state of the art tests for uniformity.
        
        With the exception of the power analysis tests in \ref{subsubseq:single_sample_power_analysis} where the samples are drawn directly from a continuous uniform distribution, the samples in the following examples are transformed to the unit interval from their respective sampling distributions through empirical PIT and are tested against the hypothesis of discrete uniformity.
        
    \subsubsection{Effect of Difference in Sample Mean}\label{subec:sample_mean}
        \begin{figure}
        \centering
            \subfloat[Histogram of PIT values]{
            \includegraphics[width=0.28\textwidth]{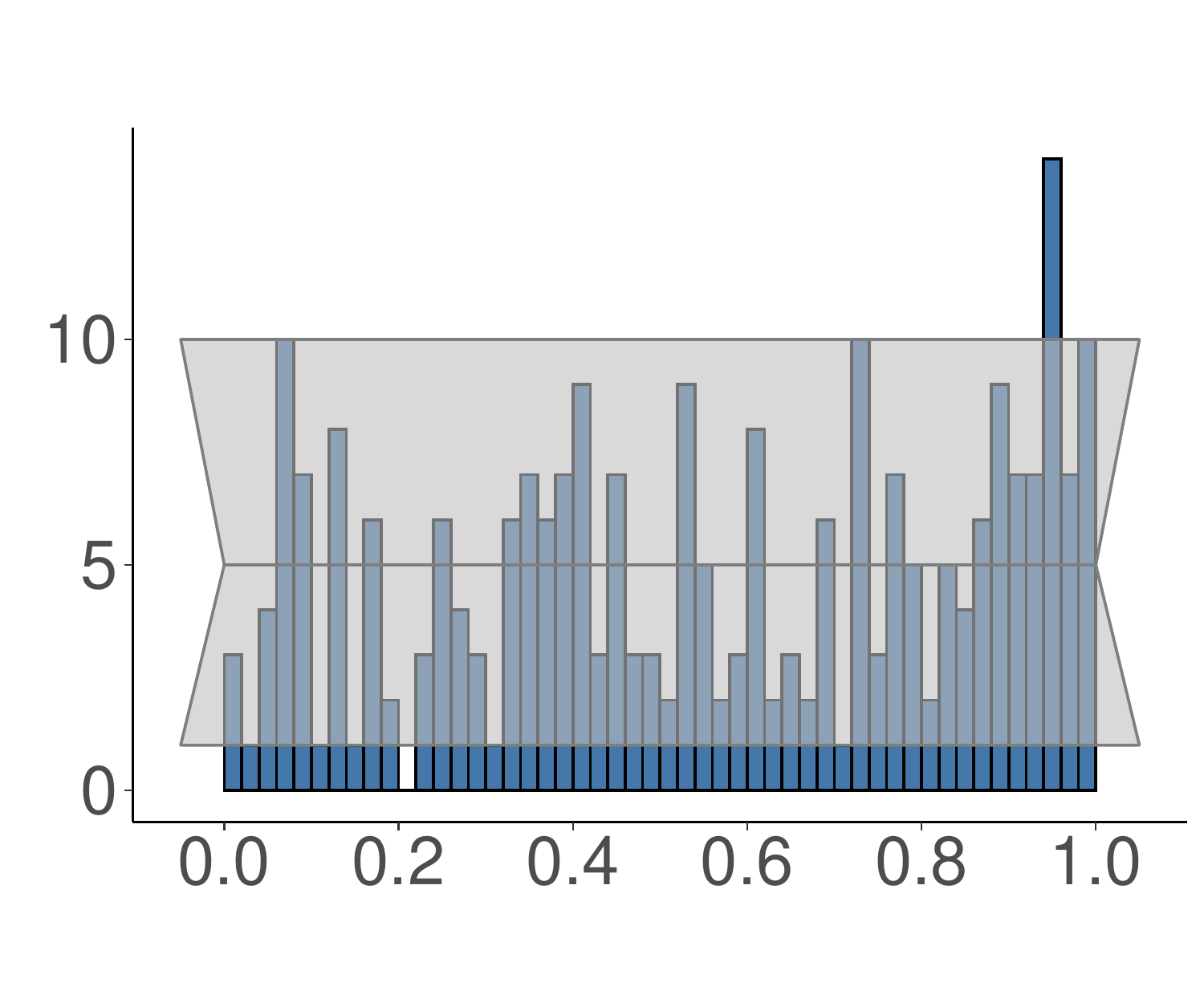}}~
            \subfloat[ECDF plot of PIT values]{
            \includegraphics[width=0.28\textwidth]{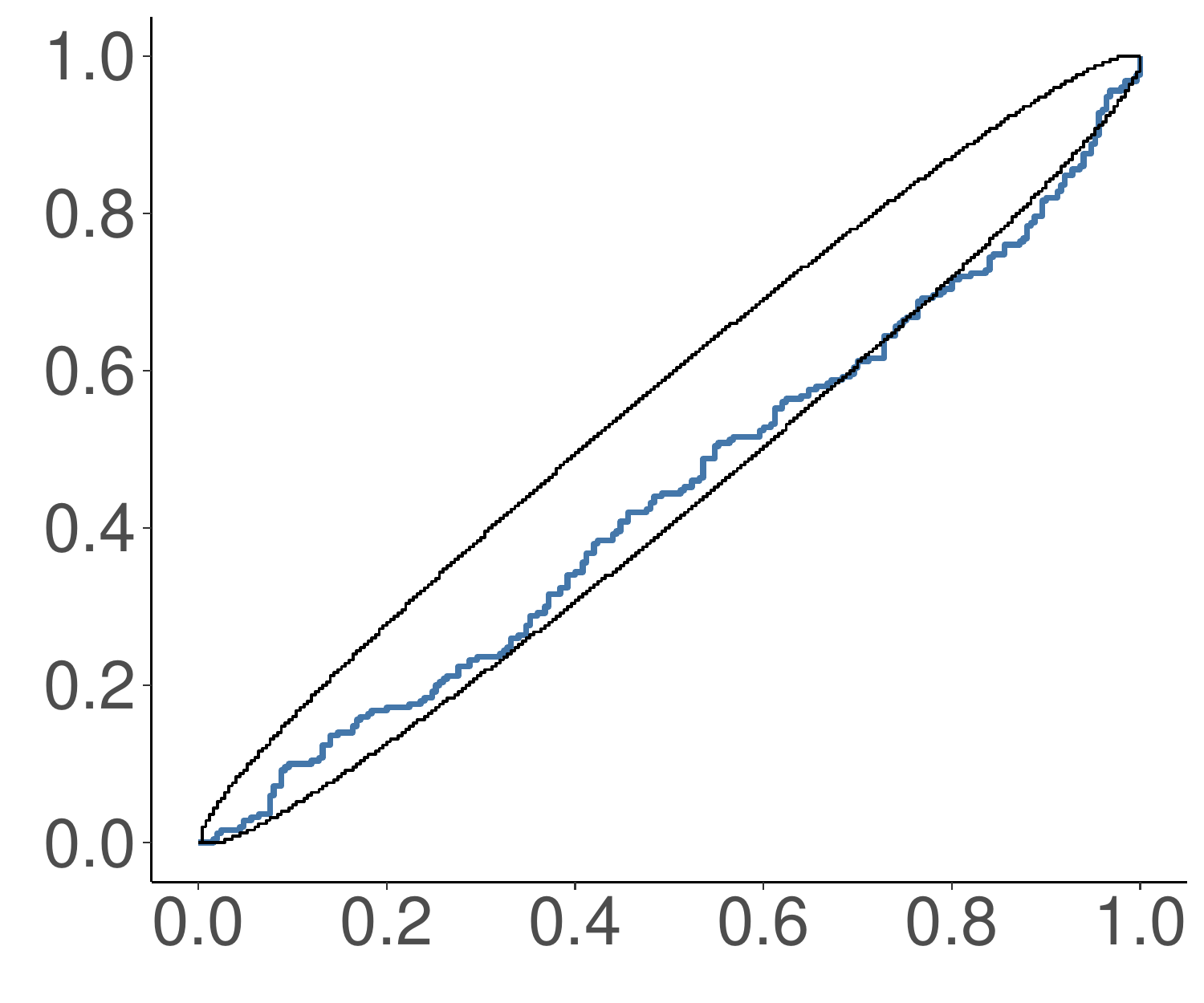}}~
            \subfloat[ECDF difference plot of PIT values]{
            \includegraphics[width=0.28\textwidth]{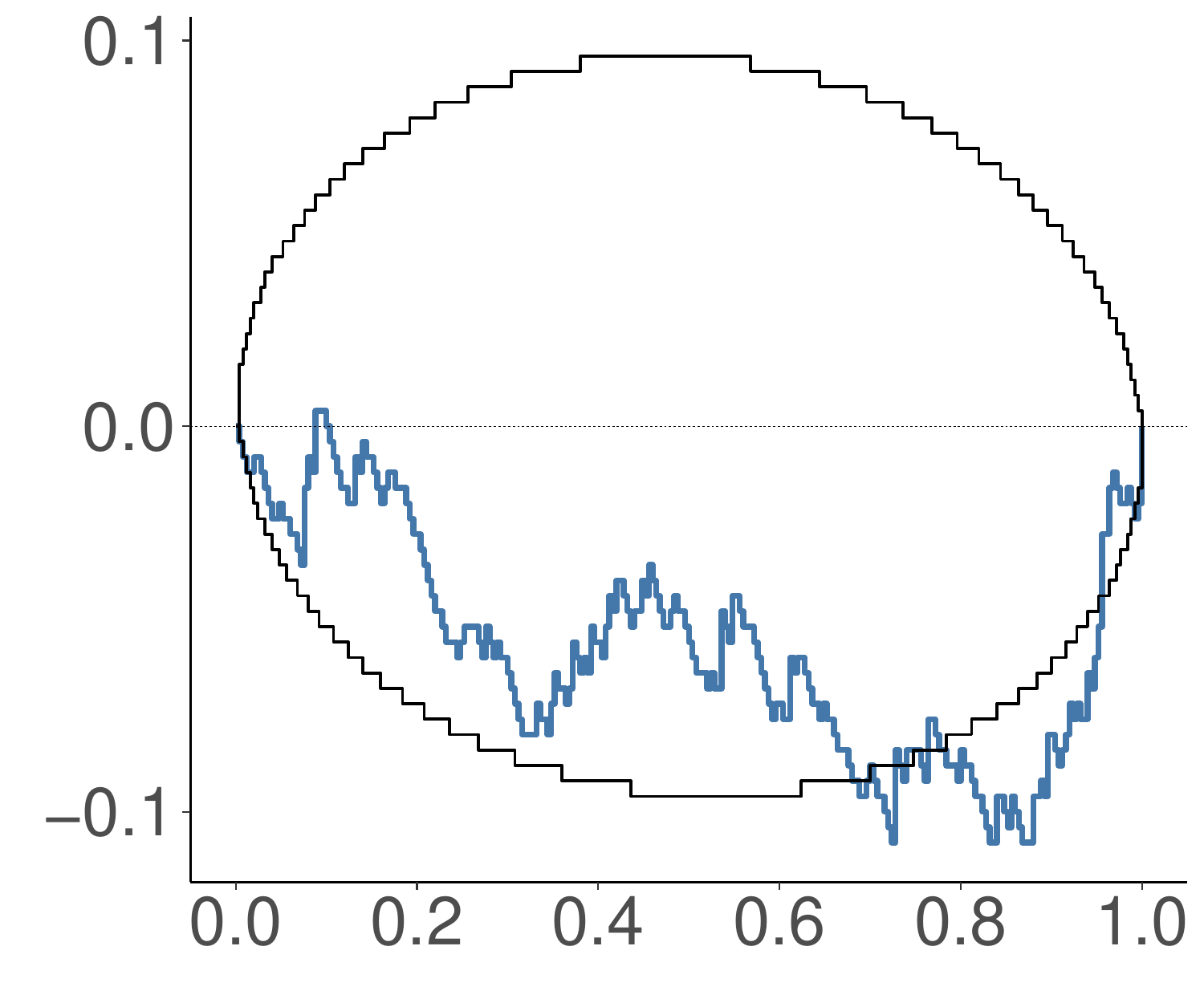}}
            \caption{Effect of the sample mean. The histogram, ECDF plot, and ECDF difference plot of the empirical PIT values of $y=y_1,\dots,y_{250}\sim\normal(0.25,1)$ with respect to $x^ i=x^ i_1,\dots,x^ i_{250}\sim\normal(0,1)$ for $i=1,\dots,N$. The larger than expected mean of the sample is somewhat visible as a slant to the right in the histogram, whereas the ECDF difference plot displays a clear $\cup$-shape. In the histogram 95\% confidence intervals are provided for each of the 50 bins and the ECDF plots show the 95\% simultaneous confidence bands.}
            \label{fig:single_moved_bands}
        \end{figure}
        To observe the typical characteristics of a sample with a mean different than that of the comparison distribution, we draw $y=y_1,\dots,y_N\sim\normal(0.25,1)$ and N independent comparison samples $x^i=x^i_1,\dots,x^ i_N\sim\normal(0,1)$ with $N=250$. We then test for $y$ being standard normal distributed by transforming the sampled values to the unit interval through empirical PIT. Figure \ref{fig:single_moved_bands}(a) shows the histogram of the transformed sample exhibiting a higher than expected mean. As seen in the figure, a shift in the sample mean leads to the histogram being slanted towards the direction of the shift.
        The ECDF plot in Figure~\ref{fig:single_moved_bands}(b), shows this shift through the ECDF of the PIT values remaining under the theoretical CDF, which is also seen in the ECDF difference plot in Figure~\ref{fig:single_moved_bands}(c).
        If the sample in question would instead have a mean lower than expected, the histogram would be slanted to the left and the behaviour of the resulting ECDF plot and ECDF difference plot would be reversed. That is, the ECDF plot would stay above the theoretical CDF as a higher than expected density is covered at low fractional ranks and the ECDF difference plot would respectively show a $\cap$-shape above the zero level.
        
        \subsubsection{Effect of Difference in Sample Variance}\label{subec:sample_var}
        \begin{figure}
            \centering
            \subfloat[Histogram of PIT values]{\includegraphics[width=0.28\textwidth]{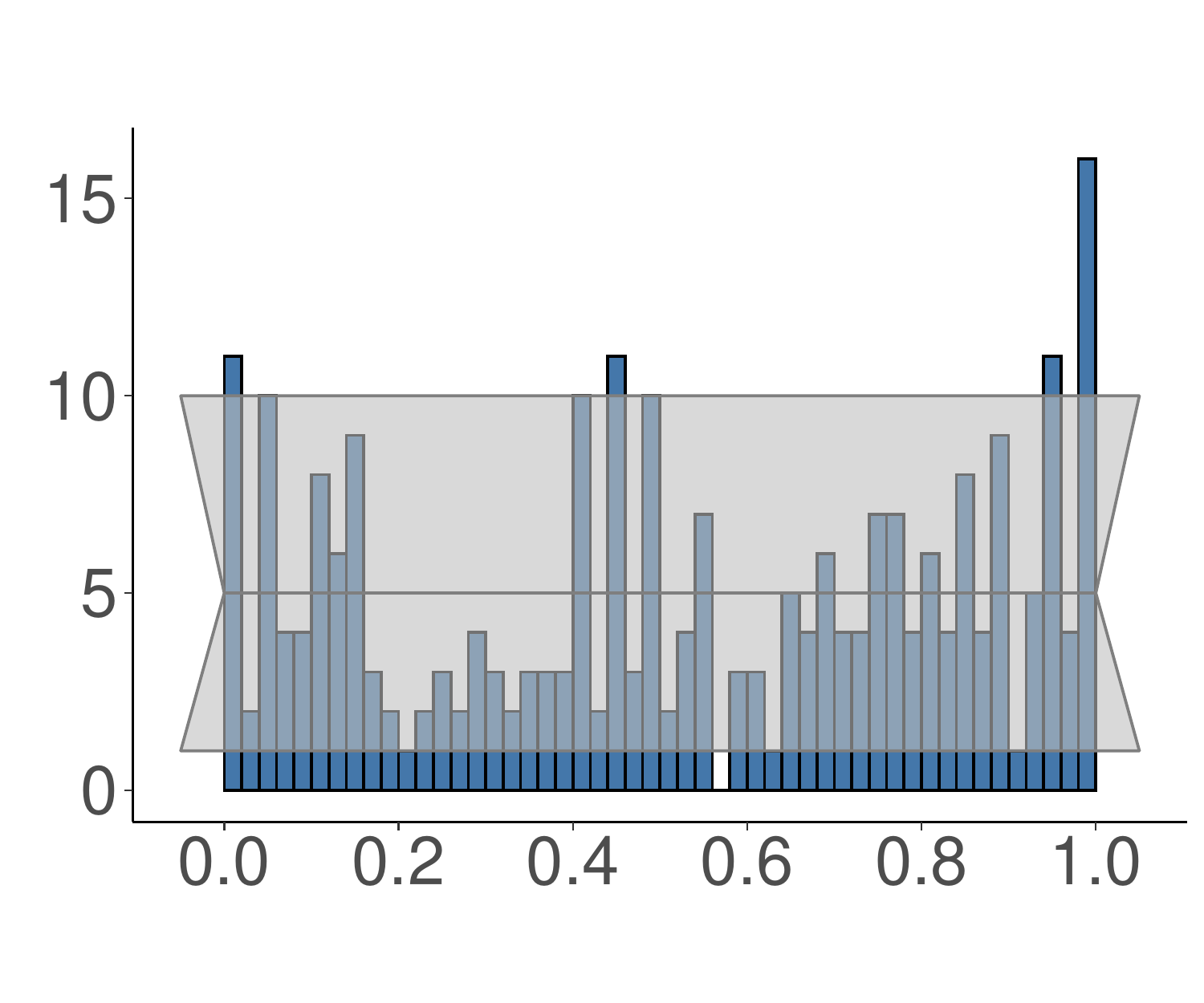}}~
            \subfloat[ECDF plot of PIT values]{\includegraphics[width=0.28\textwidth]{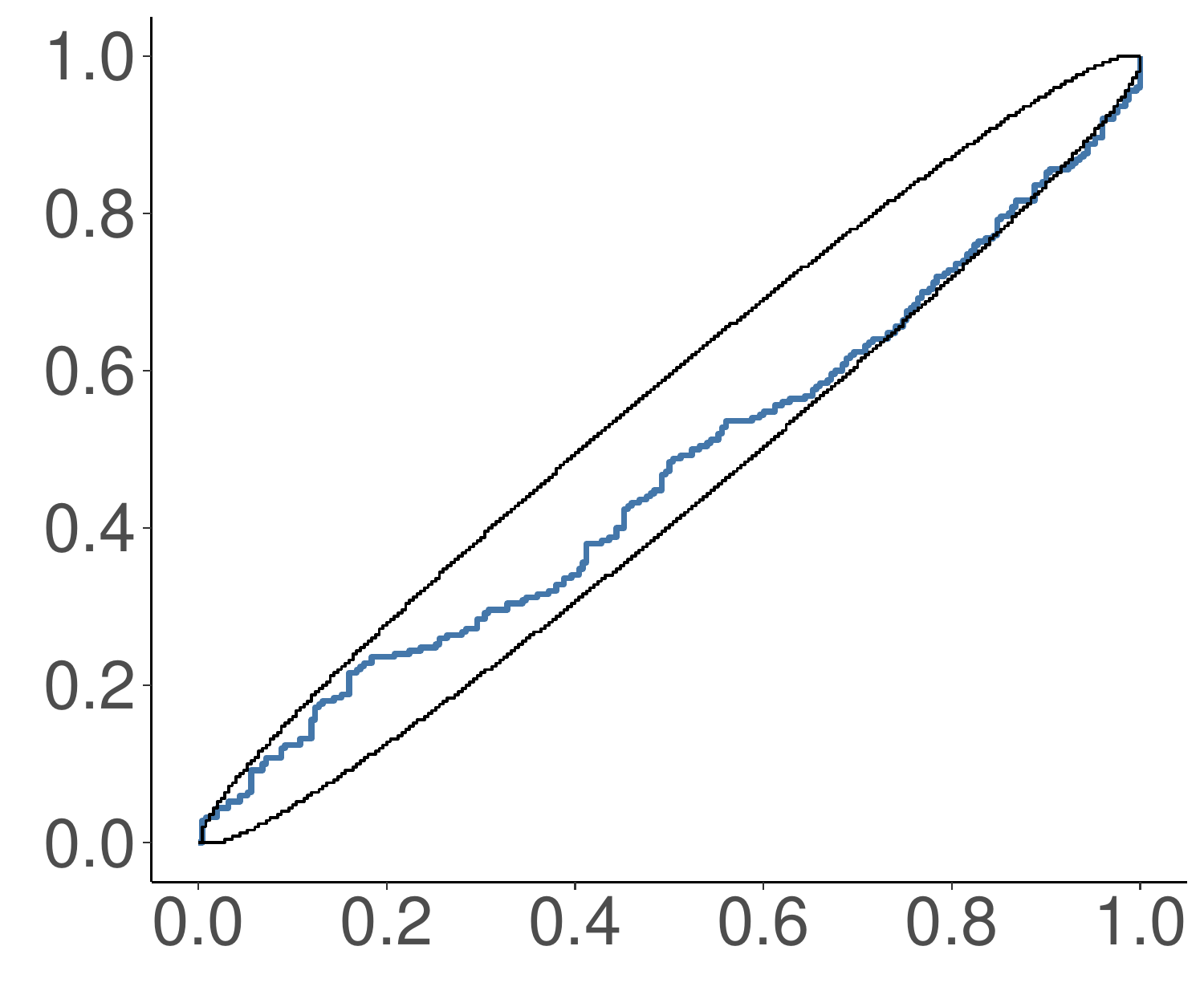}}~
            \subfloat[ECDF difference plot of PIT values]{\includegraphics[width=0.28\textwidth]{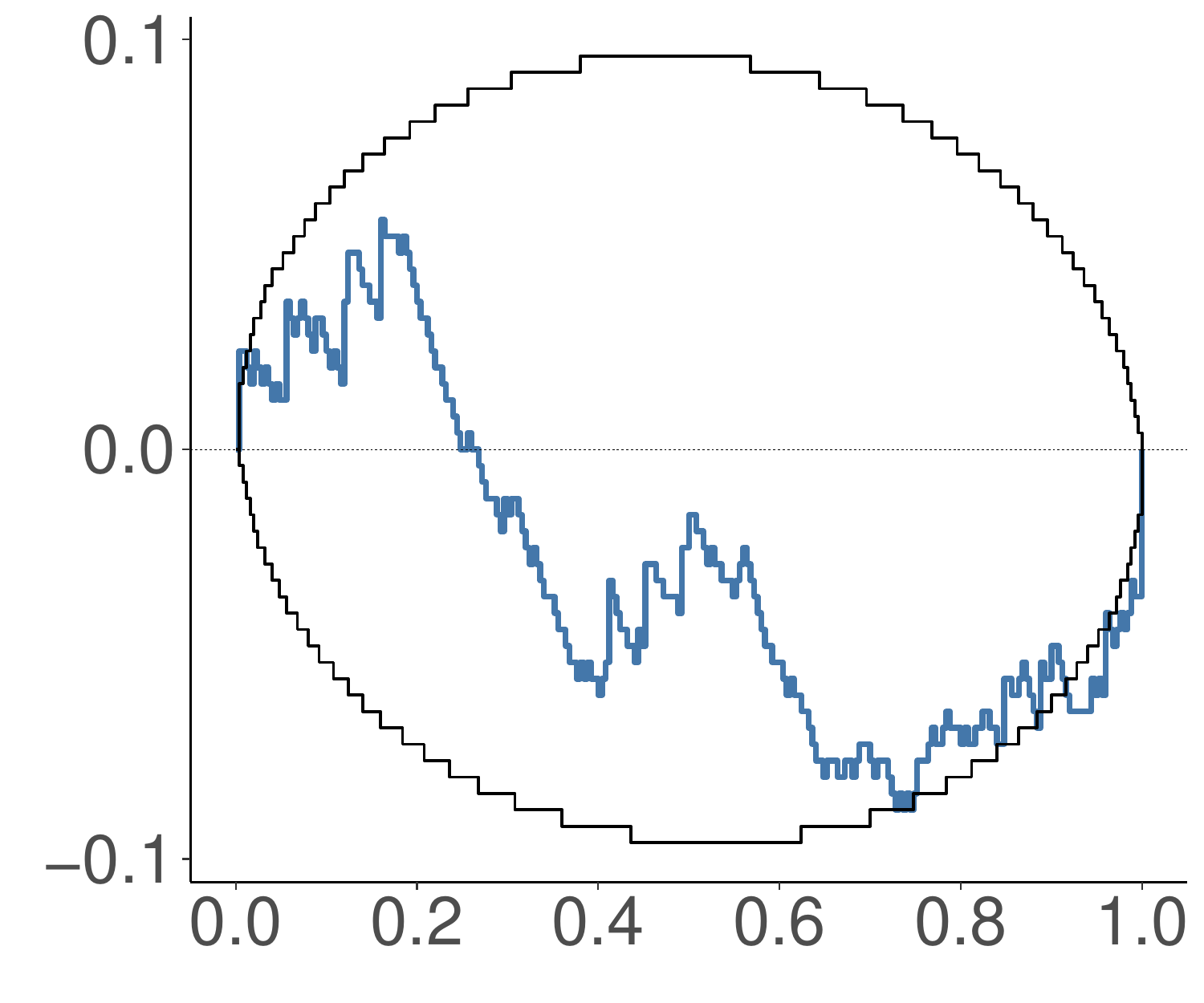}}
            \caption{Effect of the sample variance. The histogram, ECDF plot, and ECDF difference plot of the empirical PIT values of $y=y_1,\dots,y_{250}\sim\normal(0,1.25)$ with respect to $x^ i=x^ i_1,\dots,x^ i_{250}\sim\normal(0,1)$ for $i=1,\dots,,N$. The larger than expected variance of the sample is visible as a $\cup$-shape in the histogram, whereas the ECDF difference plot displays rapid growth near the ends of the interval as a larger than expected number of values is covered. In the histogram, 95\% confidence intervals are provided for each of the 50 bins and the ECDF plots show the 95\% simultaneous confidence bands.}
            \label{fig:single_scaled_bands}
        \end{figure}
        Next, we investigate an example where the sample has a higher than expected variance. To this end we draw $y = y_1,\dots,y_N \sim \normal(0,1.25)$ and for each $y_i$ a standard uniform comparison sample $x^i = x^ i_1,\dots,x^ i_N\sim\normal(0,1)$ with $N=250$.
        Figure~\ref{fig:single_scaled_bands}(a) shows the histogram of the empirical PIT values. In general, a larger than expected variance leads to a $\cup$-shaped histogram and one can indeed see some of the histogram bins breaching the 95\% confidence bounds.
        In the ECDF plot shown in Figure~\ref{fig:single_scaled_bands}(b), the larger than expected variance leads to faster than expected growth near the edges and slower than expected growth in the middle.
        The shape is more clearly seen in the ECDF difference plot in Figure~\ref{fig:single_scaled_bands}(c) depicting the difference between the ECDF and the theoretical CDF.
        If the sample would instead present a variance lower than expected, the histogram would be $\cap$-shaped and the behaviour of the resulting ECDF plot and ECDF difference plot would be reversed. 
        In the ECDF plot this is shown as faster increase near the middle. 
        In general, the ECDF difference plot is decreasing when a smaller than expected density of samples is covered, and correspondingly increases when covering a higher than expected density.

\subsubsection{Simulation Based Calibration: Eight Schools}\label{subsubsec:eight_schools_SBC}

The eight schools
\parencite{bda3}, is a classic hierarchical model example. The training course effects $\theta_j$ in eight schools are modelled using an hierarchical varying intercept model.

If the model is constructed with the centered parameterization, the posterior distribution exhibits a funnel shape contracting to a region of high curvature near the population mean $\mu$ when sampled with small values of the population standard deviation $\tau$. This property makes exploring the distribution of $\tau$ difficult for many MCMC methods. The centered parameterization $(\mu, \sigma, \mu_0, \tau)$ of the problem is as follows:
\begin{align}
    \mu_j &\sim \normal(\mu_0, \tau)\\
    y_j &\sim \normal(\mu_j, \sigma_j).
\end{align}
\textcite{Cook+Gelman+Rubin:2006} proposed a simulation-based calibration method for validating Bayesian inference software. The idea is based on the fact we can factor the joint distribution of data $y$ and parameters $\theta$ in two ways
\begin{align}
  \pi(y,\theta) = \pi(y|\theta)\pi(\theta) = \pi(\theta|y)\pi(y).
\end{align}
By considering $\theta'$ and $\theta''$ the joint distribution is 
\begin{align}
  \pi(y,\theta',\theta'') = \pi(y)\pi(\theta'|y)\pi(\theta''|y),
\end{align}
and it's easy to see that $\theta'$ and $\theta''$ have the same distribution conditionally on $y$. If write the joint distribution in an alternative way
\begin{align}
  \pi(y,\theta',\theta'') = \pi(\theta')\pi(y|\theta')\pi(\theta''|y),
\end{align}
$\theta'$ and $\theta''$ still have the same distribution conditionally on $y$. We can sample from the joint distribution $\pi(y,\theta',\theta'')$ by first sampling from $\pi(\theta')$ and  $\pi(y|\theta')$, which is usually easy for generative models. The last step is to sample from the conditional $\pi(\theta|y)$, which is usually not trivial and instead, for example, a Markov chain Monte Carlo algorithm is used. We can validate the algorithm and its implementation used to sample from $\pi(\theta''|y)$ by checking that the samples obtained have the same  distribution as $\theta'$ (conditionally on $y)$.

\textcite{Cook+Gelman+Rubin:2006} operationalize the approach by drawing $\theta'_i$ from $\pi(\theta')$, generating data $y_i \sim \pi(y_i|\theta'_i)$ and then using the algorithm to be validated to draw a sample $\theta''_1,\ldots,\theta''_S \sim \pi(\theta''|y_i)$. If the algorithm and its implementation are correct, then $\theta'_i,\theta''_1,\ldots,\theta''_S$ conditional on $y_i$ are draws from the same distribution. \textcite{Cook+Gelman+Rubin:2006} propose to compute empirical PIT valued for $\theta'_i$ that they show to be uniformly distributed given $S \to \infty$. The process is repeated for $i=1,\ldots,N$ and $N$ empirical PIT values are used for testing.  \textcite{Cook+Gelman+Rubin:2006} propose to use $\chi^2$-test for the inverse of the normal CDF of the empirical PIT values. However, with finite $S$ this approach doesn't correctly take into account the discreteness or the effect of correlated sample from Markov chain \textcite{Gelman:correction}.

By thinning $\theta_1^{''},\ldots,\theta_S^{''}$ to be approximately independent, the uniformity of empirical PIT values can be tested with the approach presented in this paper.  See Appendix \hyperref[appendix: autocorrelated samples]{A} for more on thinning.

\begin{figure}[t]
    \centering
    \subfloat[Histogram of PIT values of prior draws]{
    \includegraphics[width=0.28\textwidth]{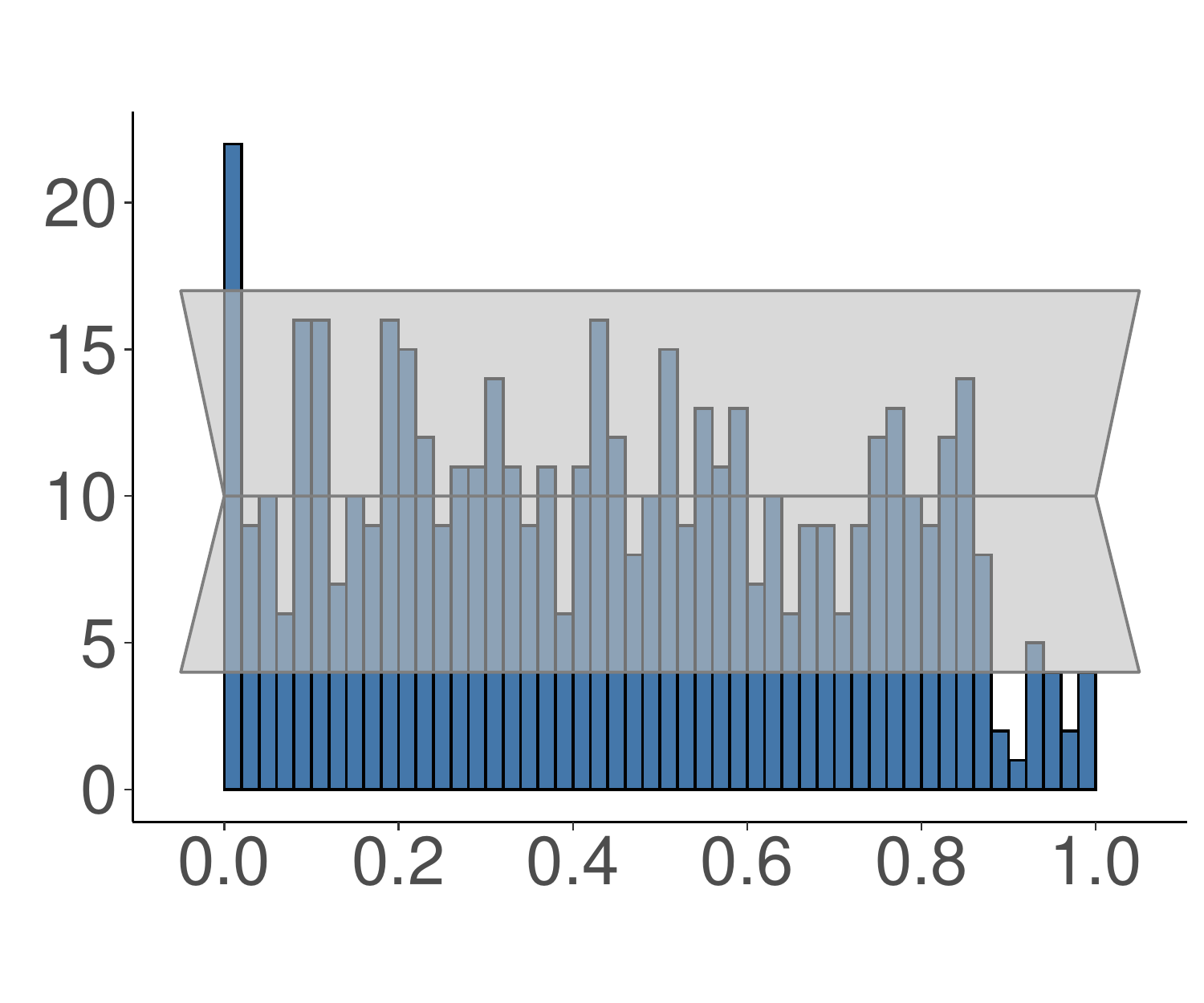}}~
    \subfloat[ECDF plot of PIT values]{
    \includegraphics[width=0.28\textwidth]{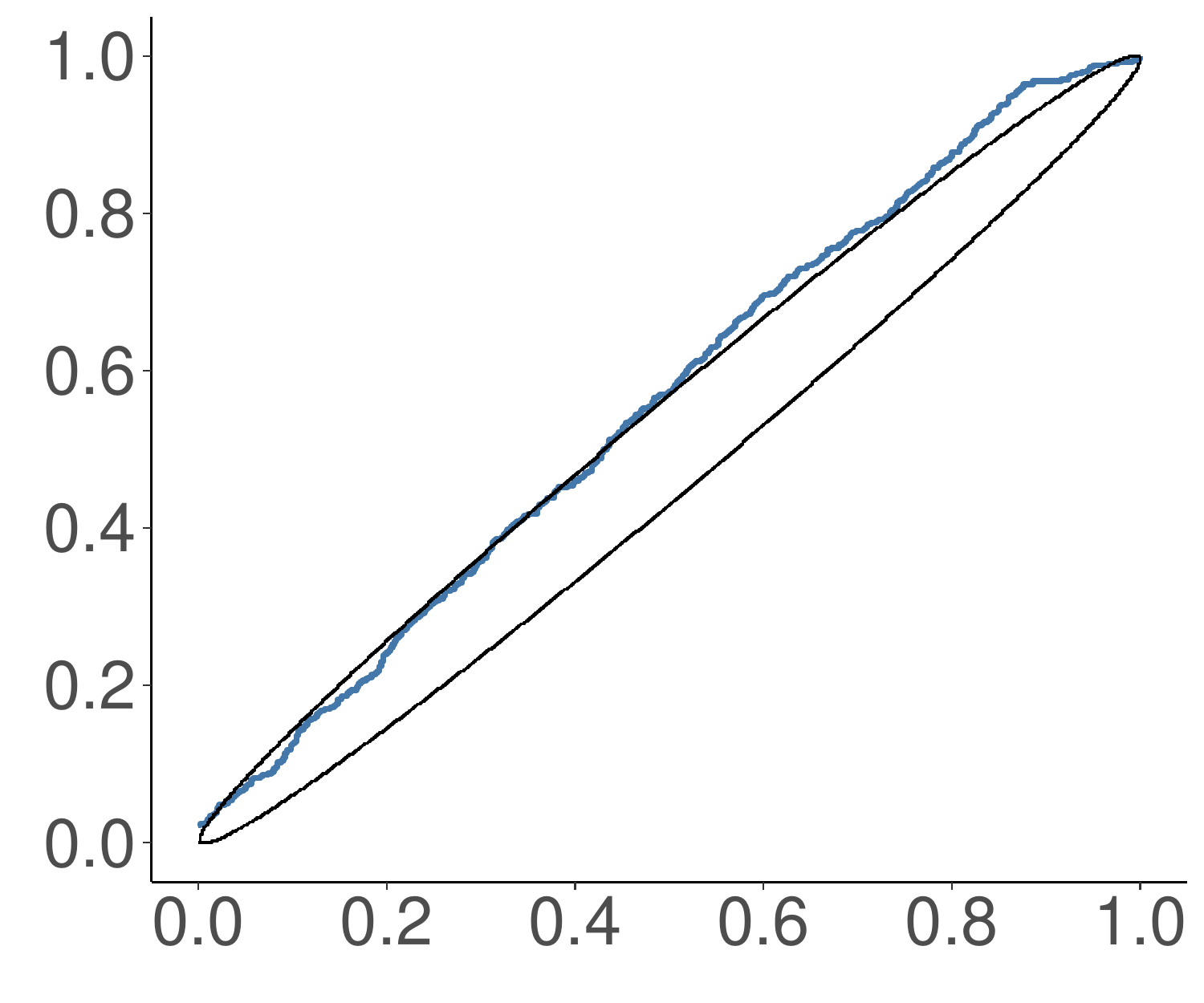}}~
    \subfloat[ECDF difference plot of PIT values]{
    \includegraphics[width=0.28\textwidth]{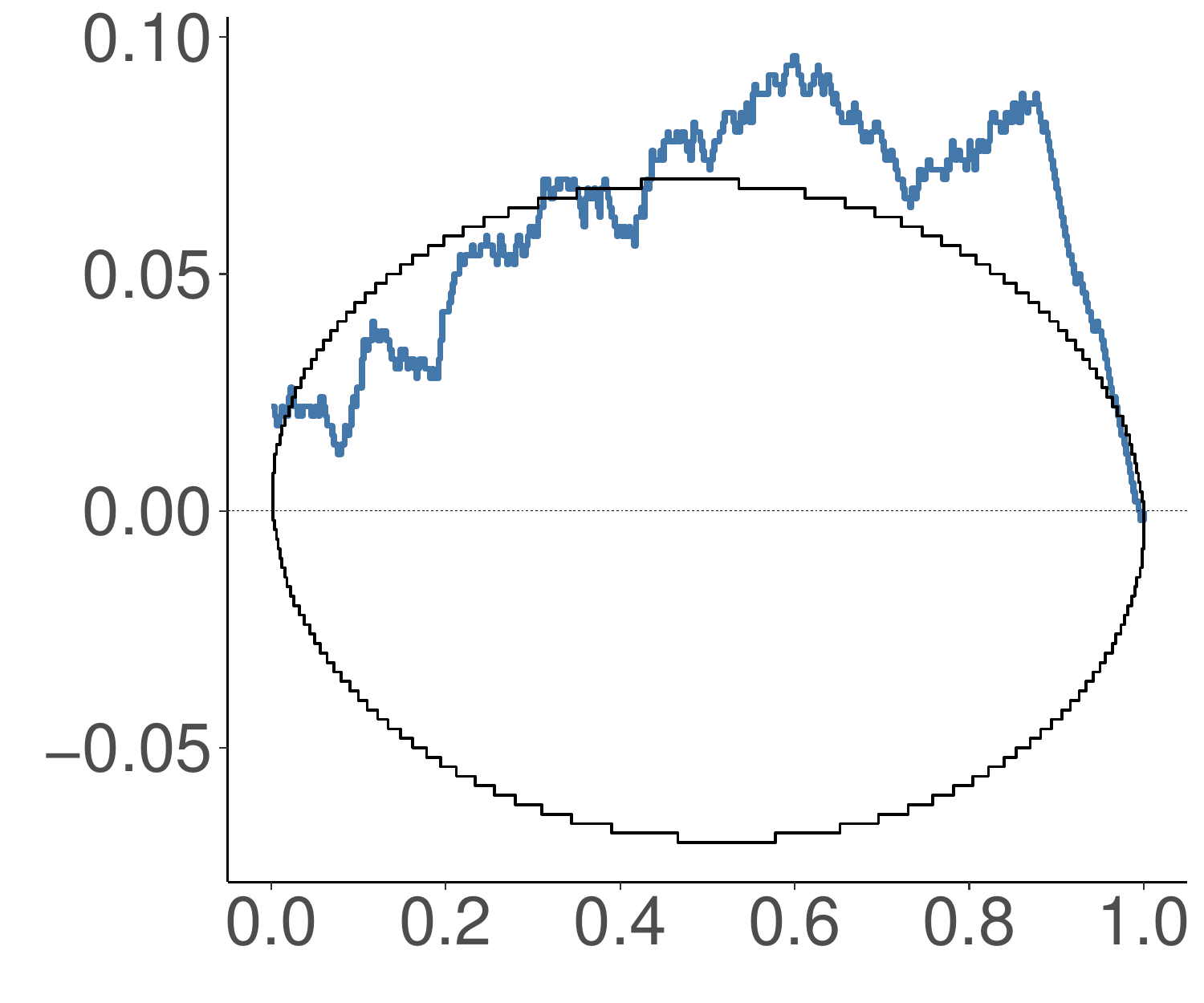}}
    \caption{SBC of the parameter $\tau$ in the centered parameterization eight schools model. The empirical PIT values of the prior draws of $\tau$, when compared to the corresponding posterior samples, show a strong underrepresentation of large PIT values in both the histogram and the ECDF difference plot. This indicates that the MCMC is not sampling correctly from the target distribution.}
    \label{fig:eight_schools_SBC}
\end{figure}

Figure \ref{fig:eight_schools_SBC} shows the histogram and ECDF plots of 500 prior draws of the population standard deviation $\tau$, each ranked based on a thinned posterior sample of 150 draws obtained from a chain of 3000 draws. The graphical test rejects the hypothesis of the prior draws being uniform, moreover the ECDF plots show that the prior draws of the parameter $\tau$ ranked in relation to the posterior samples obtained from the centered parameterization of the eight schools model are skewed to small ranks. This suggests that the MCMC is not sampling correctly from the target distribution (which in this case is known to be caused by inability to reach the narrow funnel part of the posterior). 

In Section \ref{subsubsec:eight_schools_rank_plots}, we will return to the eight schools model by providing further analysis on the convergence of individual chains in the centered parameterization case and illustrating how our method can be used to detect these convergence issues.

\subsubsection{Power analysis}\label{subsubseq:single_sample_power_analysis}
As our primary focus is on providing a graphical uniformity test, which gives the user useful information regarding the possible deviations from uniformity, we want to also ensure that the overall performance of our test is, if not the best, competitive with tests aimed at accurately detecting specific deviations from uniformity. To this end, we compare the sensitivity of our method with existing tests for uniformity, by considering the rejection rate of samples drawn from uniform distribution and then transformed according to the following three transformation families \textcite{Marhuenda2005} use in their article comparing various tests for uniformity:
\begin{align}\label{eq:transformation families}
f_{A,k}(x) &= 1-(1-x)^k, \quad 0\leq x\leq 1,\\
f_{B,k}(x) &=
\begin{cases}
2^{k-1}x^k & \text{if }0\leq x \leq 0.5\\
1-2^{k-1}(1-x)^k & \text{if }0.5<x\leq 1
\end{cases}\\
f_{C,k}(x) &= \begin{cases}
0.5 - 2^{k-1}(0.5-x)^k &  \text{if }0\leq x \leq 0.5\\
0.5 + 2^{k-1}(x-0.5)^k & \text{if }0.5 < x\leq 1.
\end{cases}
\end{align}
As \textcite{Marhuenda2005} offer an extensive comparison of tests, we limit our comparison to the test specifically recommended to target each of the transformation families in addition to the widely known Kolmogorov-Smirnov test. For each of the test statistics, a critical value is calculated and samples exceeding that value are rejected.

For transformation family A, the recommended test is the mean distance of the $i$th value of the ordered sample $u_{(i)}$ from the expected value $i/(N+1)$: 
    \begin{equation}
        T_1 = \sum_{i=1}^N\frac{|u_{(i)} - i/(N+1)|}{N}.
    \end{equation}
For family B, the smooth goodness-of-fit test, $N_h$, introduced by \textcite{Neyman1937} is recommended with the dimension $h$ chosen according to the method recommended by \textcite{Ledwina1994} resulting in the test statistic $N_S$, which also has the best overall performance across the transformation families.
The test recommended for transformation family C is the statistic recommended by \textcite{WATSON1961},
\begin{equation}
    U^2 = W^2-i\left(\bar{u}-0.5\right)^2,
\end{equation}
where $\bar{u}$ is the mean of the $u_i$ and $W^2$ is the Cramér-von Mises statistic,
\begin{equation}
    W^2 = \sum_{i=1}^N\left\lbrace u_{(i)}-\frac{2i-1}{2N}\right\rbrace^2 + \frac{1}{12N}.
\end{equation}

\begin{figure}[t]
    \centering
    \includegraphics[width=\textwidth]{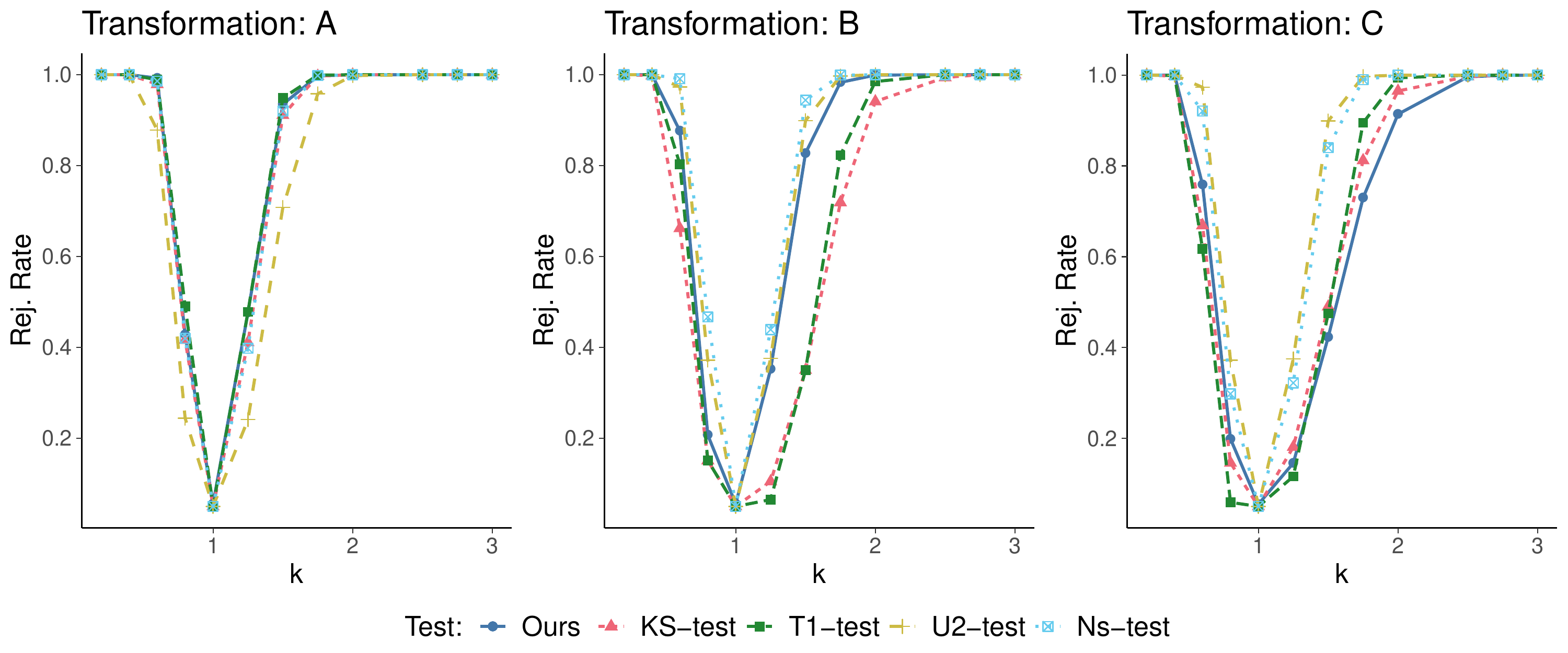}
    \caption{When compared to existing tests for uniformity, the rejection rate of the graphical test with simultaneous confidence bands performs well in all the three families of deviations introduced in \eqref{eq:transformation families}. A slightly lower rejection rate can be observed in family C for $k\in (1,3)$, which corresponds to samples biased towards the center of the unit interval.}
    \label{fig:single_sample_test_power_comparisons}
\end{figure}

The rejection rates of these tests and our test through simultaneous confidence bands are shown in Figure \ref{fig:single_sample_test_power_comparisons} for families A, B, and C with sample size $N=100$ and $k$ varying between $0.20$ and $3.00$. For each value of $k$, the rejection rate among $100,000$ samples was computed.
As seen from these results, the proposed ECDF simultaneous confidence band method performs in a manner similar to the recommended tests with the exception of family C, where our method exhibits a lower rejection rate compared to some of the other tests.

\subsection{Comparing Multiple Samples}\label{subsec:multi_sample_examples}
    When testing if two or more samples are produced from the same underlying distribution, we can compare the ranks of each sample relative to the sample obtained by combining all the samples in the comparison.
    As mentioned in Section \ref{sec:multi_sample_test}, we need to adjust the confidence bands to take into account the dependency of the ranks of the values of one sample on the values in other samples in the comparison.
    
    When using the multiple sample test for MCMC convergence diagnostics, we recommend first using existing numerical convergence statistics, such as the $\widehat{R}$ by \textcite{Vehtari2021} or the $R^*$ by \textcite{lambert2021} which can assess the convergence of all model parameters jointly and can indicate which parameters have possible convergence issues. In the case that these statistics indicate possible issues, further insight into the nature of these deviations can be obtained with the ECDF plots of fractional ranks.
    
    \subsubsection{Effect of difference in means and variances}
    We first compare two cases of MCMC sampling with four chains containing 250 independent draws, which is enough to reliably estimate the variances and autocorrelations required for $\widehat{R}$ and effective sample size (ESS) as long as the rank-normalized ESS of the sample exceeds 400 \parencite{Vehtari2021}, which is the case as the draws are independent. In each case, chains 2 to 4 were sampled from a $\normal(0,1)$ distribution. In the first case, chain 1 is sampled with a larger mean than the other chains, $\normal(0.5, 1)$. In the second case, chain 1 is sampled with a larger variance, $\normal(0,1.5)$.
    
\begin{figure}
    \centering
    \subfloat[Rank plot: Chain 1]{
    \includegraphics[width=0.28\textwidth]{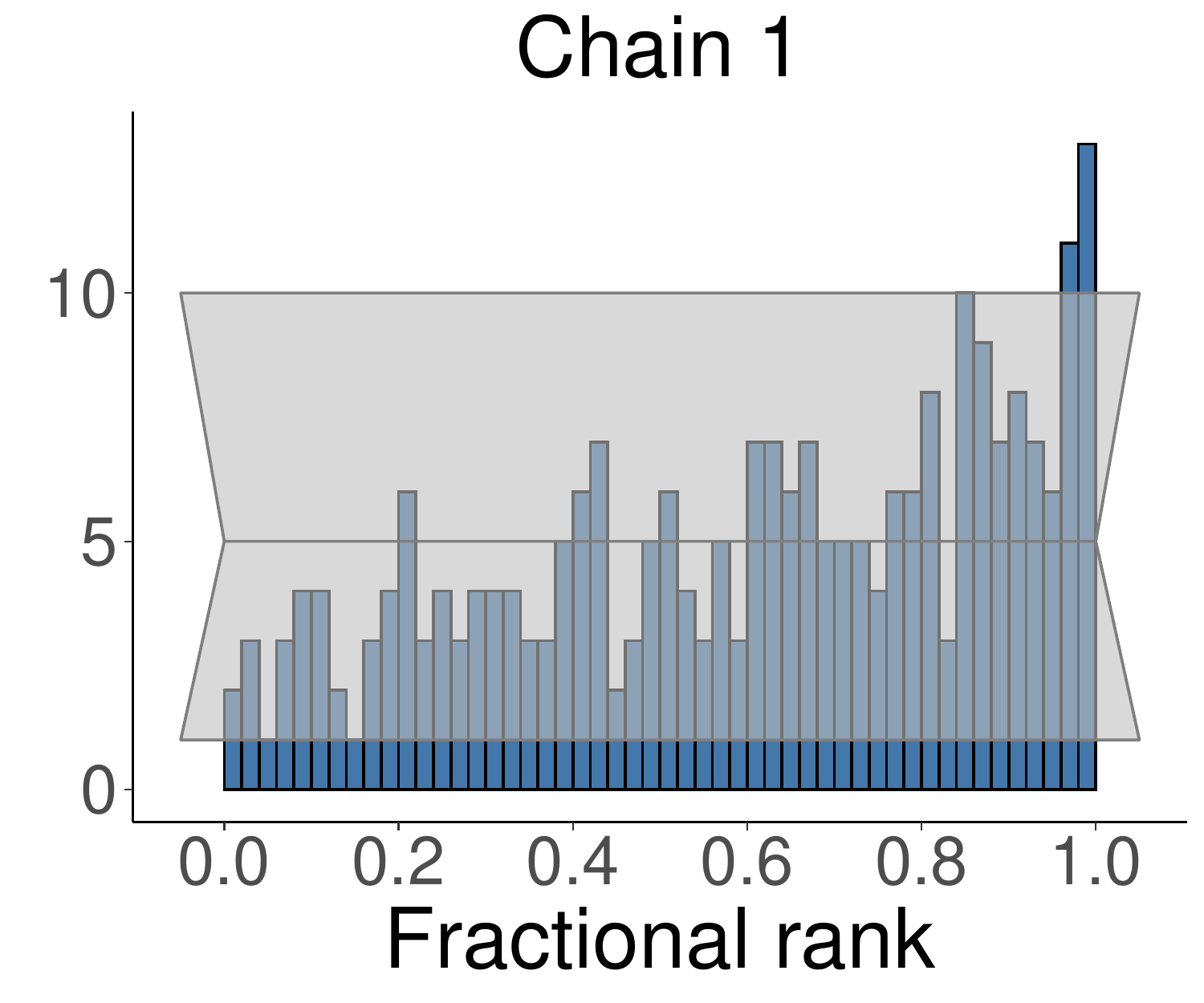}}~
    \subfloat[Rank plot: Chain 2]{
    \includegraphics[width=0.28\textwidth]{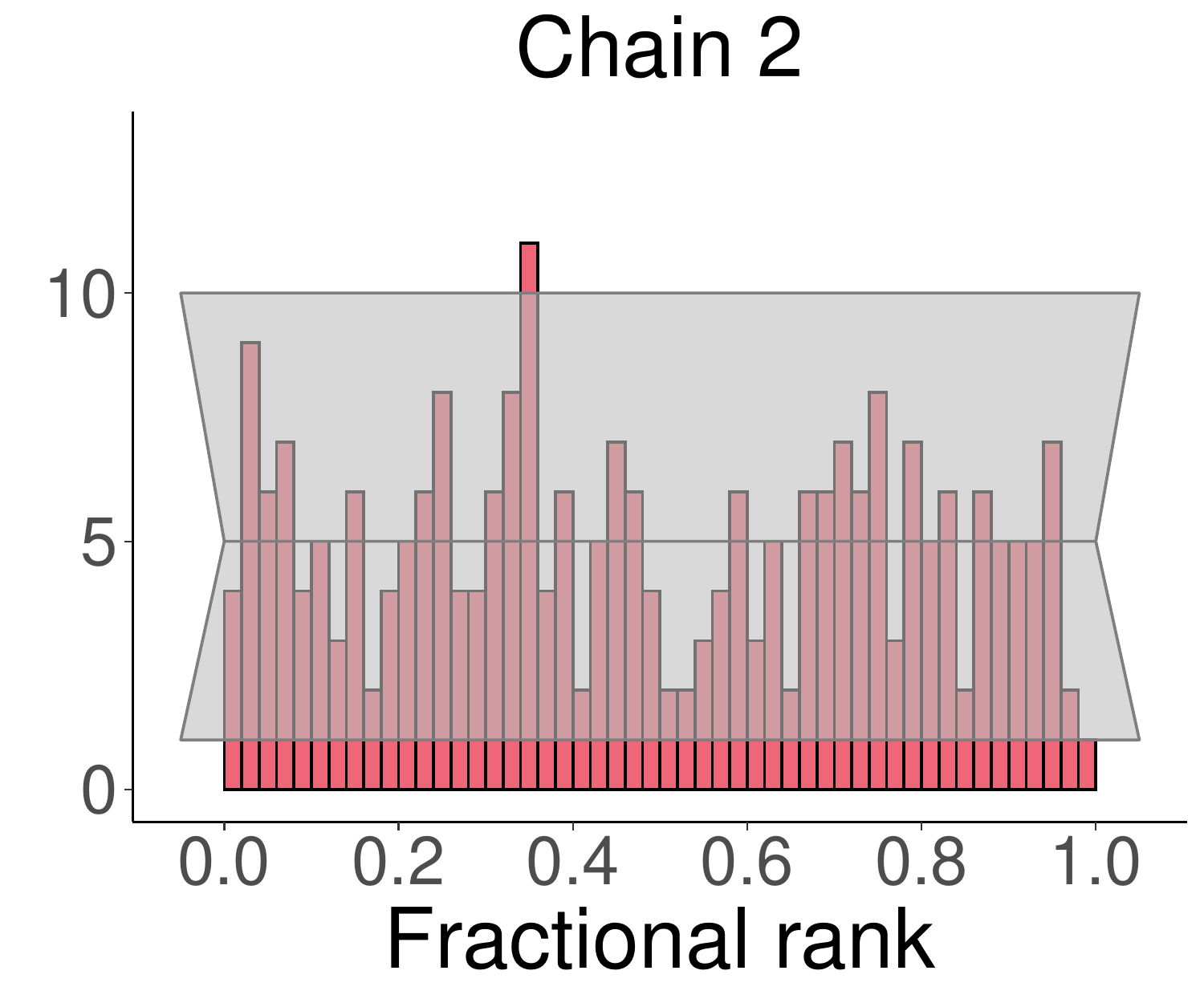}}~
    \subfloat[Rank plot: Chain 3]{
    \includegraphics[width=0.28\textwidth]{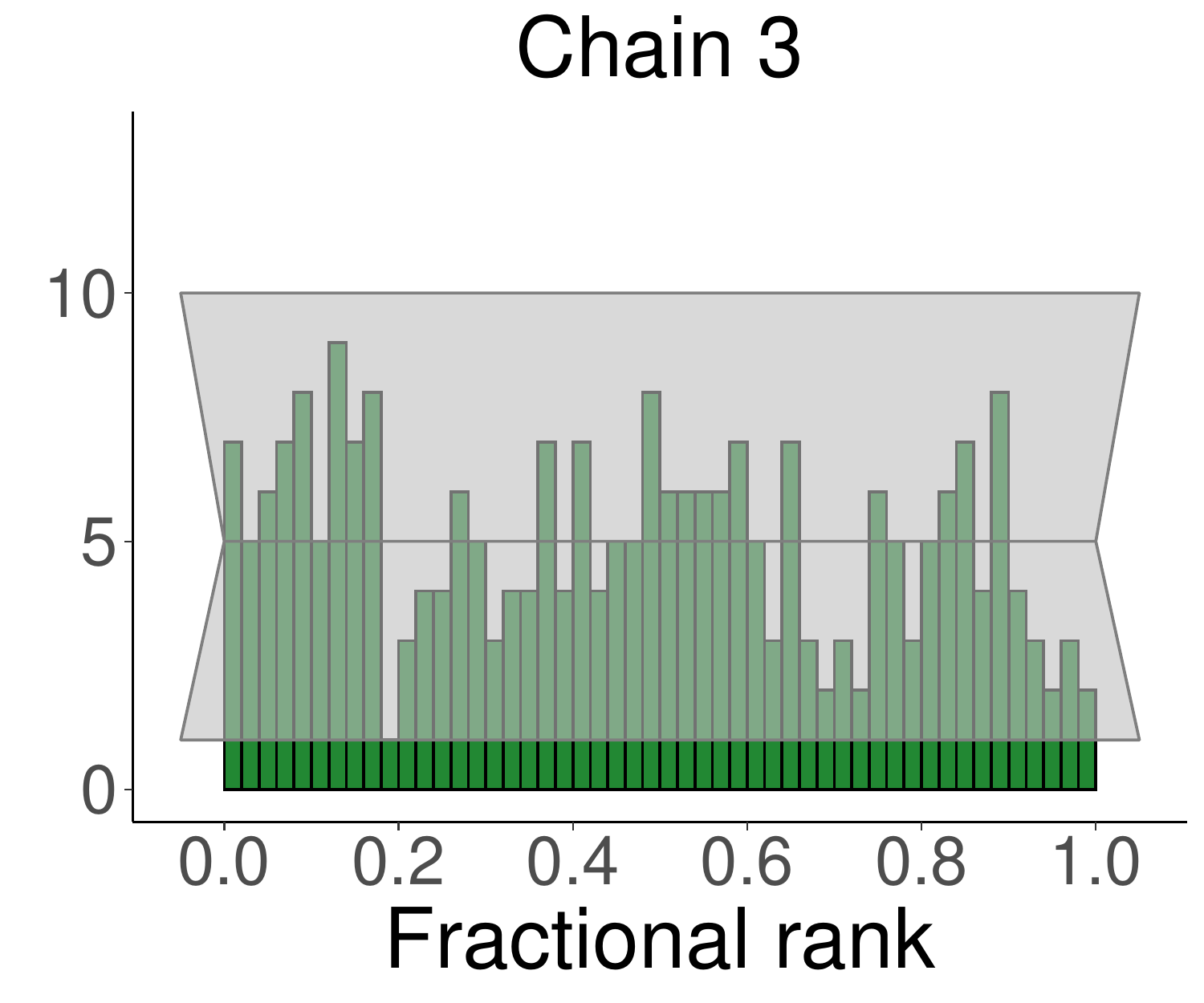}}\\
    \subfloat[Rank plot: Chain 4]{
    \includegraphics[width=0.28\textwidth]{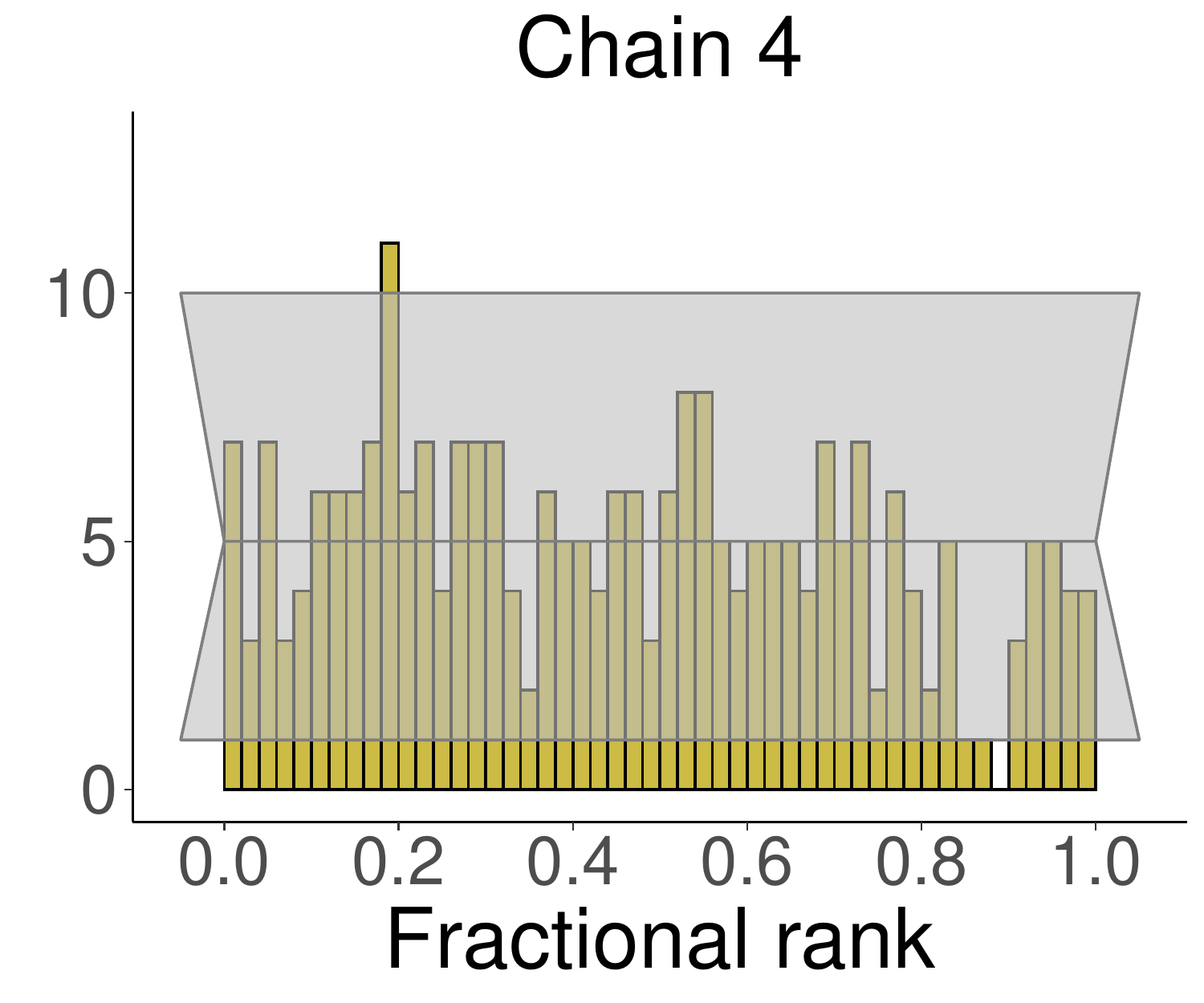}}~
    \subfloat[ECDF plot]{
    \includegraphics[width=0.28\textwidth]{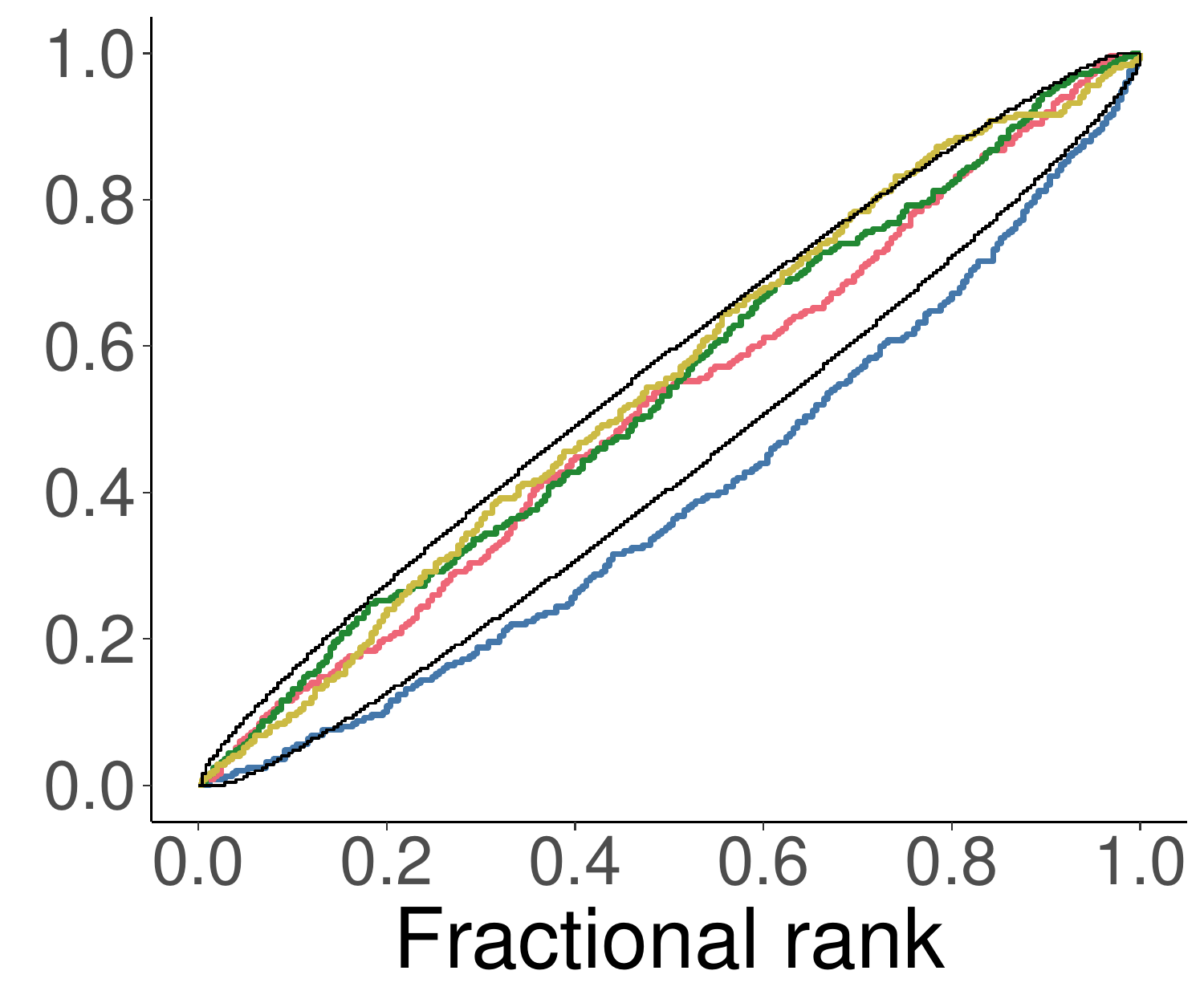}}~
    \subfloat[ECDF difference plot]{
    \includegraphics[width=0.28\textwidth]{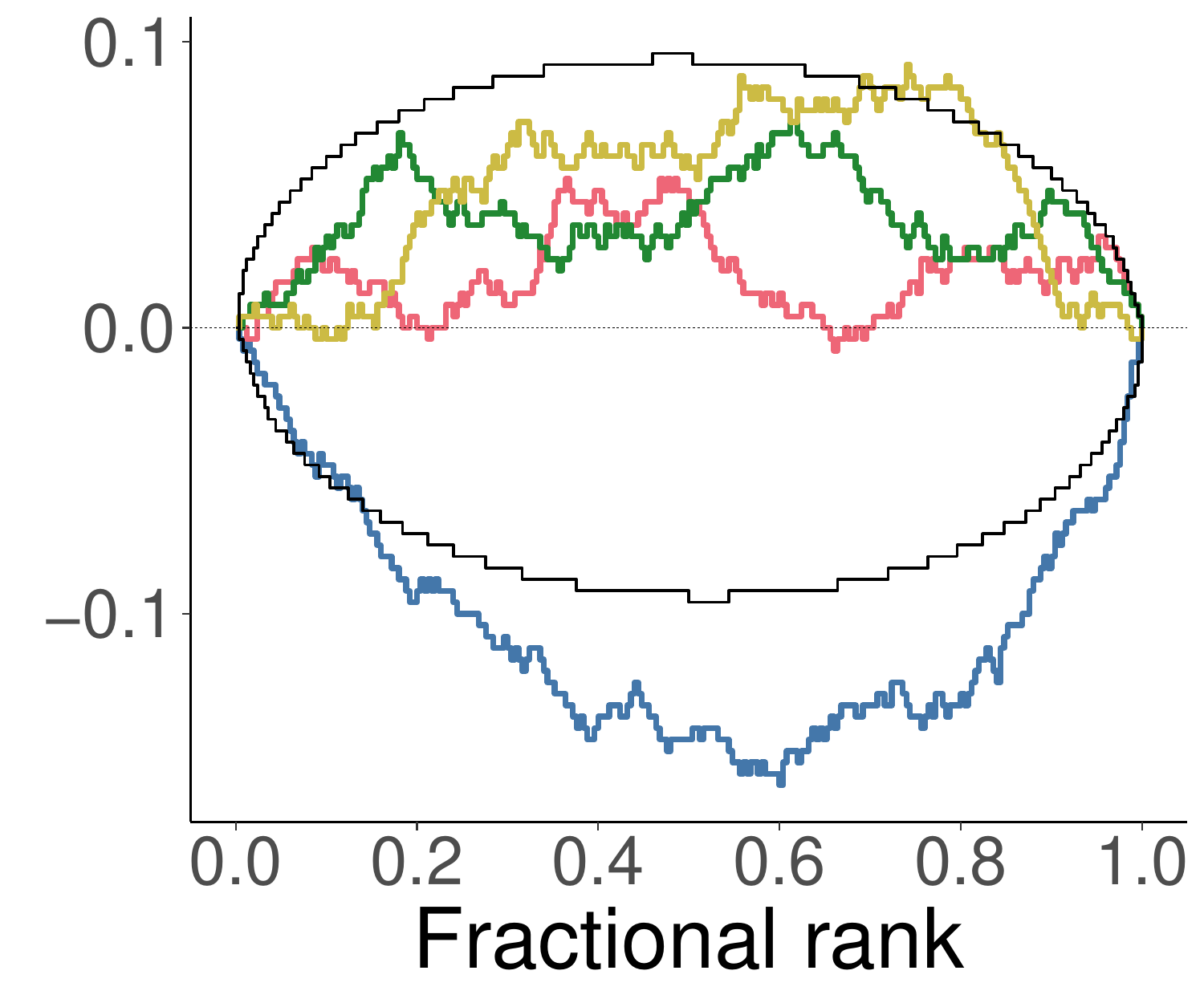}}
    \caption{Effect of differences in sample mean. While chains 2 to 4 are drawn from the standard normal distribution, chain 1 is drawn from $\normal(0.5,1)$, which can be seen as a bias towards large fractional ranks in the rank plot of chain 1 and as a slightly lowered frequency of large fractional ranks in chains 2-4. In the ECDF plot and the ECDF difference plot, the ECDF of chain 1 obtains values considerably lower than expected resulting in a clear $\cup$-shape in the ECDF difference plot.}
    \label{fig:multiple_comparison_shifted}
\end{figure}

\begin{figure}
    \centering
    \subfloat[Rank plot: Chain 1]{
    \includegraphics[width=0.28\textwidth]{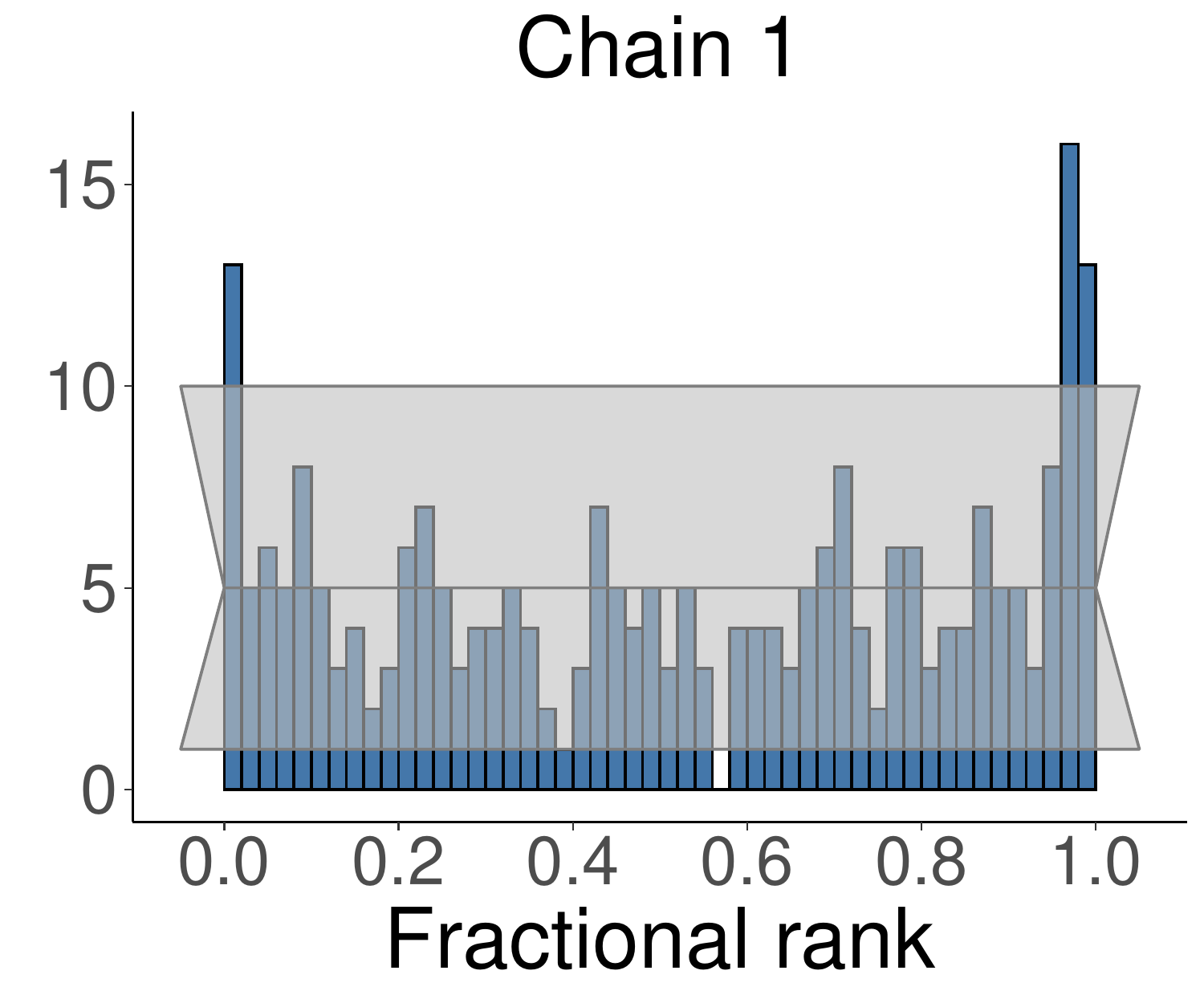}}~
    \subfloat[Rank plot: Chain 1]{
    \includegraphics[width=0.28\textwidth]{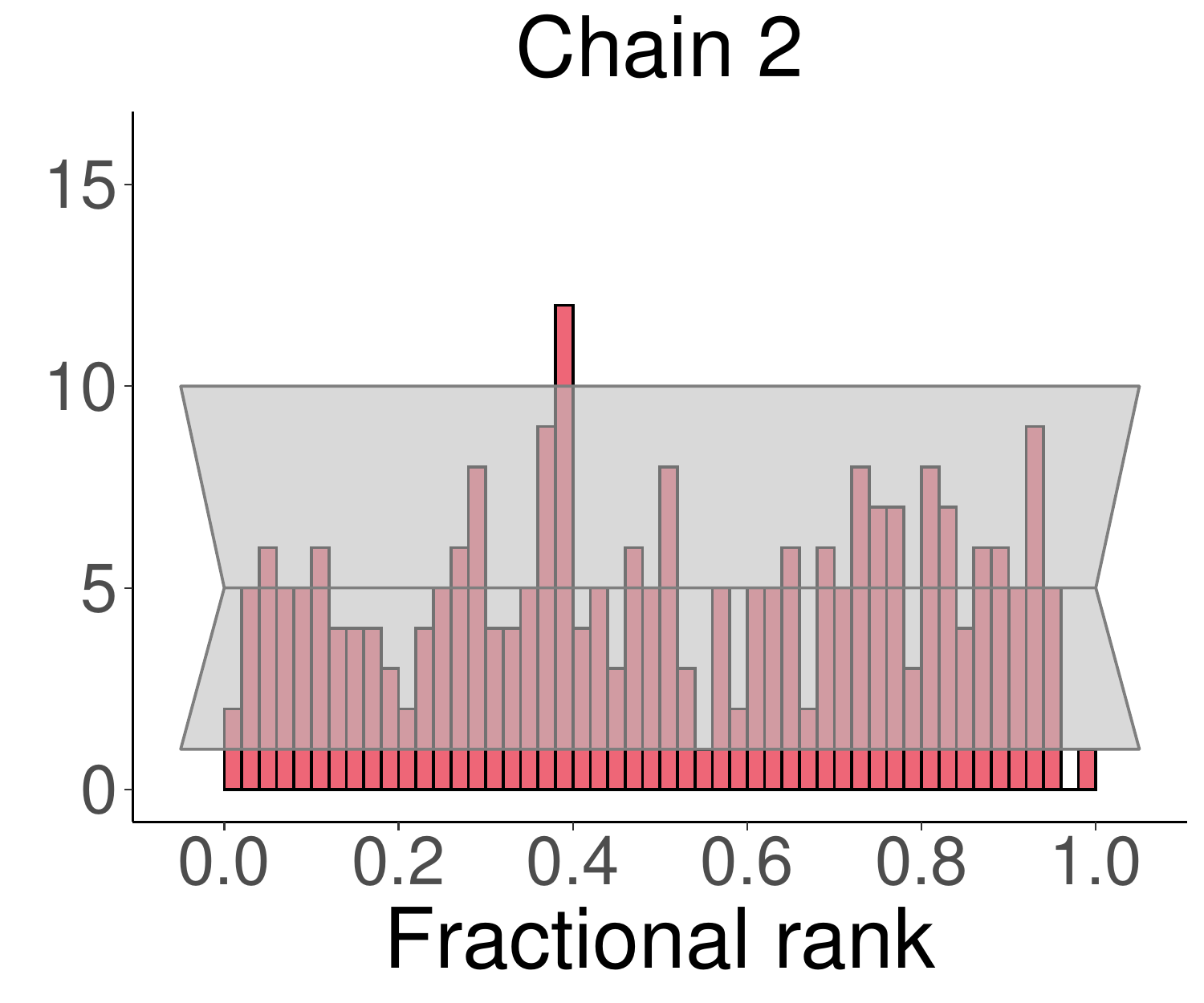}}~
    \subfloat[Rank plot: Chain 1]{
    \includegraphics[width=0.28\textwidth]{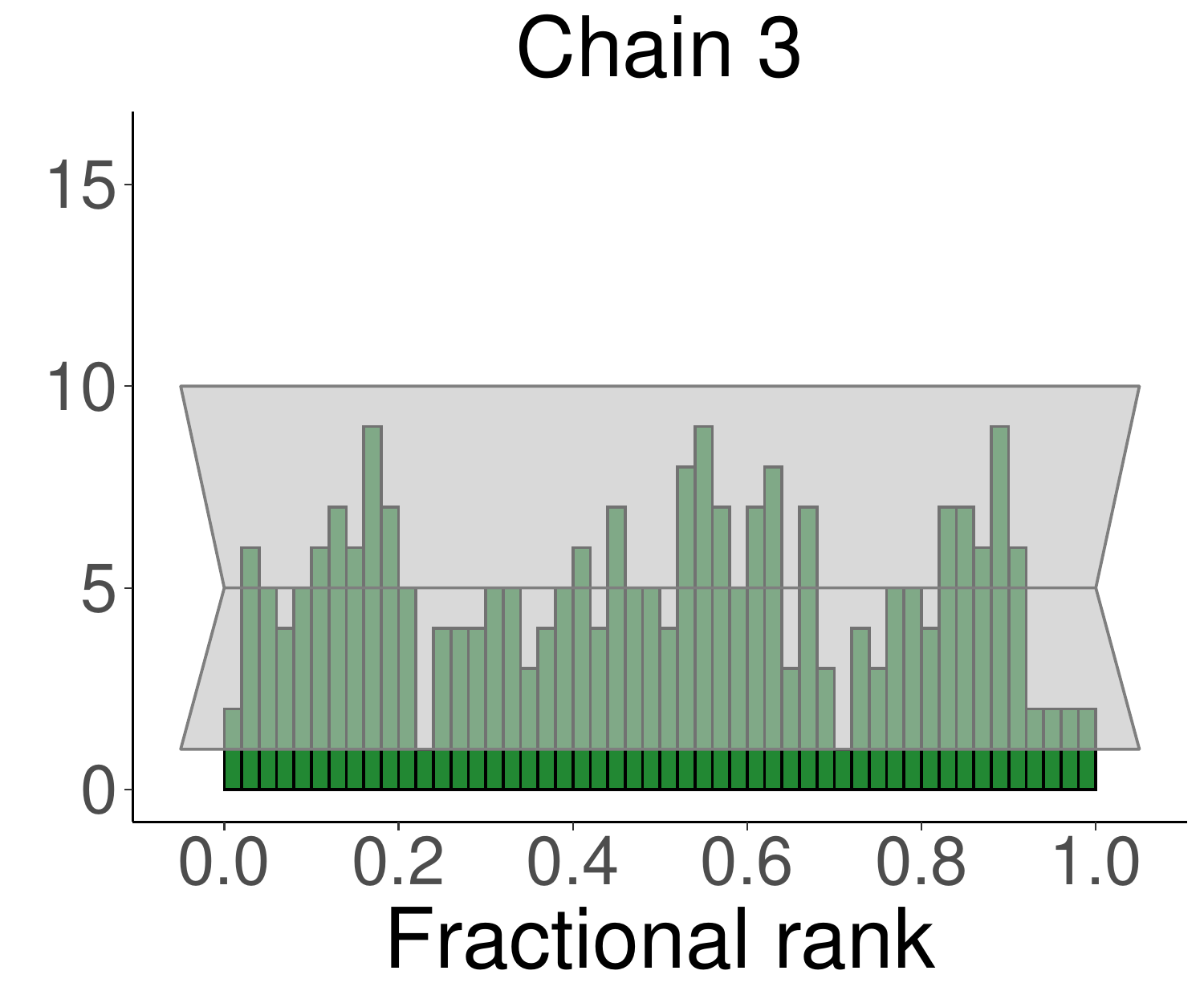}}\\
    \subfloat[Rank plot: Chain 1]{
    \includegraphics[width=0.28\textwidth]{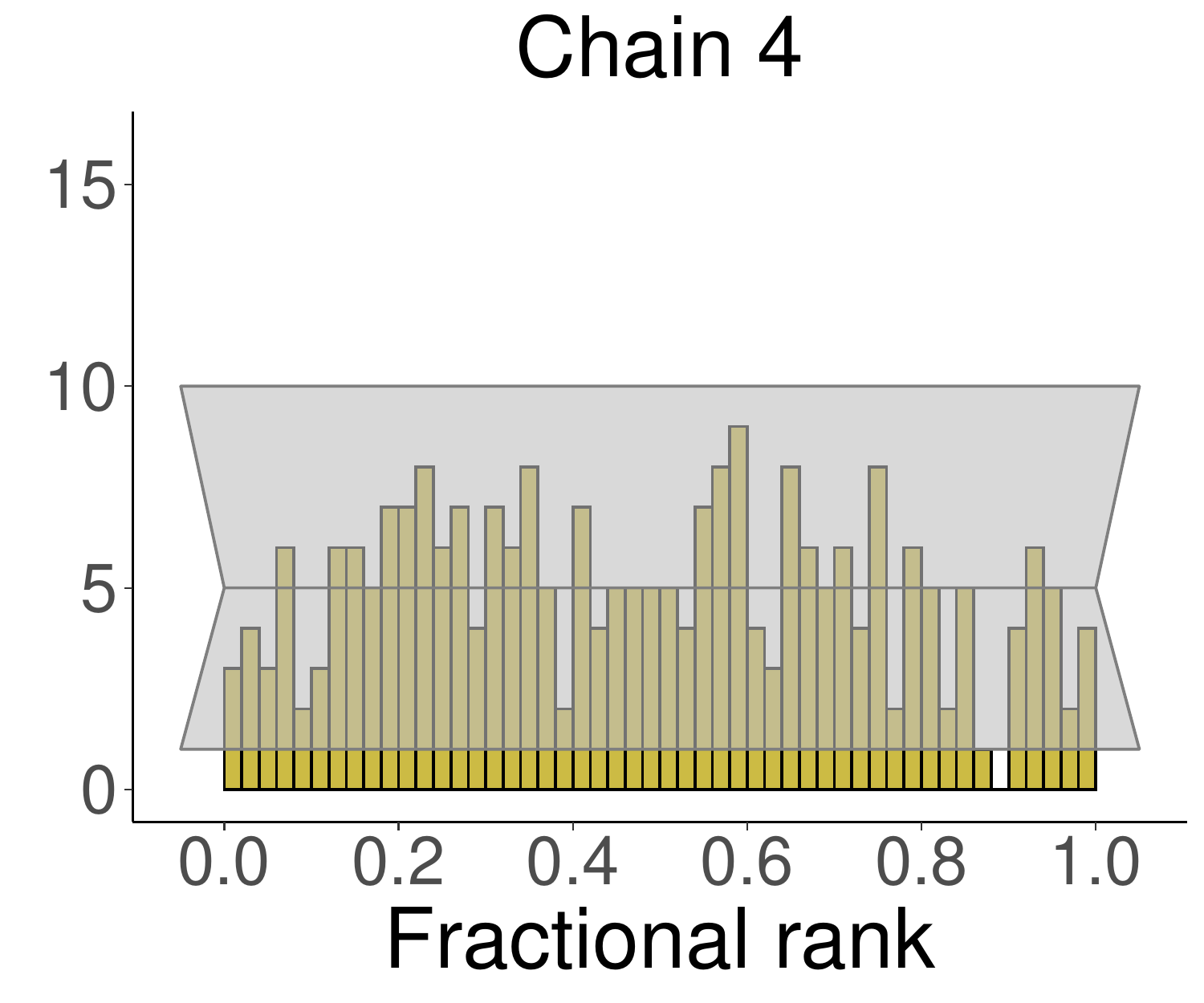}}~
    \subfloat[ECDF plot]{
    \includegraphics[width=0.28\textwidth]{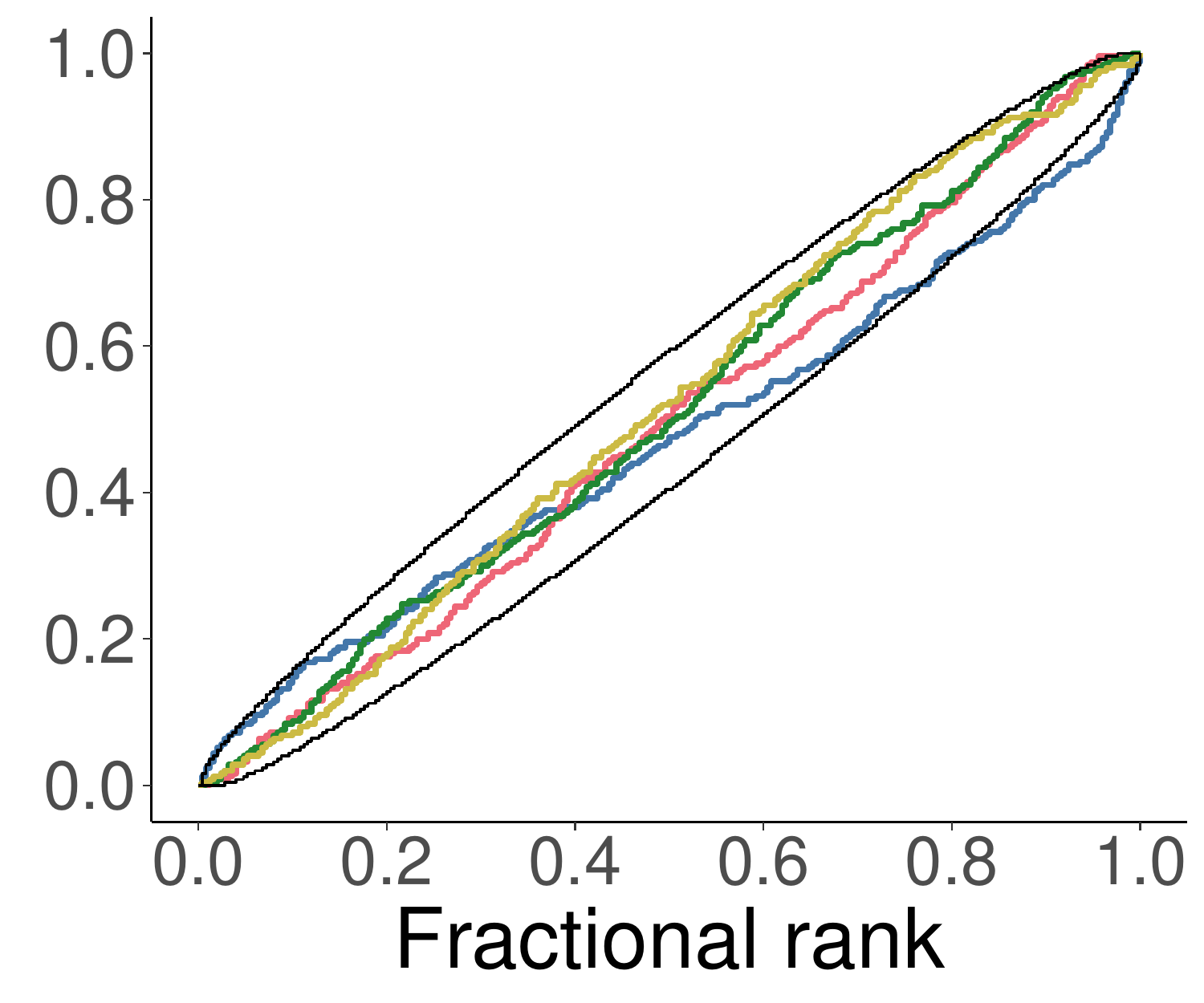}}~
    \subfloat[ECDF difference plot]{
    \includegraphics[width=0.28\textwidth]{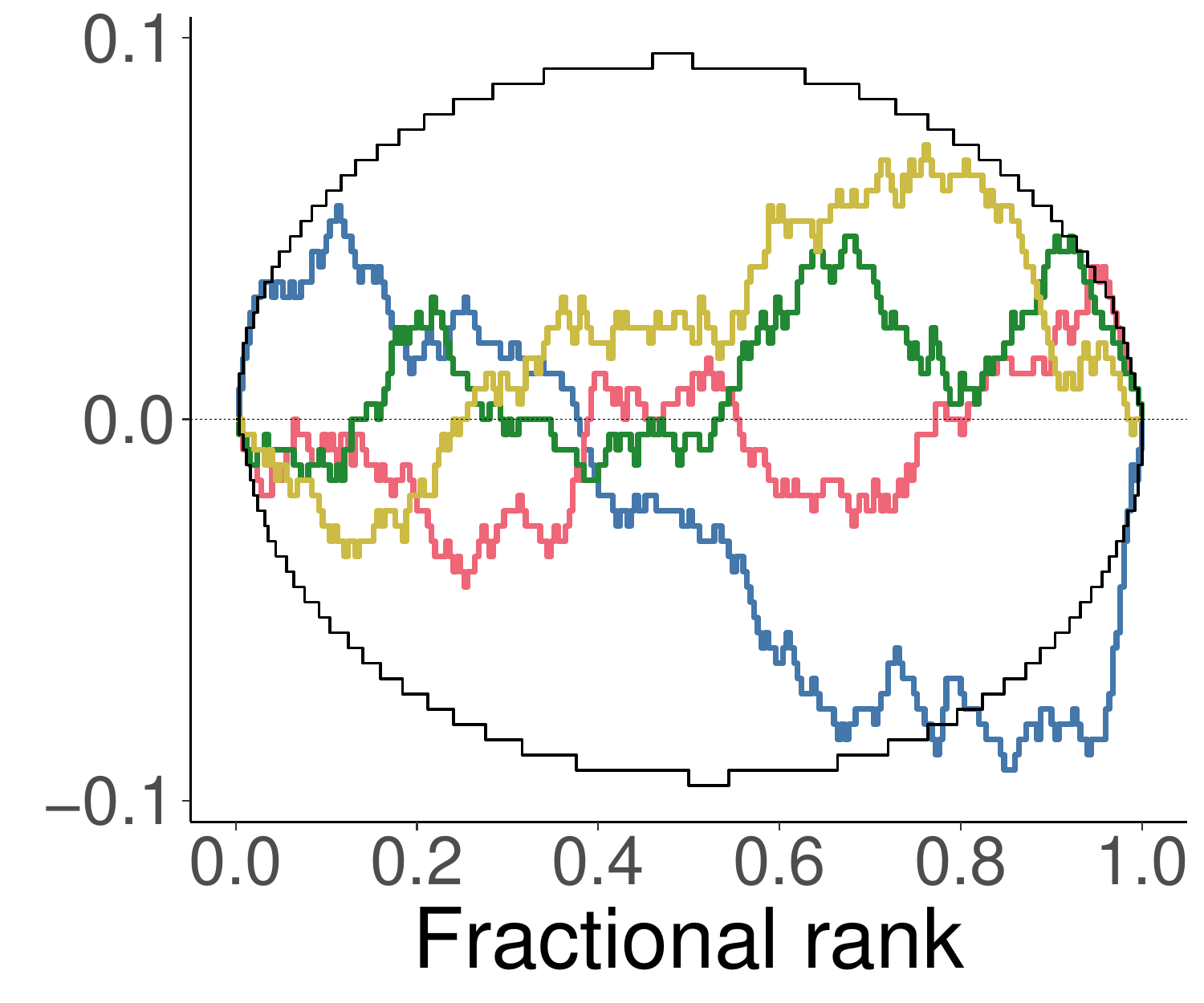}}
    \caption{Effect of differences in sample variances. While chains 2--4 are drawn from the standard normal distribution, chain 1 is drawn from $\normal(0,1.5)$, which can be seen as a $\cup$-shape in the rank plot of chain 1 and as a low frequency of small and large fractional ranks in the rank plots of chains 2--4. In the ECDF plot and the ECDF difference plot, the ECDF of chain 1 grows fast near the ends of the unit interval, where a higher than expected density of fractional ranks is covered. In the middle of the interval, the ECDF difference plot of chain 1 is decreasing, whereas chains 2-4 are increasing.}
    \label{fig:multiple_comparison_scaled}
\end{figure}

    Rank plots for the first case with one chain having a larger mean are shown in Figure \ref{fig:multiple_comparison_shifted}(a)-(d). Even though the difference in the sampling distribution of chain 1 can be seen in the histograms with 50 bins, this effect is more clearly represented in the ECDF difference plot in Figure \ref{fig:multiple_comparison_shifted} (f) where chain 1 shows the shape familiar from \ref{subec:sample_mean} and chains 2 to 4 show a reverse shape, indicating similar behaviour between these three chains.
    Similar remarks regarding the behaviour of the chains can be made from the ECDF plot in Figure \ref{fig:multiple_comparison_shifted}(e), but the more dynamic range of the ECDF difference plot in Figure \ref{fig:multiple_comparison_shifted}(f) makes the difference in the behaviour of the chains clearer.
    In the second case, where chain 1 is sampled with a higher variance, we can see a $\cup$-shape  in the rank plot of chain 1 in Figure \ref{fig:multiple_comparison_scaled}(a), but the behaviour stands out more clearly in the ECDF difference plot in Figure \ref{fig:multiple_comparison_scaled}(f).
    
    When compared to commonly used convergence diagnostics not offering graphical insight into the nature of the possible underlying problems, both the classical $\widehat{R}$ diagnostic by \textcite{Gelman1992} and the improved $\widehat{R}$ diagnostic proposed by \textcite{Vehtari2021} indicate convergence issues as they give and estimated $\hat{R}$ values of $1.05$ and $1.04$ respectively to both the mean and variance related examples above. \textcite{Vehtari2021} suggest that $\widehat{R} > 1.01$ is an indication of potential convergence issues or too short chains.

\subsubsection{Test performance under common deviations}
\begin{figure}
    \centering
    \includegraphics[width=.9\textwidth]{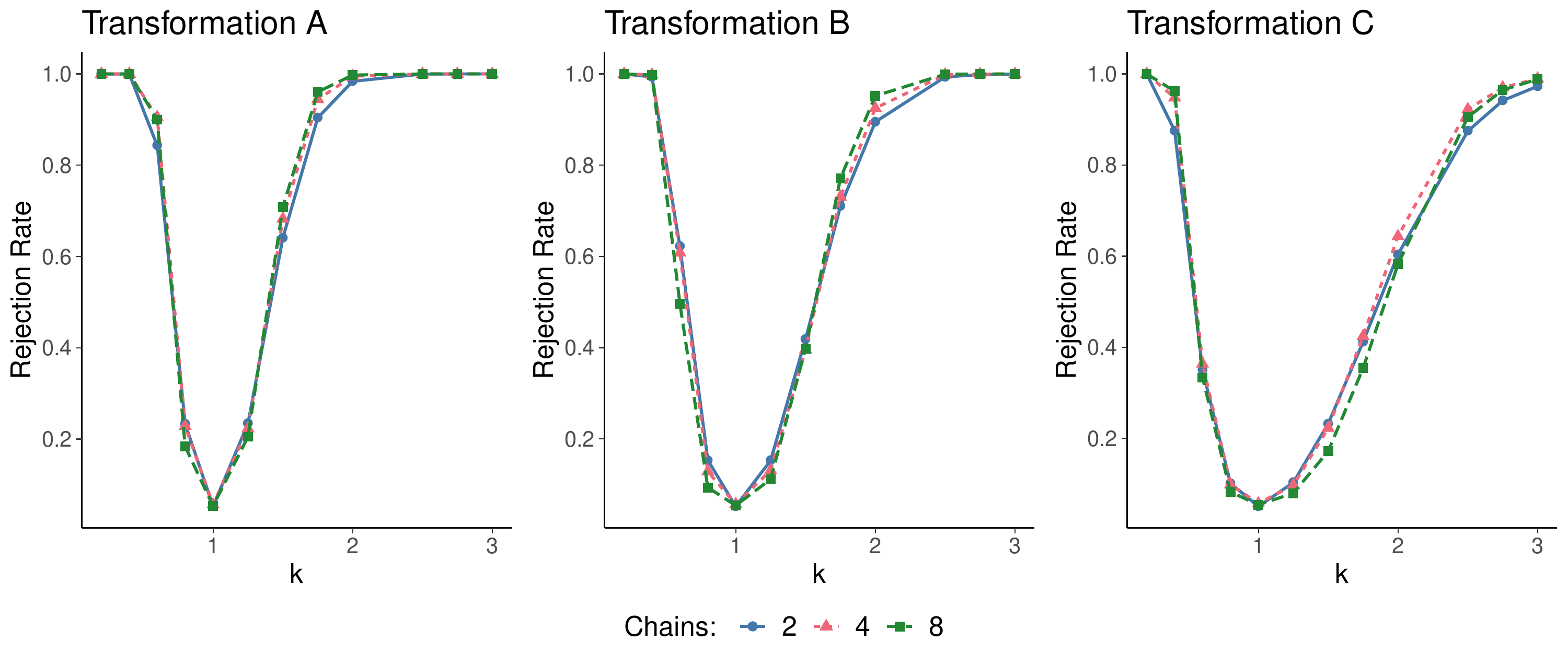}
    \caption{An inspection of the rejection rates of the sample comparison test when one of the chains is transformed according to one of the three transformation families in \eqref{eq:transformation families}, the test performance shows independence of the number of chains and demonstrates similar dependency on the extremity of the transformation.}
    \label{fig:multi_chain_power_analysis}
\end{figure}

To evaluate the performance of the multiple sample comparison test under a set of common deviations, one of the samples was transformed according to the three transformation families defined in equation \eqref{eq:transformation families}. In the analysis $2$, $4$, and $8$ chains of length $100$ were simulated from $U(0,1)$ after which one of the chains was transformed according to the transformations $f_{A,k}$, $f_{B,k}$, and $f_{C,k}$. The rejection rates of the multiple sample comparison test when varying the power, $k$, of the transformation were estimated from $10,000$ simulations and are recorded in Figure \ref{fig:multi_chain_power_analysis}.
The observed test performance is independent of the number of chains used in the sample comparison. When compared to the rejection rates observed in the single sample power analysis in \ref{subsubseq:single_sample_power_analysis}, the rejection rates show that the test sensitivity depends in a similar way on the transformation.

\subsubsection{Chains with autocorrelation}\label{subsec:autocorrelated chains}
\begin{figure}
    \centering
    \begin{tabularx}{.8\textwidth}{XX}
    \multicolumn{2}{c}{\includegraphics[width=0.7\textwidth]{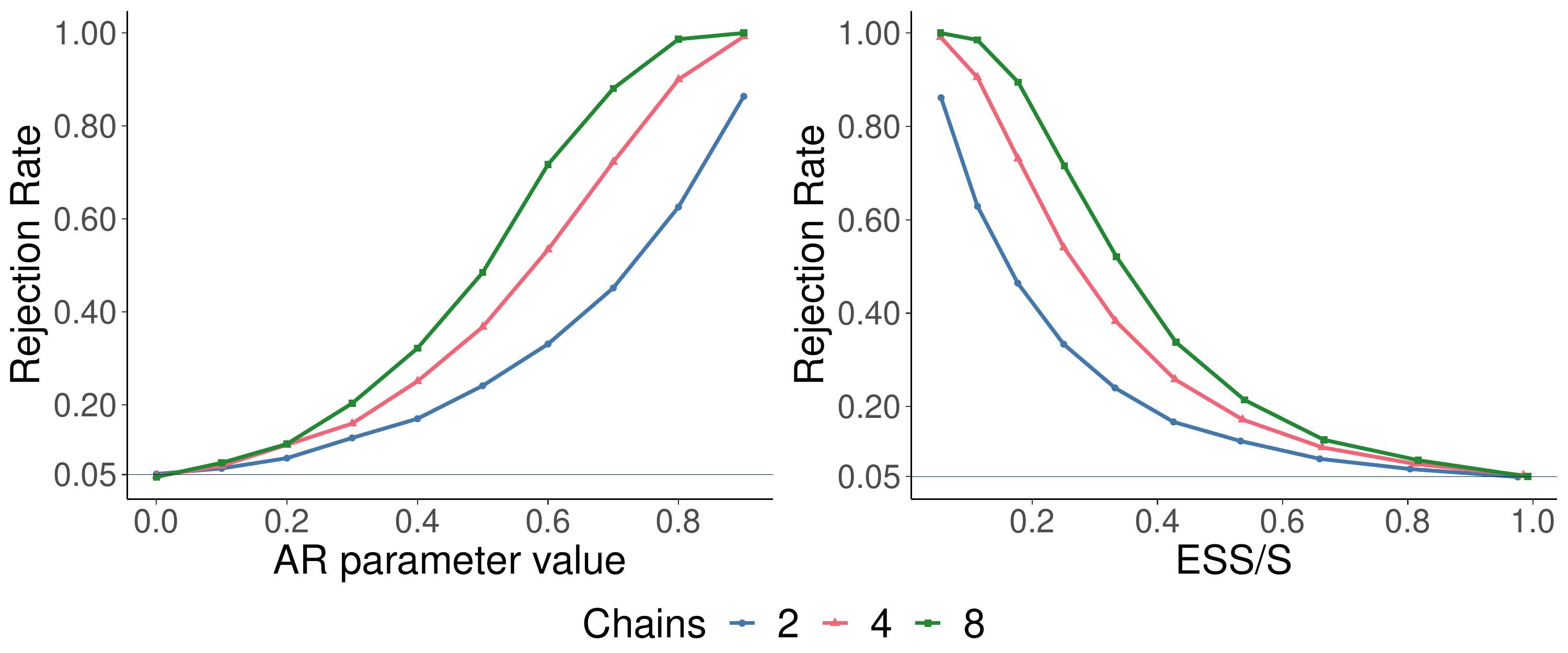}}
    \end{tabularx}
    \caption{Test rejection rate when comparing chains with autocorrelation. On the left as a function of the AR-parameter value and on the right as a function of the ratio between the bulk effective sample size, as defined by \textcite{Vehtari2021}, and the total sample size. The nominal rejection rate 0.05 is shown with a vertical line in both plots. As the test expresses sensitivity to autocorrelation, we recommend thinning the samples, in order to reduce autocorrelation, before using the sample comparison test.}
    \label{fig:autocorrelated_samples_ar}
\end{figure}

As samples generated by MCMC processes are typically autocorrelated, it is essential to analyse the performance of the sample comparison test under autocorrelated samples. In Figure \ref{fig:autocorrelated_samples_ar}, rejection rates of simulated multiple sample test 2, 4, and 8 chains produced by autoregressive models of order 1 (i.e., AR(1) models) with varying AR-parameter values are presented. Each rejection rate is computed as the mean of $100,000$ simulations. As seen in the figure, the higher the autocorrelation in the samples is and the more chains are sampled, the more likely the test is to reject the hypothesis that the samples are drawn from the same underlying distribution. Thus, before using the graphical illustration or the corresponding test, the chains should be thinned to have negligible autocorrelation. The same holds for other common uniformity tests as well, as they rely on the assumption of pairwise independence of draws.

\subsubsection{Detecting model sampling issues: eight schools}\label{subsubsec:eight_schools_rank_plots}
\begin{figure}
    \centering
    \subfloat{
    \includegraphics[width=0.28\textwidth]{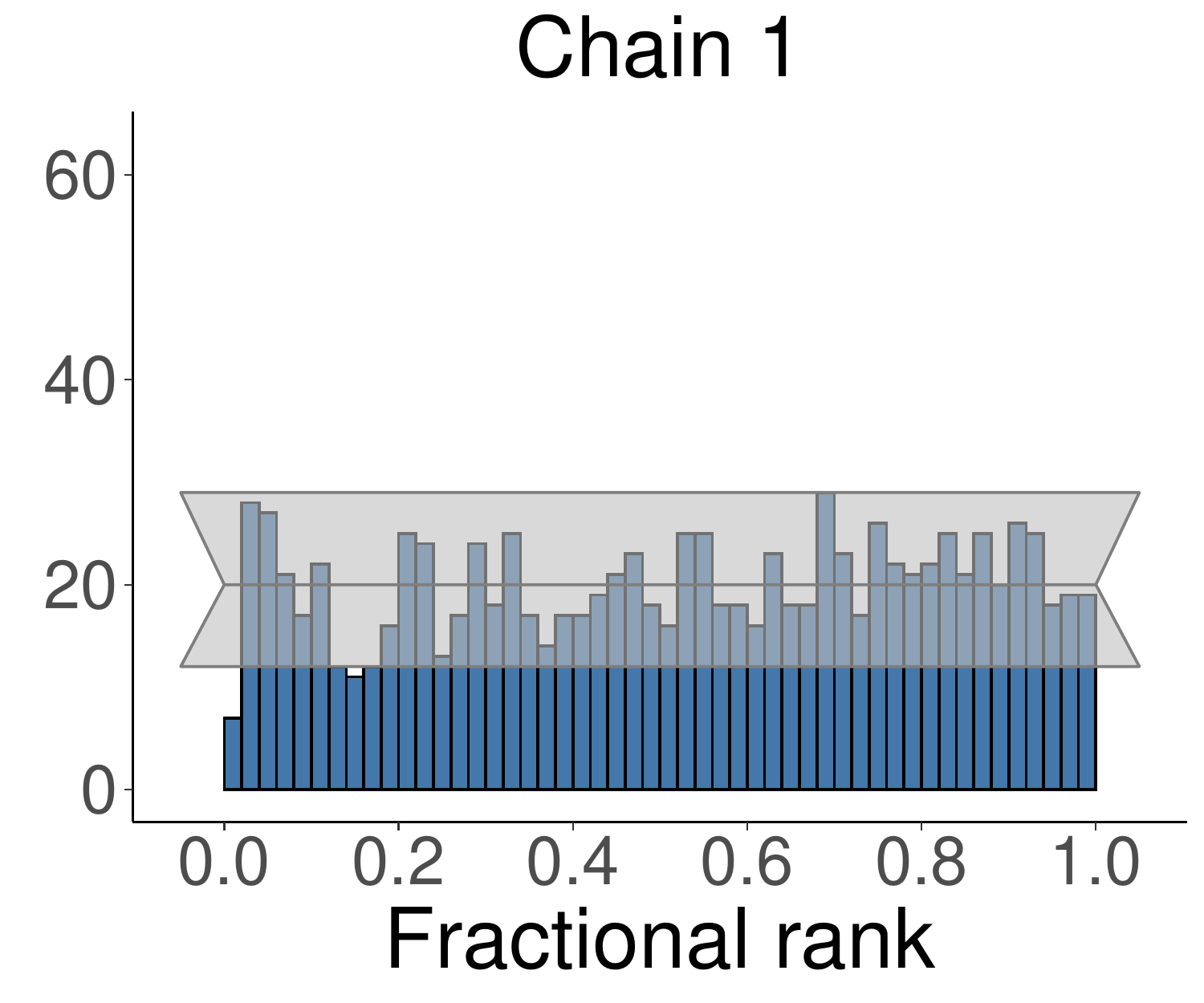}}
    \subfloat{
    \includegraphics[width=0.28\textwidth]{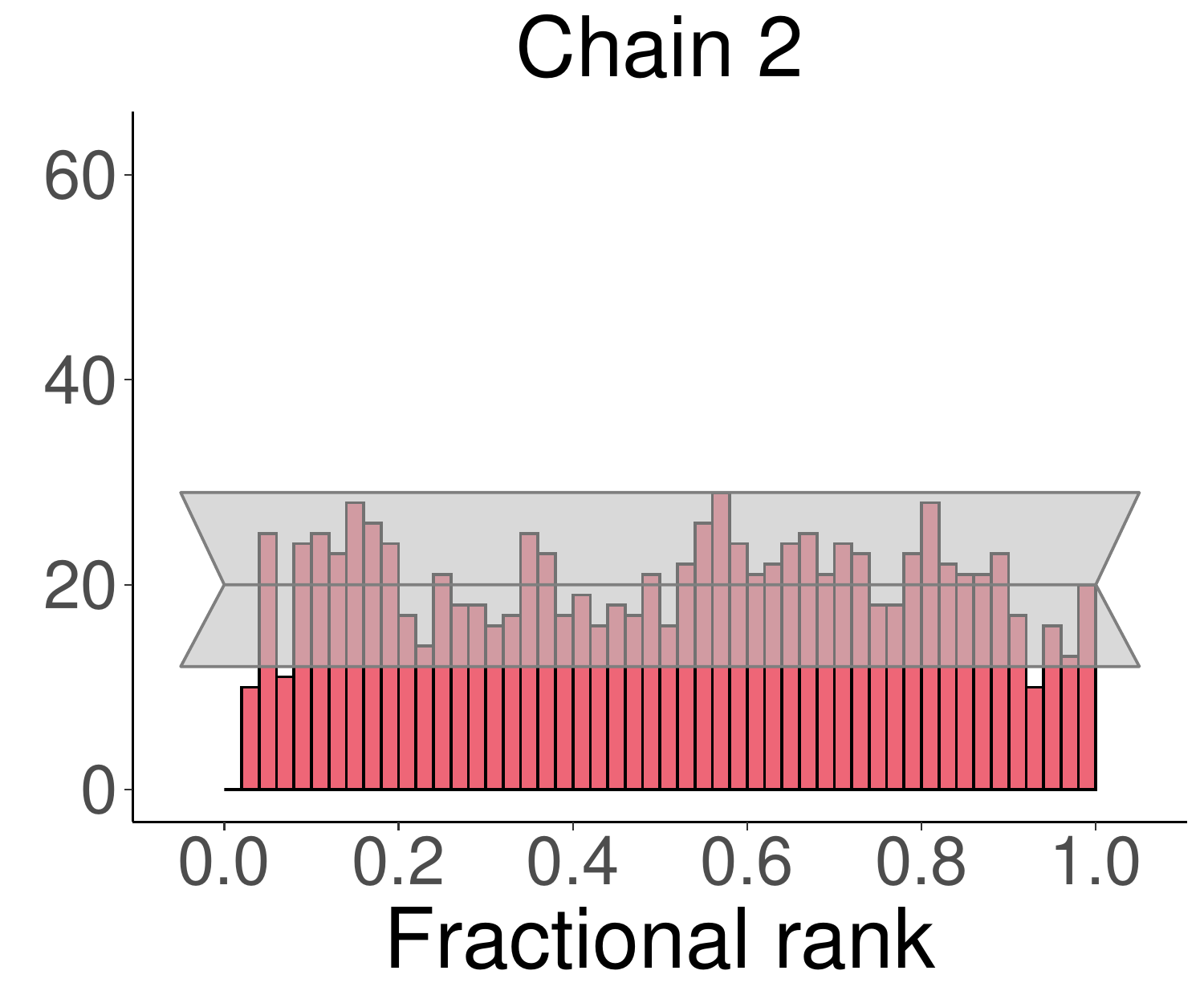}}
    \subfloat{
    \includegraphics[width=0.28\textwidth]{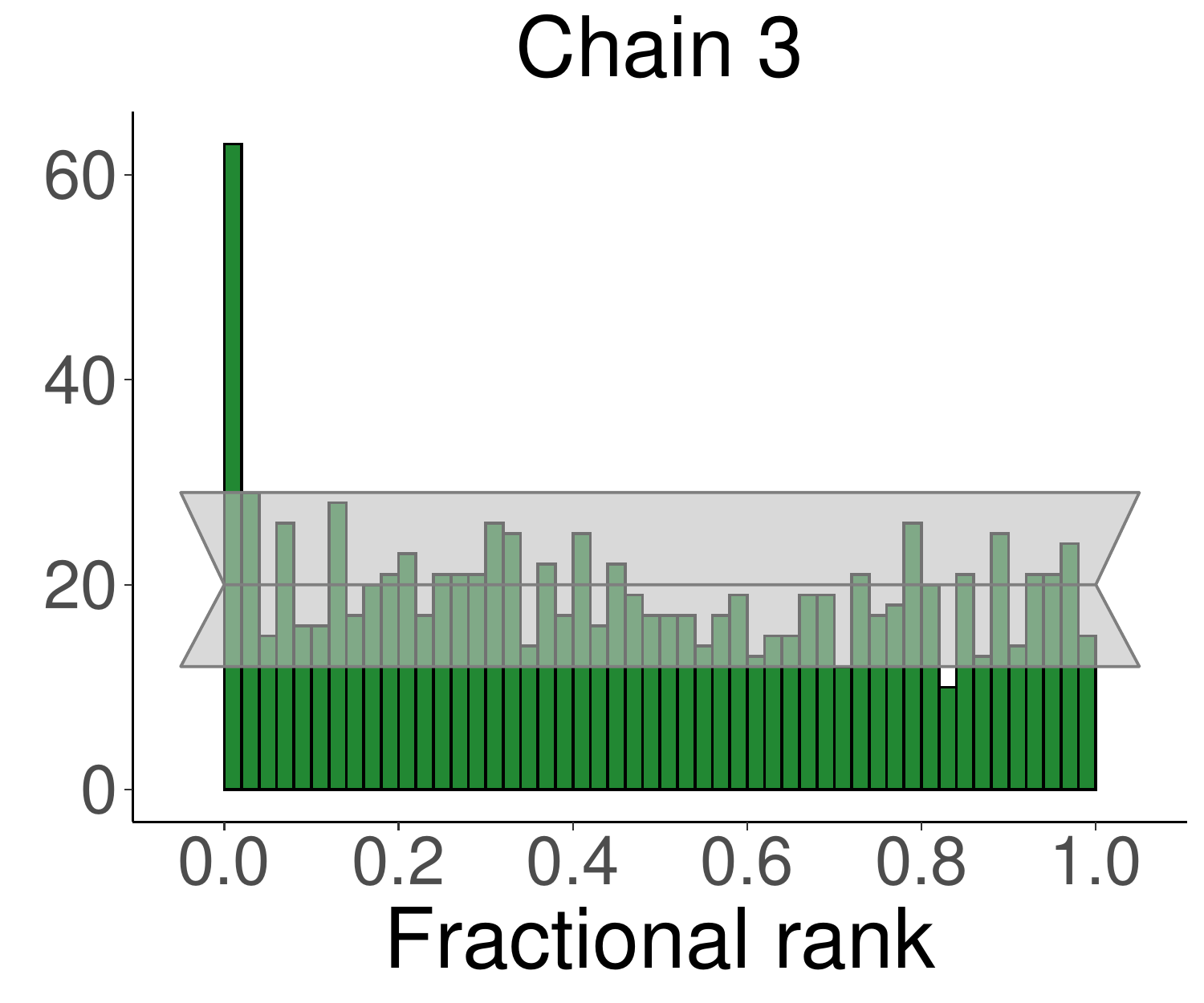}}\\
    \subfloat{
    \includegraphics[width=0.28\textwidth]{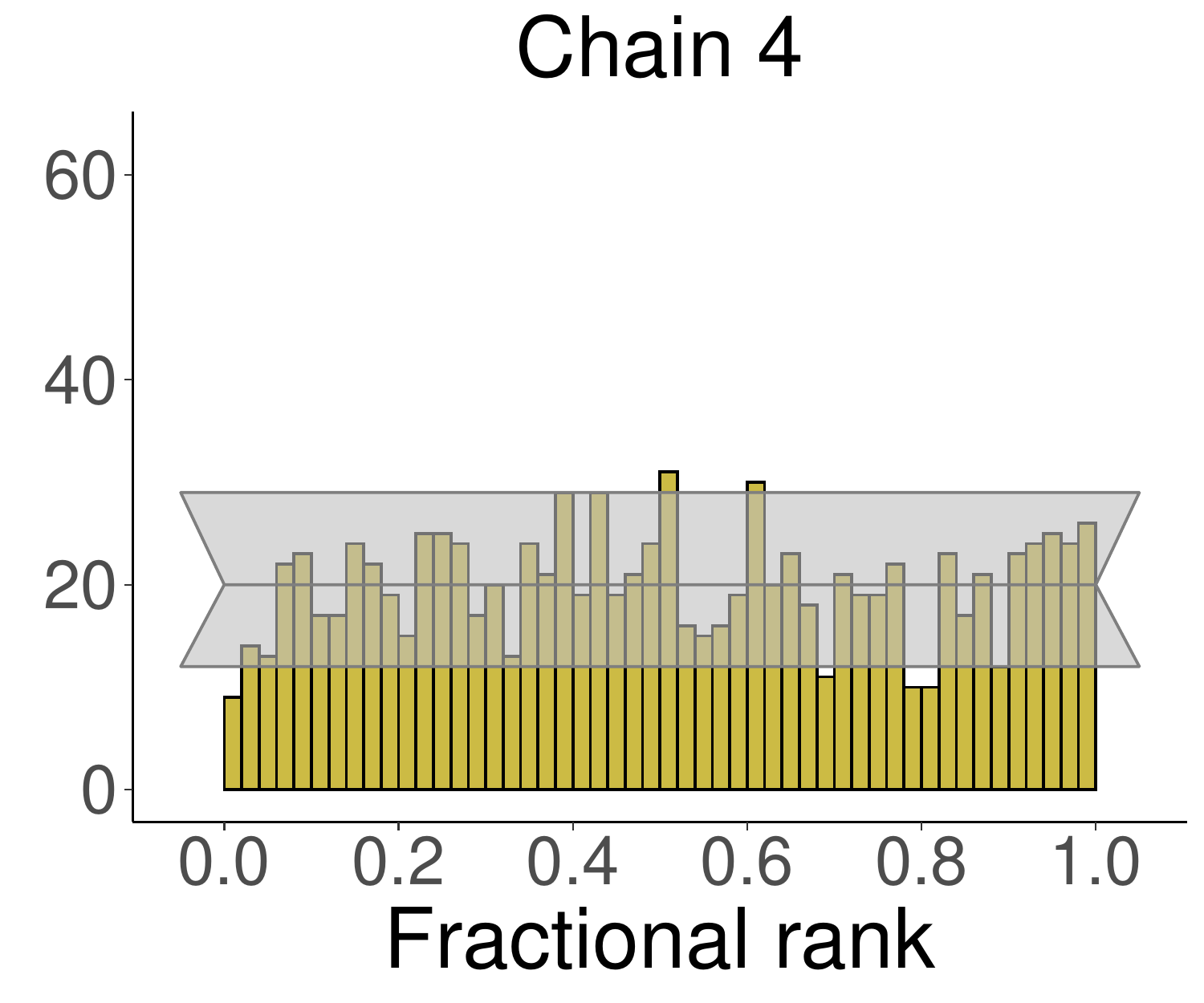}}
    \subfloat{
    \includegraphics[width=00.28\textwidth]{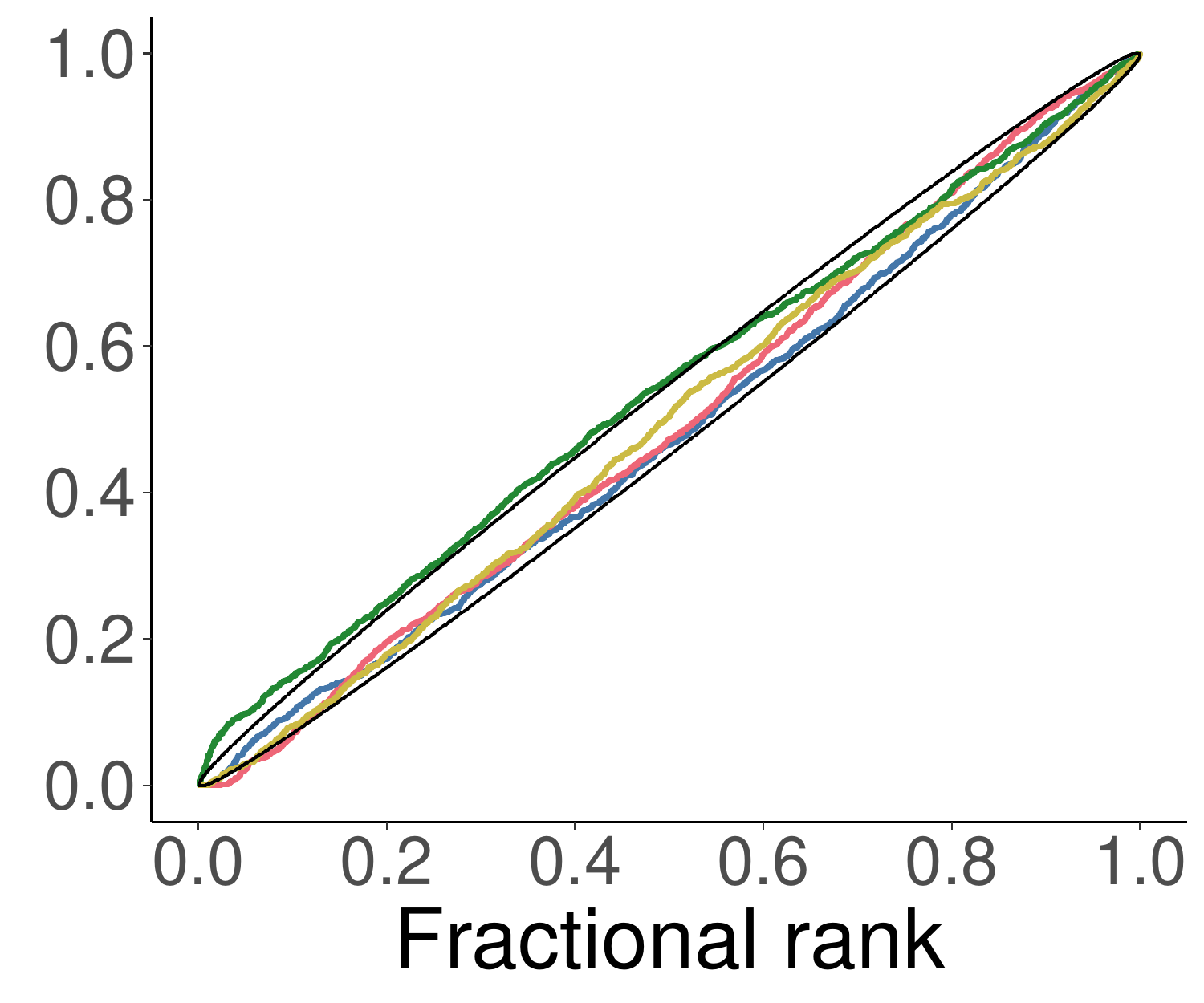}}
    \subfloat{
    \includegraphics[width=00.28\textwidth]{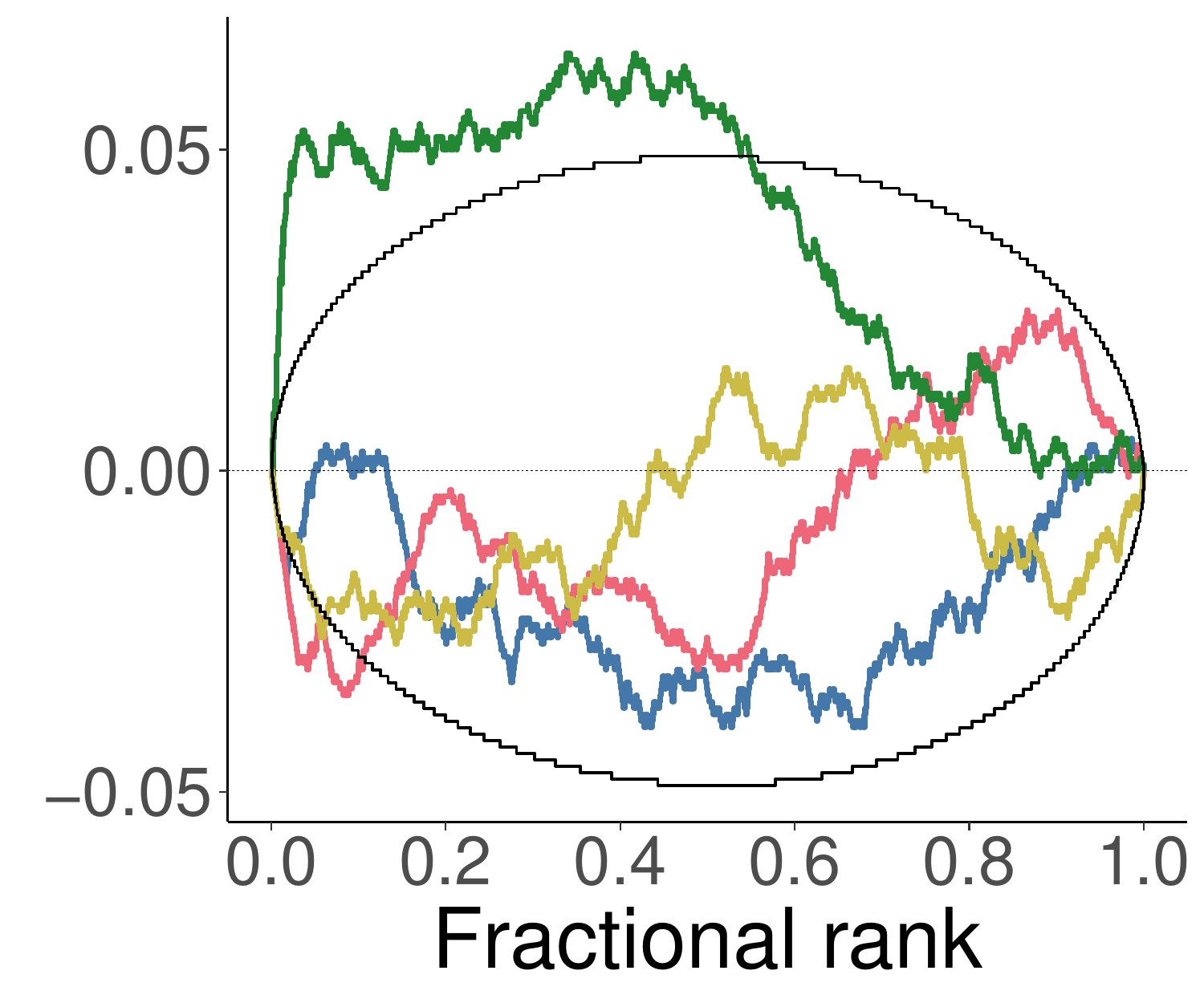}}
    \caption{Detecting model sampling issues: centered parameter eight schools model. When inspecting the sampling of parameter $\tau$, both the rank plot of chain 3 and the ECDF difference plot in the centered parameter model indicate a convergence issue with chain 3 including small values of the parameter at a rate considerably higher than the rest of the sampled chains.}
    \label{fig:schools_cp}
\end{figure}

We return to the eight schools model used to demonstrate SBC in Section \ref{subsubsec:eight_schools_SBC}. The issues detected with SBC earlier are apparent when multiple sample comparison is used to inspect the rank distribution between the four individual chains, each containing 1000 posterior draws after a warm-up period of 1000 steps.
Even when sampled with more conservative settings of the sampler, we see from Figure~\ref{fig:schools_cp} that the chains are not properly exploring the posterior and thus the realized rank transformed chains have clearly different ECDFs.

While the classical $\widehat{R}$ is estimated at $1$, the improved $\hat{R}$ diagnostic gives a value of $1.02$ indicating possible convergence issues. One should also note that the sampling efficiency for $\tau$ in the model is very low, as both the bulk-ESS and the tail-ESS by \textcite{Vehtari2021} are under 150 for the combined sample.

\begin{figure}
    \centering
    \subfloat{
    \includegraphics[width=0.28\textwidth]{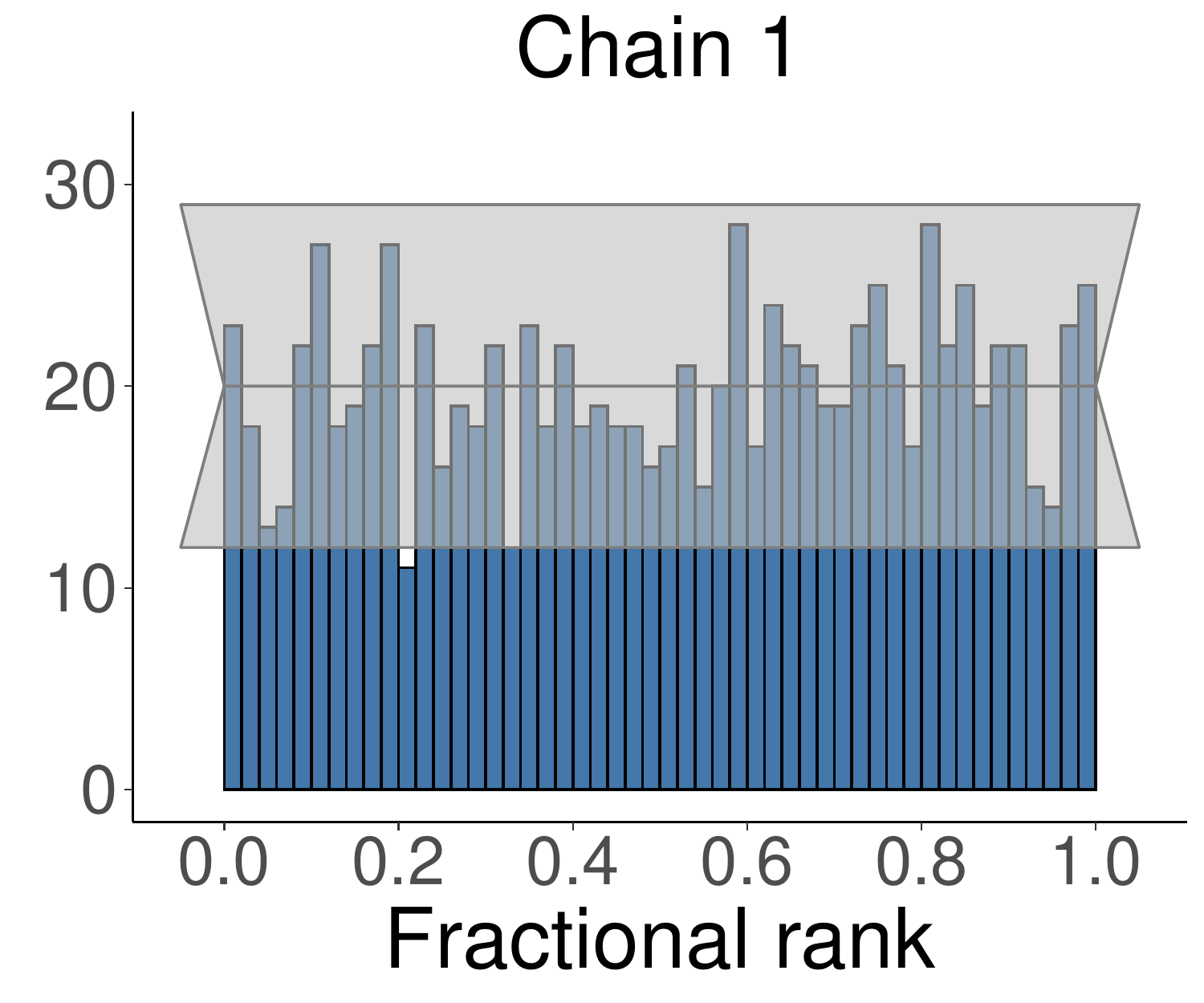}}~
    \subfloat{
    \includegraphics[width=0.28\textwidth]{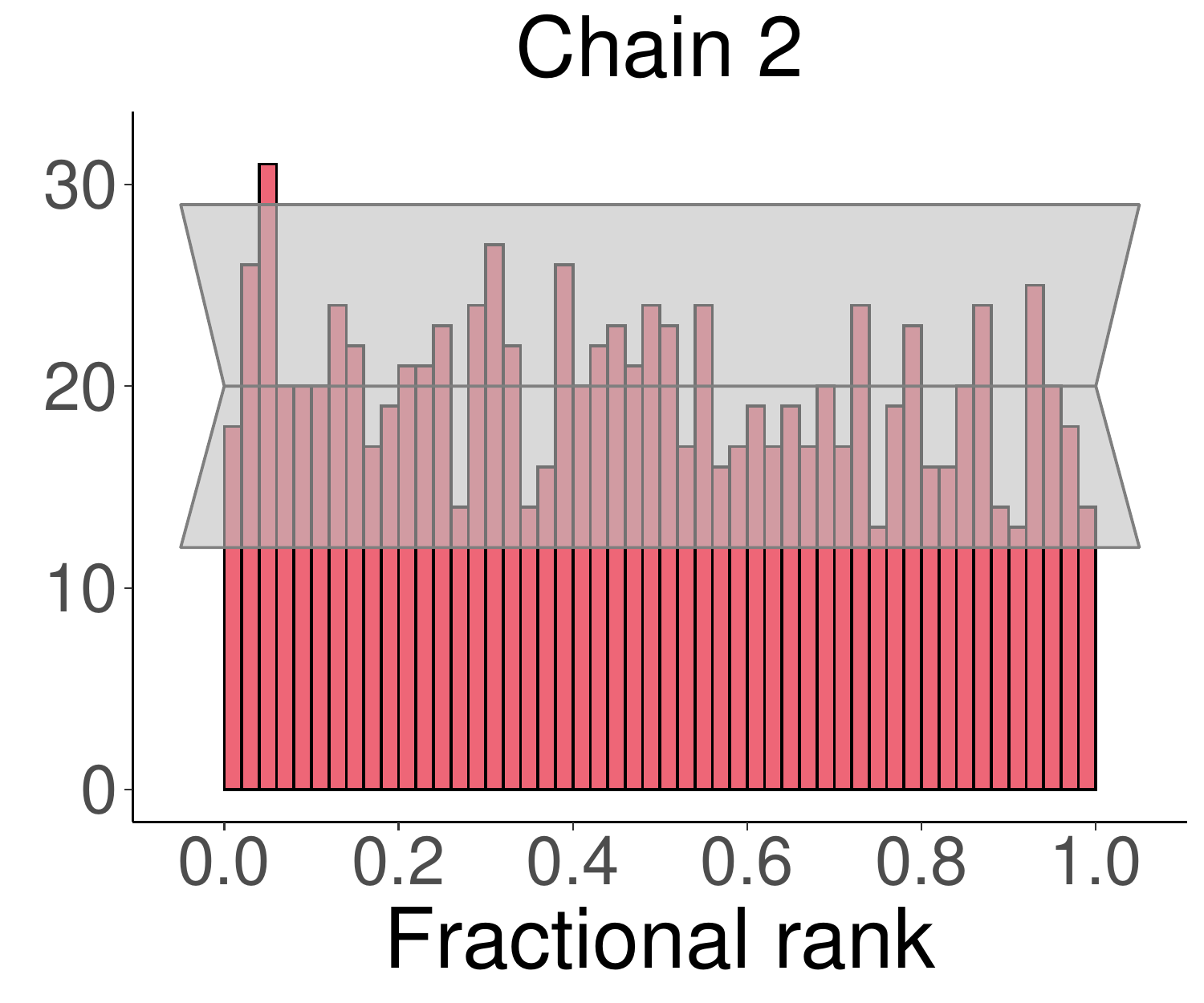}}~
    \subfloat{
    \includegraphics[width=0.28\textwidth]{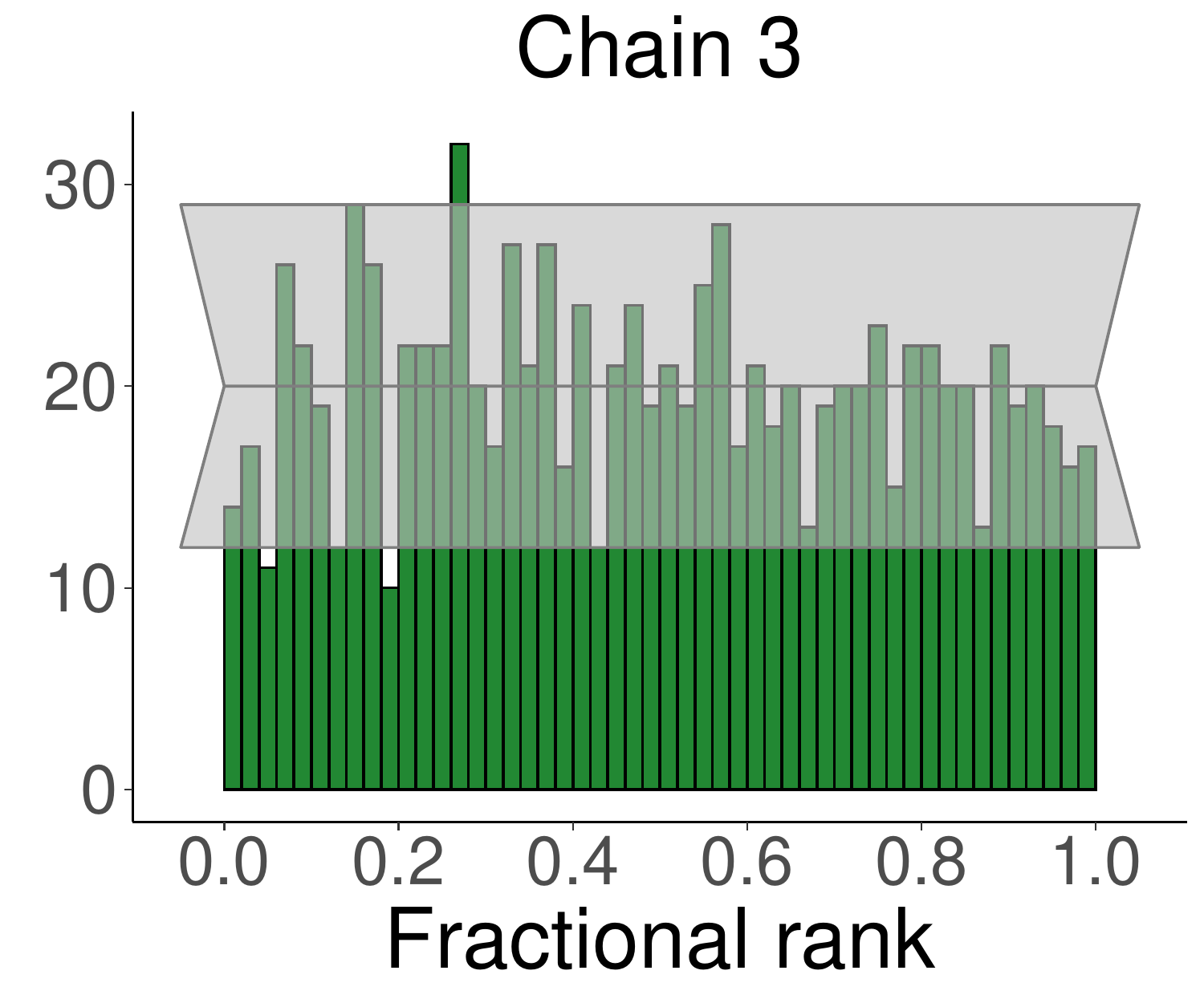}}\\
    \subfloat{
    \includegraphics[width=0.28\textwidth]{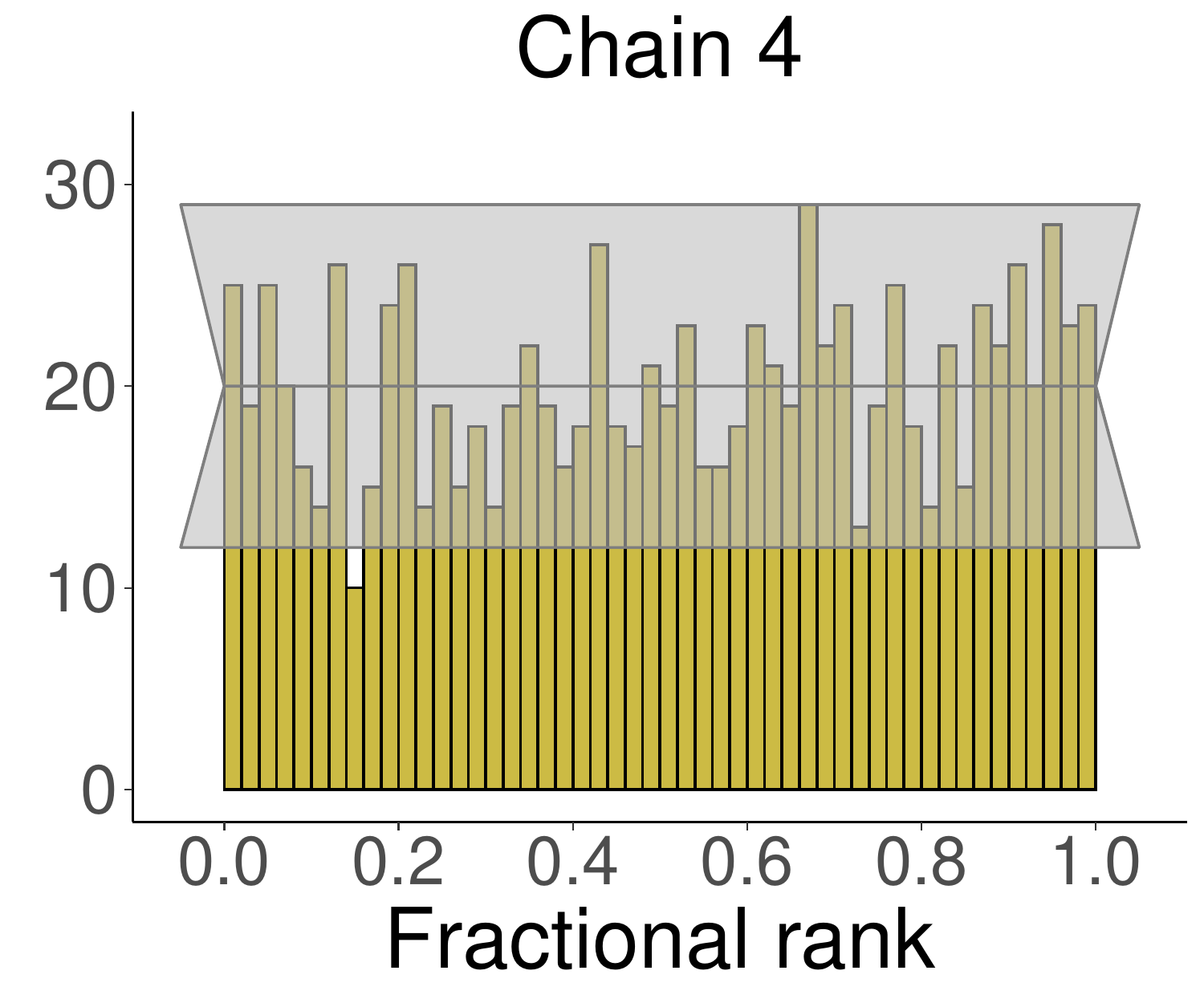}}~
    \subfloat{
    \includegraphics[width=00.28\textwidth]{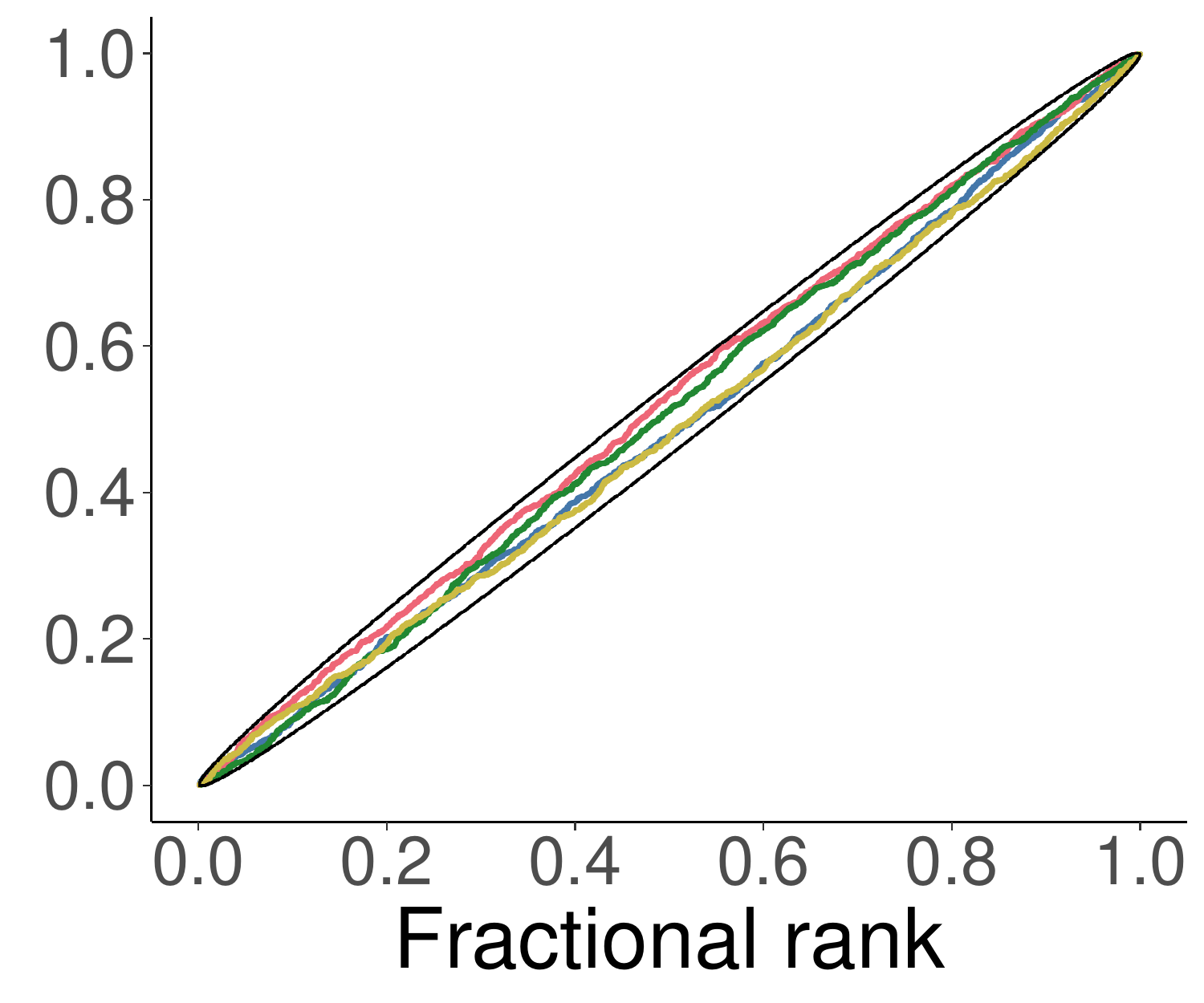}}~
    \subfloat{
    \includegraphics[width=00.28\textwidth]{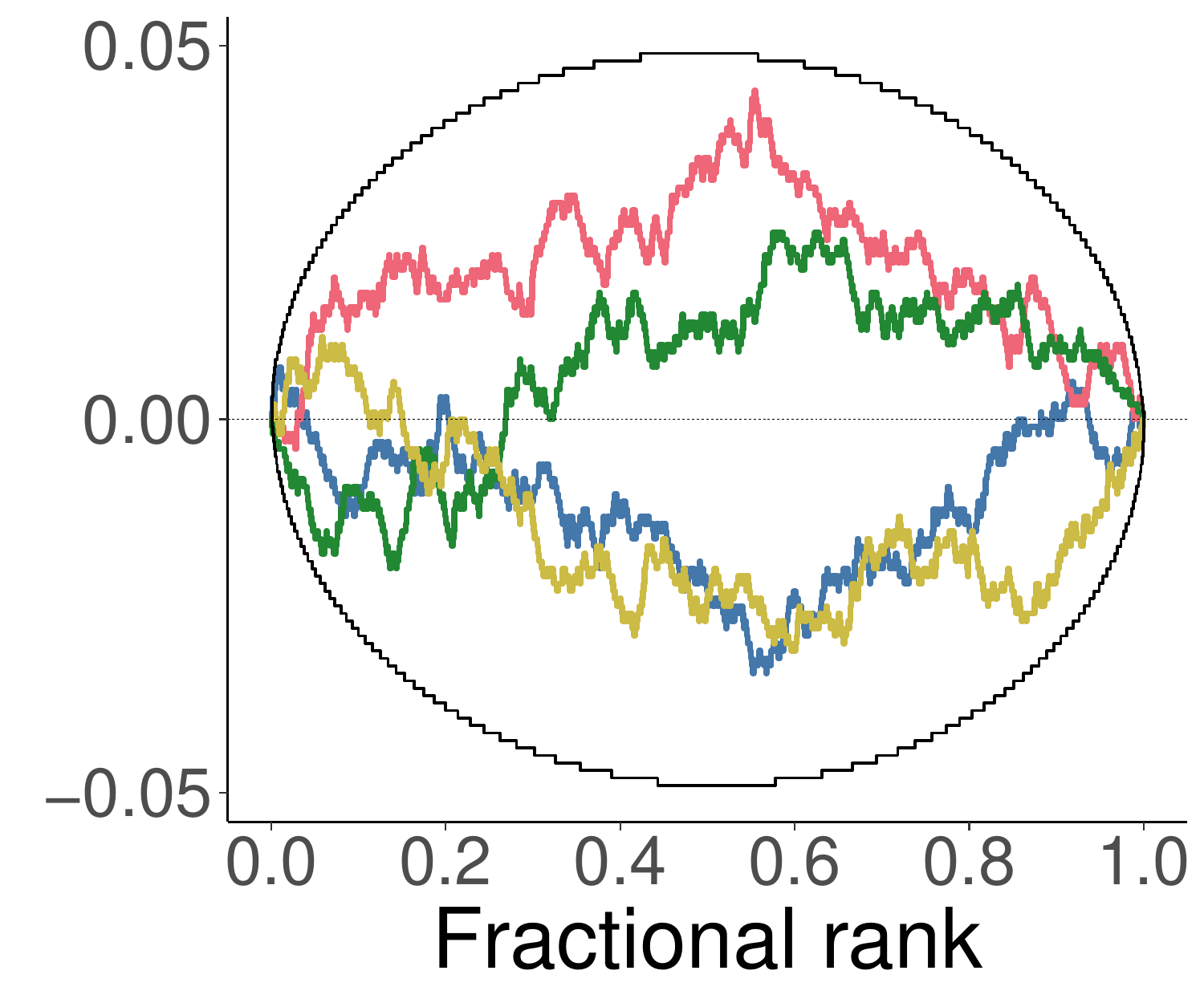}}
    \caption{Detecting model sampling issues: non-centered parameter eight schools model. When inspecting the sampling of parameter $\tau$, even when the $95\%$ confidence bands of the rank plots of chains 2 and 3 are exceeded by one bin each, the ECDF plot and the ECDF difference plot of the non-centered parameterization eight schools model indicate no mixing issues as the ECDF of the fractional ranks of each chain stay between the $95\%$ simultaneous confidence bands.}
    \label{fig:schools_ncp}
\end{figure}

As recommended in Section 22.7 of the Stan User's Guide \parencite{StanUG}, these observed sampling issues of a hierarchical model with weak likelihood contribution can often be avoided by using the non-centered parameterization $(\tilde{\theta}, \mu, \tau, \sigma)$ of the model:
\begin{align}
    \tilde{\theta}_j &\sim \normal(0, 1)\\
    \theta_j &= \mu + \tau\tilde{\theta}_j\\
    y_j &\sim \normal(\theta_j, \sigma_j)
\end{align}
In the above parameterization, the treatment effect $\theta_j$ is derived deterministically from the other parameter values and instead $\tilde{\theta}_j$ is sampled. 
To keep the models comparable, we use the same conservative sampling options for the non-centered model although this is not required to obtain well mixing chains. In Figure \ref{fig:schools_ncp}, we see an improvement in the sampling  compared to the centered parameterization, as the sample ranks are distributed approximately uniformly among the four chains implying that the chains are mixing well.

Now, both of the $\hat{R}$ diagnostics agree on convergence with the graphical test, yielding values close to $1.00$, while also the sampling efficiency issues detected in the centered parameterization model have disappeared giving samples with bulk-ESS and tail-ESS reaching 2200 and 1600 respectively.

\section{Discussion}\label{sec:discussion}
By providing a graphical test for uniformity and comparison of samples, we offer an accessible tool to be used in many parts of practical statistical workflow. %

For assessing the uniformity of a single sample, 
we recommend the optimization-based adjustment method, as it is efficient even for large sample sizes.
For comparing multiple samples, the simulation-based method is likely to be computationally more efficient than the optimization-based method. To speed up the computations, we recommend pre-computing adjusted $\gamma$ values for a set of sample size and number of samples (chains) and then interpolate (in log-log space) the adjustment as needed.

In the examples, we used empirical PIT with SBC, where the uniformity is expected by construction if the inference algorithm works correctly. PIT has also been used to compare predictive distributions. Specifically, in the LOO-PIT approach, PIT has been used to compare leave-one-out (LOO) cross-validation predictive distributions to the observations \parencite[e.g.][]{Gneiting+etal:2007,Czado2009}. Although the graphical LOO-PIT test is useful for visualization of model-data discrepancy, exact uniformity of LOO-PIT values can be expected only asymptotically given the true model. For example, if the data comes from a normal distribution and is modeled with a normal distribution with unknown mean and scale, the posterior predictive distribution is a Student's $t$ distribution that approaches normal only asymptotically. Thus use of graphical LOO-PIT tests needs further research.

We have assumed that distributions $g$ and $p$ are continuous and only the fractional rank statistics $u_i$ from Eq.~\eqref{eq: rank statistic} are discrete. Our proposed methods do not work directly if $g$ and $p$ are discrete, as values obtained through PIT are no longer uniform. Also, in the multiple sample comparison case, the rank statistics are no longer mutually distinct as ties are possible. The potential approach to handling discrete $g$ and $p$ is to use randomized or non-randomized modifications of PIT values for discrete distributions, as discussed by  \textcite{Czado2009}. However, developing proven and efficient algorithms for this purpose requires further work, which is left for future research.

\section*{Acknowledgments}

We thank the Academy of Finland (grant 298742), the Finnish Center for Artificial Intelligence, and the Technology Industries of Finland Centennial Foundation (grant 70007503; Artificial Intelligence for Research and Development) for partial support of this research. We also acknowledge the computational resources provided by the Aalto Science-IT project.

\printbibliography

\begin{appendices}
\section{Autocorrelated Samples}\label{appendix: autocorrelated samples}

In this appendix, we highlight the effect autocorrelated draws have when they are used to estimate the extreme rank statistics of the target distribution. Accounting for autocorrelation is important when inspecting the distribution of order statistics, including the PIT values in Section \ref{sec:confidence_bands} or the between chain fractional ranks in Section \ref{sec:multi_sample_test}.

\begin{figure}
    \centering
    \subfloat{
    \includegraphics[width=0.35\textwidth]{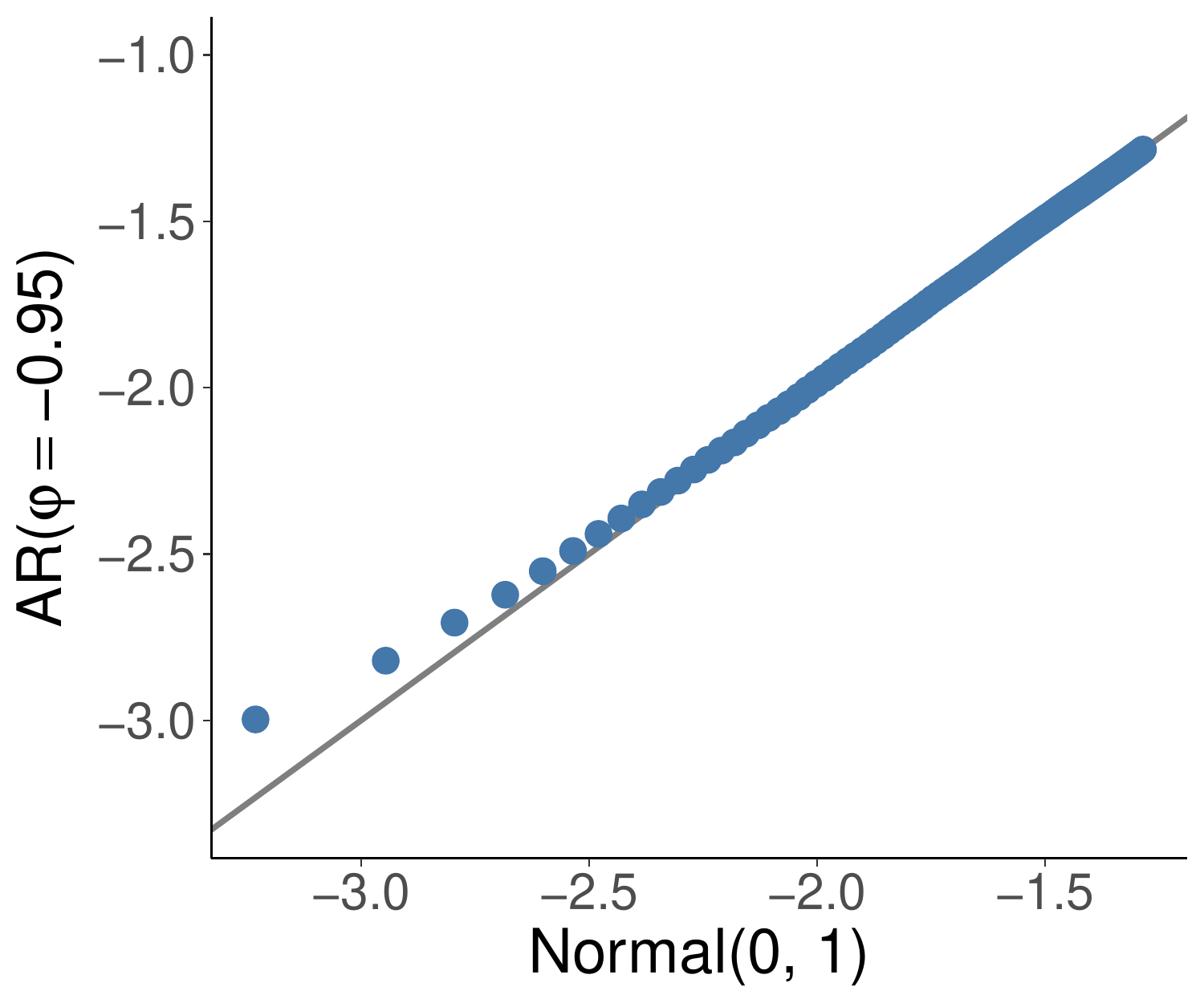}}~\hspace{0.05\textwidth}
    \subfloat{
    \includegraphics[width=0.35\textwidth]{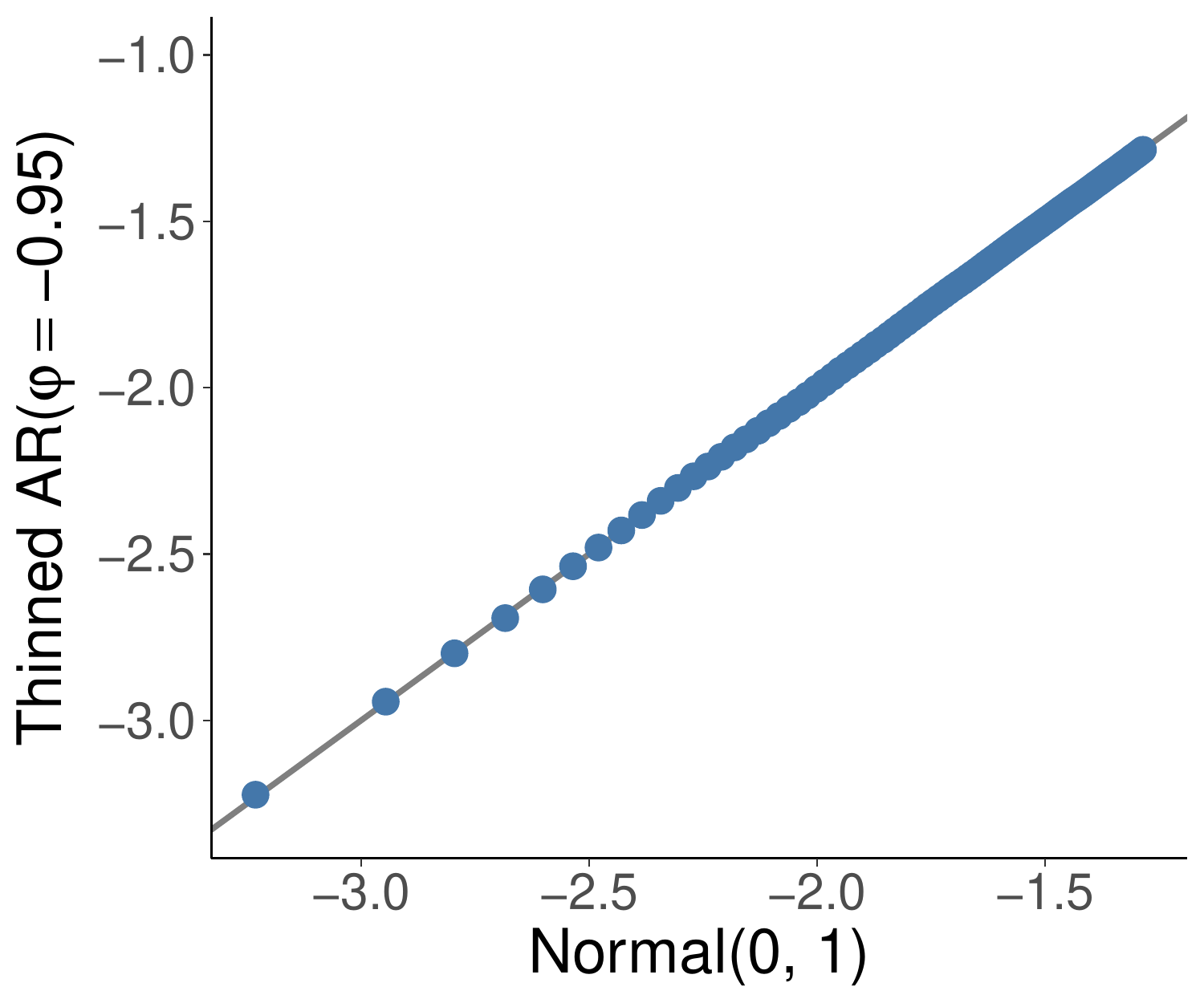}}\\
    \subfloat{
    \includegraphics[width=0.35\textwidth]{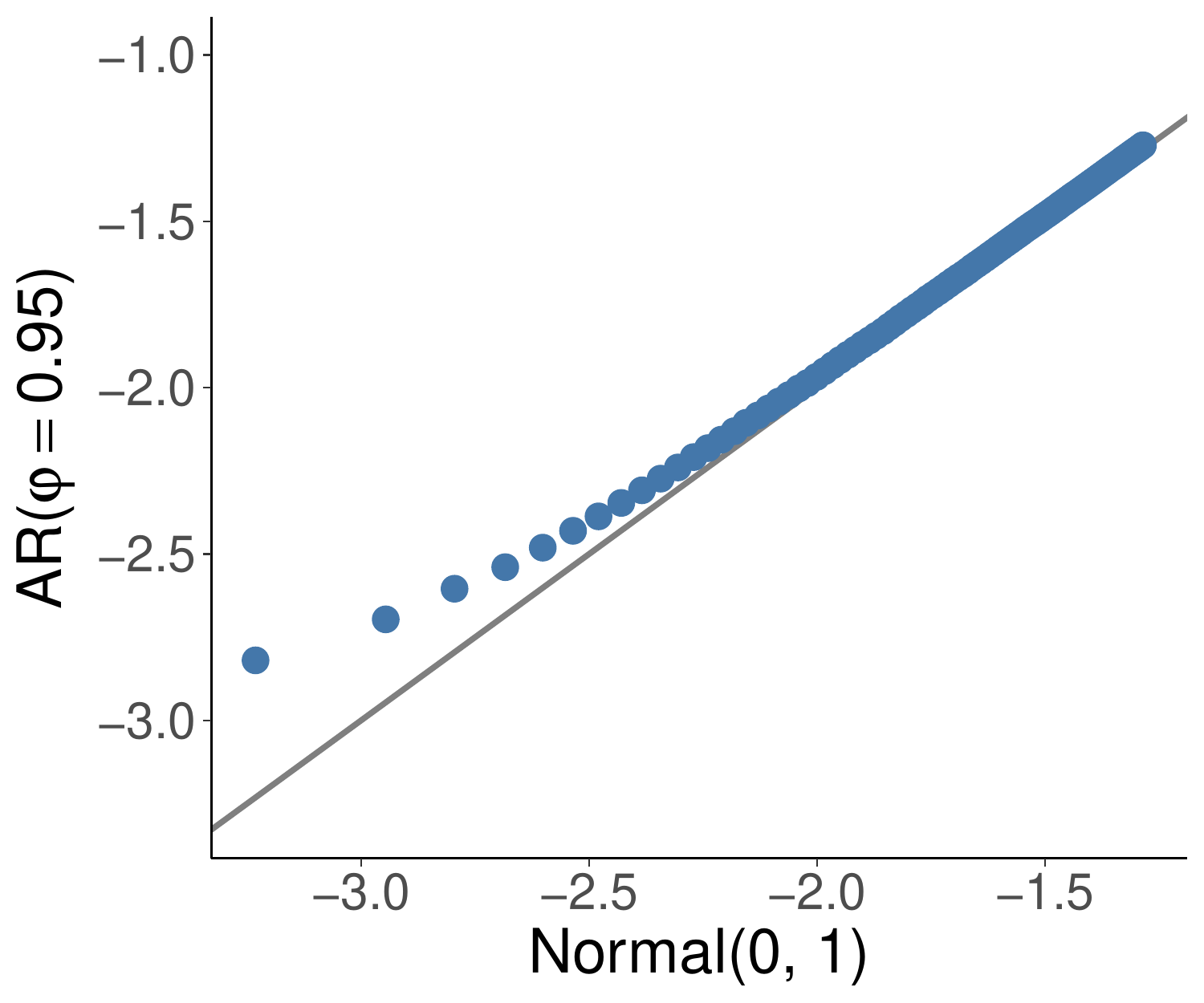}}~\hspace{0.05\textwidth}
    \subfloat{
    \includegraphics[width=0.35\textwidth]{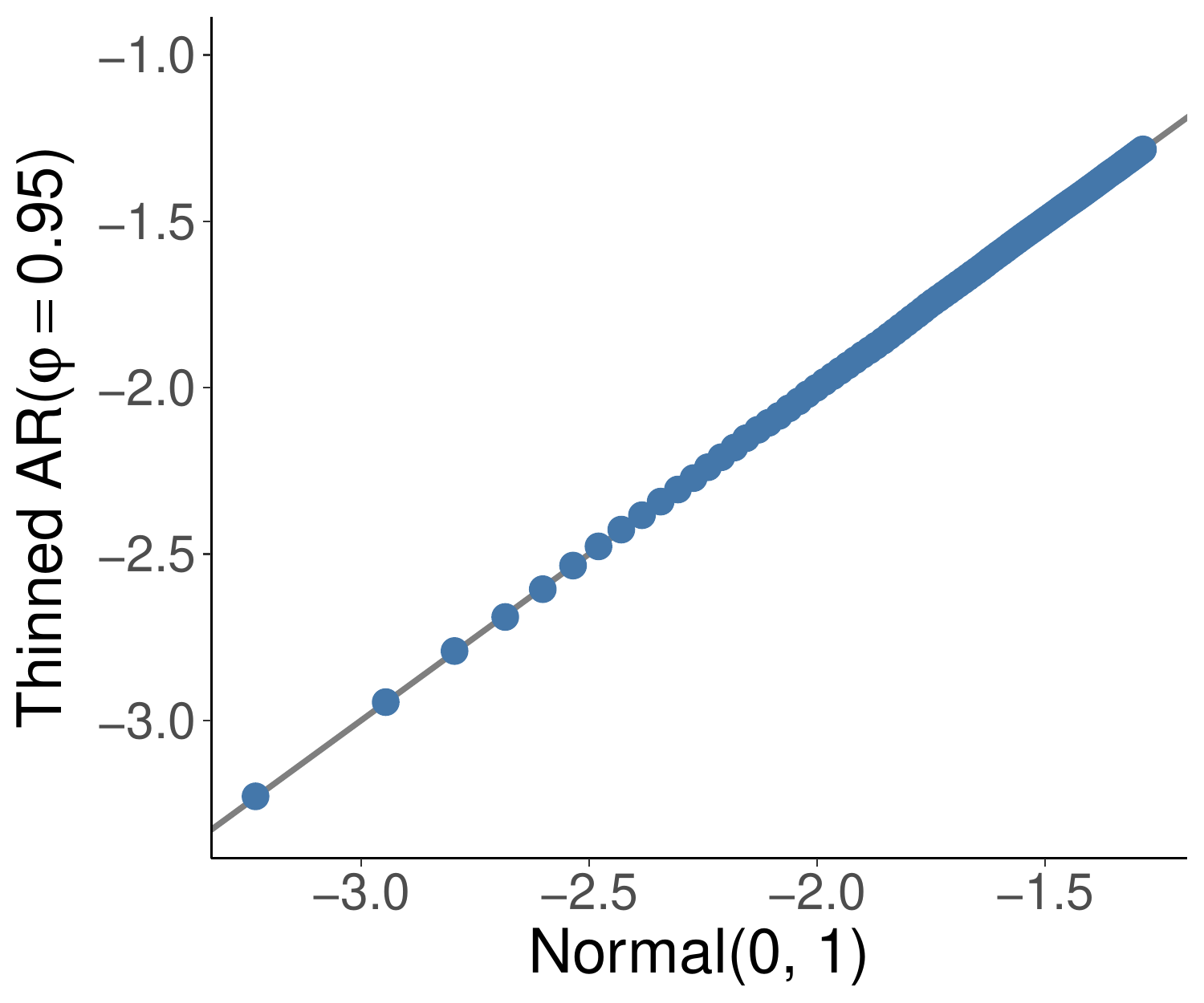}}
    \caption{The difference between the expectation of the first 100 ordered statistics of a sample of size 1000. The expectations are computed from 1000 simulations. One can see that both the AR process with strong positive autocorrelation, $\phi = 0.95$, and the process with negative autocorrelation, $\phi = -0.95$, produce samples with expected ordered statistics that are biased towards the center of the distribution. When thinned according to the tail-ESS, resulting in thinning by 18 and 7 respectively, the samples align well with the expected ordered statistics of the target distribution.}
    \label{fig:appendix_order_stats}
\end{figure}

Given finite variance, central limit theorem holds also for correlated samples and many useful expectations can be estimated with desired accuracy by increasing the sample size. However, the bias in extreme ordered statistics can be non-negligible. This manifests in the expected value of the smallest and largest order statistics of an autocorrelated sample being less extreme than expected. This phenomenon is demonstrated with AR(1) processes in Figure \ref{fig:appendix_order_stats}, which shows expected values of the 100 smallest ordered statistics computed from a sample of length 1000. The bias is smaller with less extreme ordered statistics, and, for example, estimates of $p(x<-1.5)$ or 10\% quantile in this case are likely to have negligible bias. In the uniformity test, extreme PIT estimates can have non-negligible bias, increasing the probability that ECDF steps out of the simultaneous confidence band.

A standard approach to reduce sample autocorrelation is to thin the sampled chains by keeping only every $T$ values in the sample. Below, we compare three thinning strategies.
First, the traditional approach, where $T = S / \mathrm{ESS}$, where ESS is computed for the posterior mean (without rank-normalization) \parencite{Vehtari2021}.
Second, an approach recommended by \textcite{Talts2018}, where the above ESS is computed for estimating the ECDF, $P(y < y^*)$ where $y^*$ are empirical quantiles of the sample $y$. The authors recommend using 19 quantiles $(0.05,0.1,\ldots,0.95)$ and thinning the sample based on the ESS, which would result in the largest thinning factor. This method is targeted to address differences in sampling efficiency between the distribution quantiles.
The third method, we introduce, is calculating the tail-ESS and bulk-ESS as defined by \textcite{Vehtari2021}, and picking the one resulting into the stricter thinning. This method aims to address possible differences in sampling efficiency between the 
central $90\%$ quantile and the two $5\%$ tail quantiles. R package `posterior` was used for all ESS computations.

In Figure \ref{fig:appendix_order_stats} we additionally show the first 100 order statistics of the standard normal distribution compared to AR processes thinned according to the tail-ESS, as our focus here is on the tails of the distribution.  In order to arrive at a thinned sample of equal length, an expected Tail-ESS was obtained by averaging over 10,000 simulations and the sample length was chosen accordingly to yield thinned samples of length 1000.
After the thinning, the order statistics closely match those drawn independently from the standard normal distribution.

\begin{figure}
    \centering
    \subfloat{
    \includegraphics[width=0.9\textwidth]{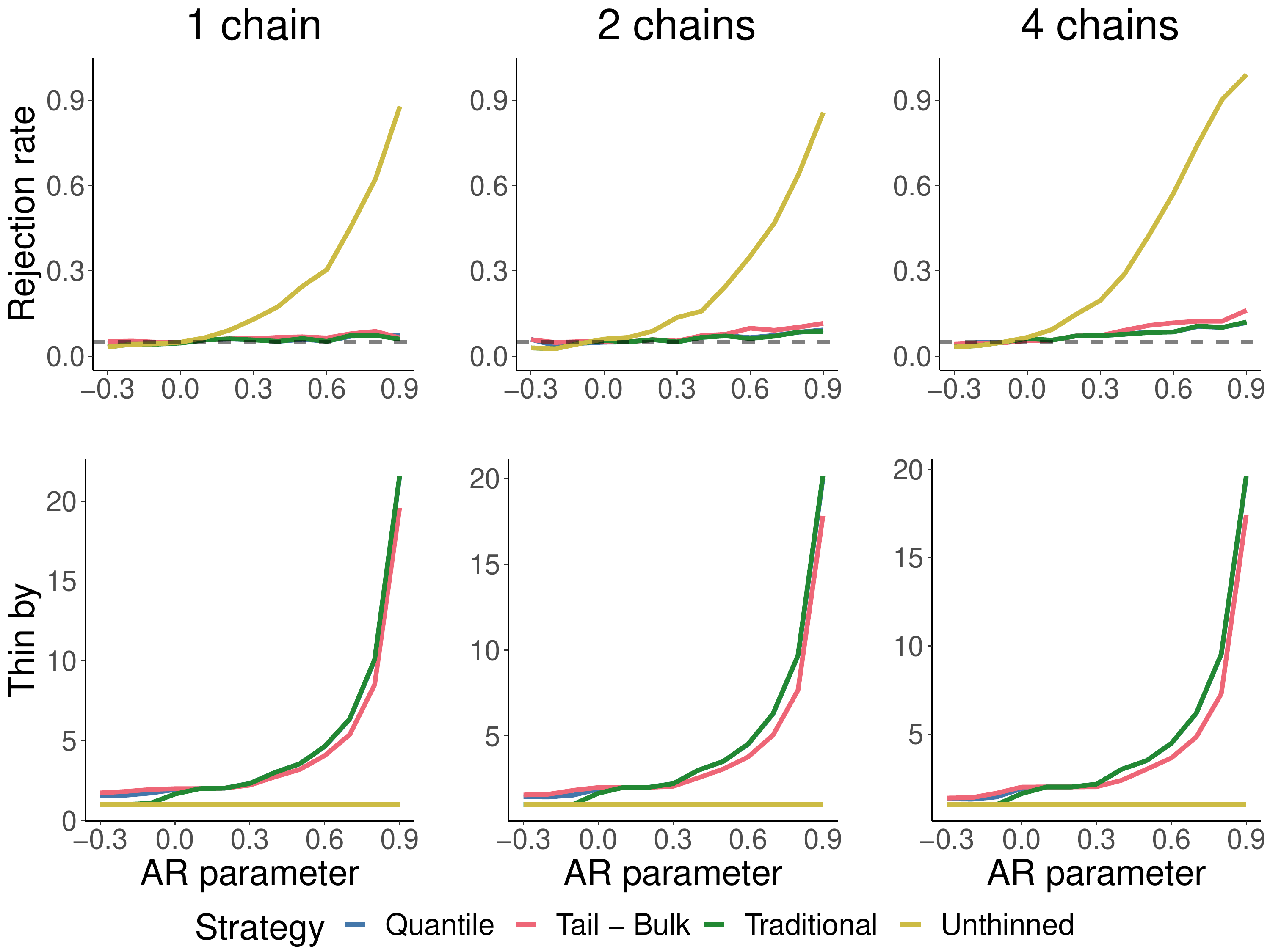}}
    \caption{The behaviour of the three thinning strategies when applied to a sample consisting of 1, 2, or 4 chains created with an AR(1) process. With AR(1) process, the three strategies agree on the recommended strategy and are fairly successful at recovering the desired confidence level, while the rejection rate of the unthinned chains grows as more autocorrelation is introduced. For large values of the AR parameter, $\phi$, the thinning amounts are quite high, which would require larger samples to reliable estimate properties of the target distribution.}
    \label{fig:appendix_thinning_results_chains}
\end{figure}

We inspected how the three above mentioned thinning strategies manage to reduce the autocorrelation which, as shown in Section \ref{subsec:autocorrelated chains}, the ECDF based test is sensitive to. In this experiment, 1, 2, and 4 chains of length 1000 were drawn from the AR(1) process with varying values of the AR parameter $\phi$. The results of this experiment are displayed in Figure \ref{fig:appendix_thinning_results_chains} and as one can see, all three of the methods produce very similar thinning recommendations, and thus also test results, managing to reduce the rejection rate near the desired $5\%$. 

If after using some default thinning approach, there are still many extreme PIT estimates, it is possible that there is still substantial autocorrelation in the sample and more careful investigation of the remaining autocorrelation is warranted. There is certainly a trade-off between the computation time and how accurately the behavior of extreme tails need to be examined. Often the major issues can be seen with less accurate computation, and natural workflow can include iterative refinement of the diagnostic accuracy.

Although thinning may be needed for uniformity test as part of SBC or PPC, when estimating quantities of interest that are not related to extreme tails, better efficiency is obtained by using all the posterior draws.

\end{appendices}

\end{document}